\def\cameraready{}
\def\arxivversion{}
\PassOptionsToPackage{table}{xcolor}
\documentclass[10pt]{article}

\usepackage[accepted]{tmlr}
\def\month{10}
\def\year{2026}
\def\openreview{\url{https://openreview.net/forum?id=iwYk6keMAM}}

\ifdefined\cameraready
  \newcommand{\red}[1]{#1}
  \newcommand{\blue}[1]{#1}
\else
  \newcommand{\red}[1]{\textcolor{red}{#1}}
  \newcommand{\blue}[1]{\textcolor{blue}{#1}}
\fi

\usepackage[utf8]{inputenc} 
\usepackage[T1]{fontenc}    
\usepackage{hyperref}       
\usepackage{url}            
\usepackage{booktabs}       
\usepackage{amsmath}        
\usepackage{amsfonts}       
\usepackage{nicefrac}       
\usepackage{microtype}
\usepackage{graphicx}
\usepackage{natbib}
\usepackage{doi}
\usepackage{tabularx}
\usepackage{array}
\usepackage{booktabs}
\usepackage[table]{xcolor}
\colorlet{covGap}{blue!10}
\usepackage{booktabs}
\usepackage{array}
\usepackage{tabularx}
\usepackage{xltabular}
\usepackage{makecell}
\usepackage{graphicx}
\usepackage{pifont}
\usepackage{multirow}
\usepackage{hhline}
\usepackage{placeins}
\usepackage[most]{tcolorbox}

\definecolor{defbg}{HTML}{FBF9F2}
\newtcolorbox{termbox}{
  colback=defbg, colframe=black!80,
  boxrule=0.7pt, arc=5pt,
  left=10pt, right=10pt, top=8pt, bottom=8pt}

\makeatletter
\@ifundefined{T1+lmr}{\input{t1lmr.fd}}{}
\makeatother
\DeclareFontShape{T1}{lmr}{bx}{sc}{<->ssub * cmr/bx/sc}{}
\DeclareFontShape{T1}{lmr}{m}{scit}{<->ssub * lmr/m/scsl}{}
\DeclareFontShape{T1}{lmr}{bx}{scit}{<->ssub * cmr/bx/sc}{}

\AtBeginEnvironment{table}{\setlength{\belowcaptionskip}{2pt}}

\definecolor{linknavy}{HTML}{1F4E79}
\ifdefined\cameraready
  \hypersetup{colorlinks=true, linkcolor=linknavy, citecolor=linknavy, urlcolor=linknavy}
\else
  \hypersetup{hidelinks}
\fi
\hypersetup{
  pdftitle={From Agent Traces to Trust: A Survey of Evidence Tracing and Execution Provenance in LLM Agents},
  pdfauthor={Yiqi Wang, Jiaqi Zhang, Zhangkai Wu, Taotao Cai, Zirui Liu, Zequn Sun, Mingkai Zheng, Manqing Dong, Yu-Xuan Qiu, Xuefei Yin, Yanming Zhu},
  pdfsubject={Transactions on Machine Learning Research},
  pdfkeywords={LLM agents, agent traces, evidence tracing, execution provenance, provenance graphs, tool-use safety, memory provenance, agent observability, runtime guardrails, trustworthy AI}}

\definecolor{covStrong}{HTML}{2F80A8}   
\definecolor{covPartial}{HTML}{8ECAE6}  
\definecolor{covLimited}{HTML}{DDF4FF}  
\definecolor{covHeader}{HTML}{CFEFFF}   
\definecolor{covRow}{HTML}{F4F6F9} 
\newcommand{\subrowrule}{\hhline{~>{\arrayrulecolor{black!20}}--->{\arrayrulecolor{black}}}}
\newcommand{\subrowruleshaded}{\hhline{>{\arrayrulecolor{covRow}}->{\arrayrulecolor{black!20}}--->{\arrayrulecolor{black}}}}

\newcolumntype{C}[1]{>{\centering\arraybackslash}m{#1}}
\newcolumntype{L}[1]{>{\raggedright\arraybackslash}m{#1}}

\newcommand{\Strong}{\cellcolor{covStrong}\textcolor{white}{\textbf{S}}}
\newcommand{\Partial}{\cellcolor{covPartial}\textcolor{black}{\textbf{P}}}
\newcommand{\Limited}{\cellcolor{covLimited}\textcolor{black}{\textbf{L}}}
\newcommand{\MemYes}{\textcolor{covStrong}{\ding{51}}}
\newcommand{\MemPartial}{\textcolor{orange!85!black}{\ding{115}}}
\newcommand{\MemLimited}{\textcolor{gray!75!black}{\ding{109}}}
\newcommand{\MemNo}{\textcolor{red!70!black}{\ding{55}}}
\newcommand{\MemAttack}{\textcolor{purple!75!black}{\ding{72}}}
\newcommand{\BenchYes}{\ding{51}}
\newcommand{\BenchPartial}{\textbullet}
\newcommand{\BenchNo}{\ding{55}}
\newcommand{\BenchTask}{T}
\newcommand{\BenchState}{S}
\def\keywordname{{\bfseries \emph{Keywords}}}%
\def\keywords#1{\par\addvspace\medskipamount{\rightskip=0pt plus1cm
\def\and{\ifhmode\unskip\nobreak\fi\ $\cdot$
}\noindent\keywordname\enspace\ignorespaces#1\par}}

\title{From Agent Traces to Trust: A Survey of Evidence Tracing and Execution Provenance in LLM Agents}

\ifdefined\arxivversion
\author{%
\name Yiqi Wang$^{1,*}$, Jiaqi Zhang$^{2}$, Zhangkai Wu$^{3}$, Taotao Cai$^{1,*}$, Zirui Liu$^{4}$, Zequn Sun$^{5}$, Mingkai Zheng$^{6}$, Manqing Dong$^{7}$, Yu-Xuan Qiu$^{8}$, Xuefei Yin$^{9}$, Yanming Zhu$^{9}$\\[6pt]
\parbox{\textwidth}{\raggedright\normalfont\small \mbox{$^{1}$University of Southern Queensland, Australia} \quad
\mbox{$^{2}$Jiangsu University, China} \quad
\mbox{$^{3}$The University of Sydney, Australia} \quad
\mbox{$^{4}$Peking University, China} \quad
\mbox{$^{5}$Nanjing University, China} \quad
\mbox{$^{6}$Southern University of Science and Technology, China} \quad
\mbox{$^{7}$The University of New South Wales, Australia} \quad
\mbox{$^{8}$Beijing Institute of Technology, Zhuhai, China} \quad
\mbox{$^{9}$Griffith University, Australia}\\[4pt]
$^{*}$\texttt{yiqi.wang.jennie@gmail.com}, \texttt{taotao.cai@unisq.edu.au}}
}
\makeatletter
\def\@maketitle{\vbox{\hsize\textwidth
{\LARGE\bf\sffamily \@title\par}\vskip \aftertitskip
\@startauthor \@author \@endauthor
\vskip 0.3in minus 0.1in}}
\makeatother
\else
\author{%
\name Yiqi Wang \email Yiqi.Wang@unisq.edu.au \\
\addr University of Southern Queensland, Australia
\AND
\name Jiaqi Zhang \email 3230602065@stmail.ujs.edu.cn \\
\addr Jiangsu University, China
\AND
\name Zhangkai Wu \email zhangkai.wu@sydney.edu.au \\
\addr The University of Sydney, Australia
\AND
\name Taotao Cai\thanks{Corresponding author.} \email taotao.cai@unisq.edu.au \\
\addr University of Southern Queensland, Australia
\AND
\name Zirui Liu \email zirui.liu@pku.edu.cn \\
\addr Peking University, China
\AND
\name Zequn Sun \email sunzq@nju.edu.cn \\
\addr Nanjing University, China
\AND
\name Mingkai Zheng \email zhengmk@sustech.edu.cn \\
\addr Southern University of Science and Technology, China
\AND
\name Manqing Dong \email manqing.dong@unswalumni.com \\
\addr The University of New South Wales, Australia
\AND
\name Yu-Xuan Qiu \email qiu@bit.edu.cn \\
\addr Beijing Institute of Technology, Zhuhai, China
\AND
\name Xuefei Yin \email x.yin@griffith.edu.au \\
\addr Griffith University, Australia
\AND
\name Yanming Zhu \email yanming.zhu@griffith.edu.au \\
\addr Griffith University, Australia
}

\fi

\date{}

\begin{document}
\maketitle

\begin{abstract}
\blue{Large language model (LLM)-based agents are moving beyond text generation to planning, tool use, retrieval, memory access, environmental interaction, and multi-agent collaboration. Because their outputs depend on information and execution steps across an agent run, their execution processes are difficult to verify, debug, and audit. Final-answer evaluation alone cannot resolve these difficulties: even a correct answer does not reveal which evidence supported its claims, whether tool calls were justified, how memory shaped decisions, or where execution failures originated. Relevant research addresses these questions across evidence attribution, tool-use safety, agent memory, and observability, but these areas emphasize different parts of the execution process. This survey brings them into a common framework through \emph{execution provenance}, a typed graph of an agent run, and \emph{evidence tracing}, its projection onto evidence-support relations. We define what this projection retains, discards, and adds, distinguishing evidence that \emph{supports} a conclusion from execution that \emph{influences} it. We introduce a six-dimensional taxonomy covering trace sources, evidence and execution units, provenance relations, tracing granularity and timing, representation forms, and trust functions. We complement this taxonomy with operational definitions of nine provenance relations and two cross-cutting attributes: provenance origin, describing how records are produced, and trace assurance, describing their protection against modification. Applying this framework to the surveyed systems and benchmarks reveals three gaps: limited integration of fine-grained provenance with multiple trust functions, no benchmark in our catalog that evaluates the complete provenance chain, and little reporting of tamper-evident or attested traces. These findings highlight why final-answer correctness alone is insufficient to establish process-level trust and motivate research on unified trace schemas, semantic provenance, memory and multi-agent lineage, provenance-aware safety and recovery, realistic benchmarks, trace assurance, and privacy-aware auditing. Related resources are collected and continuously updated at \url{https://github.com/xiaoqi-7/Agent-Tracing-Survey}.}
\end{abstract}


\section{Introduction}
\label{sec:introduction}
Large language model (LLM)-based agents are emerging as a powerful paradigm for complex task execution, moving beyond standalone text generation toward interactive execution pipelines that support reasoning, tool use, retrieval, memory access, environment interaction, and multi-agent collaboration~\citep{schick2023toolformer, yao2023react, shinn2023reflexion}. As these capabilities expand, recent systems and benchmarks show that LLM agents increasingly operate across tools, external knowledge sources, code execution, web or GUI environments, and multi-agent workflows, creating complex execution contexts in which many intermediate sources, actions, and observations may jointly shape the final output~\citep{qin2024toolllm, liu2024agentbench, zhou2024webarena, yao2025taubench}.

This creates a process-level accountability gap in LLM agent systems. Final-answer accuracy evaluates only the endpoint of an execution; it does not explain how an output was produced, which evidence supported each claim, whether tool calls were justified, whether memory items were relevant and trustworthy, or which execution step caused a failure. Recent studies make this limitation concrete: tool-safety benchmarks show that risks can arise from intermediate tool-use decisions, adversarial tool interactions, realistic tool environments, and MCP-based workflows~\citep{ruan2024toolemu,debenedetti2024agentdojo,vijayvargiya2026openagentsafety,zong2025mcpsafetybench}, while trace-oriented and multi-agent failure studies show that errors often require localizing responsible steps, components, agents, or error modes within long executions~\citep{deshpande2025trail,cemri2025mast,zhang2025agentracer,kong2025aegis}. These findings suggest that trustworthy agent systems require mechanisms that can record, connect, and reason over the evidence and execution steps underlying agent behavior.

\blue{Relevant work exists across several research areas, but each emphasizes different relations: attribution links claims to supporting sources, usually treating each source as given~\citep{gao2023alce,schreieder2025attribution}; observability records execution events~\citep{dong2024agentops,alsayyad2026agenttrace}; information-flow defenses track the propagation of untrusted data~\citep{debenedetti2025camel,costa2025fides}; and memory systems manage what is stored, updated, and retrieved across interactions~\citep{packer2023memgpt,zhong2024memorybank}. These perspectives address complementary questions, yet their connections are often left implicit. Connecting them requires a framework that distinguishes what happened, what influenced an outcome, and what evidence supports it.}

\begin{figure*}[t]
\centering
\includegraphics[width=0.95\textwidth]{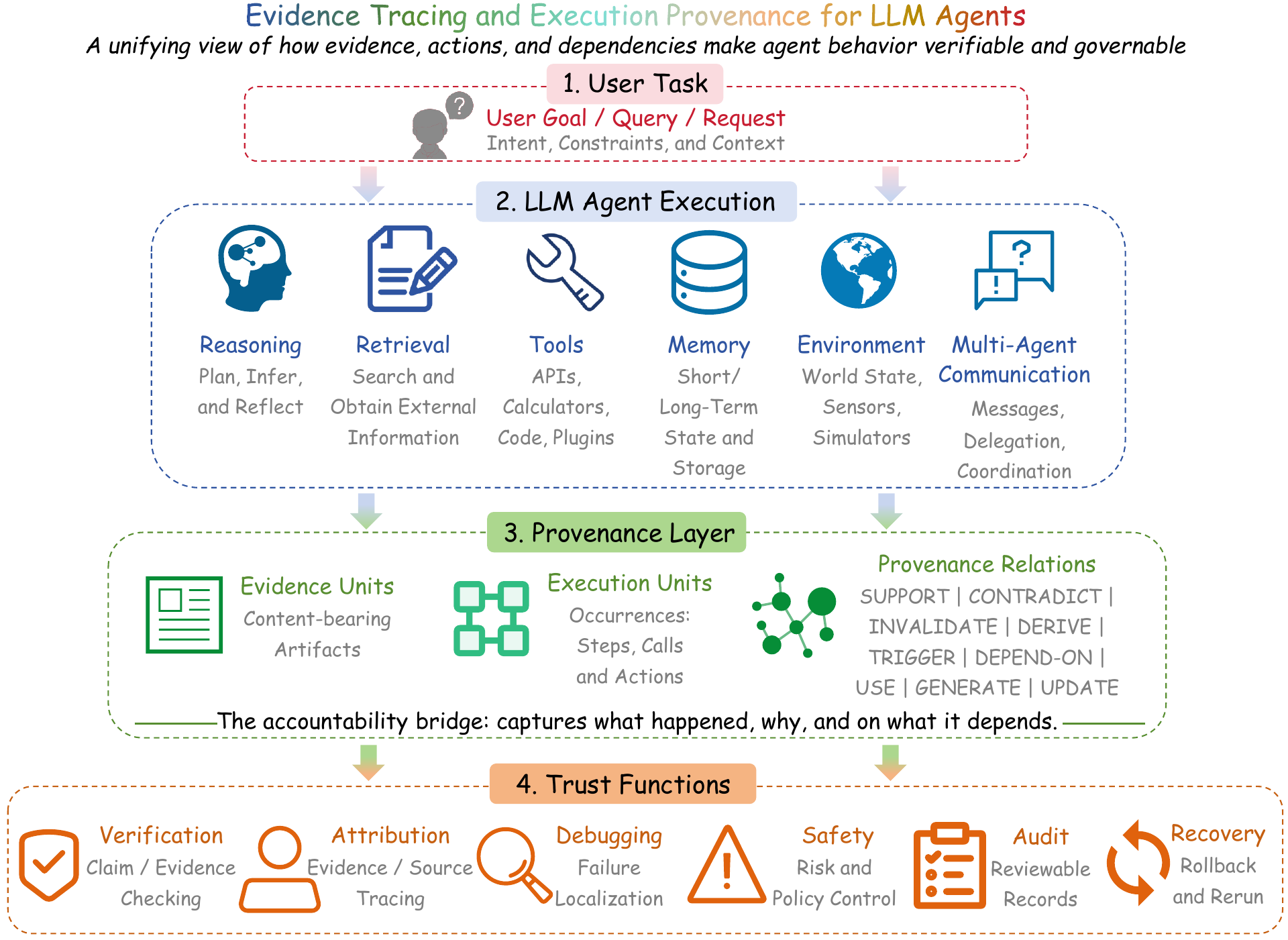}
\caption{Overview of the proposed provenance framework. A user task initiates a heterogeneous agent execution involving reasoning, retrieval, tool use, memory, environment interaction, and multi-agent communication. The provenance layer records evidence units, execution units, and typed relations  produced during this process, connecting agent execution to downstream trust functions such as  verification, attribution, debugging, safety, audit, and recovery.}
\label{fig:overview}
\end{figure*}

In this survey, we examine this need through the lens of \emph{evidence tracing} and \emph{execution provenance} \blue{(Figure~\ref{fig:overview})}.\footnote{This perspective is informed by provenance and trace modeling in software and distributed systems, including W3C PROV-DM for representing entities, activities, agents, and provenance relations~\red{\citep{w3c2013provdm}}, OpenTelemetry for distributed execution traces~\red{\citep{opentelemetry2026traces}}, and recent agent observability systems such as AgentOps and AgentTrace~\red{\citep{dong2024agentops,alsayyad2026agenttrace}}.}

\noindent This perspective starts from \textbf{agent traces}: the recorded artifacts generated or consumed during an agent run, including instructions, retrieval queries,  retrieved documents, tool calls, tool outputs, memory operations, observations, intermediate claims,  inter-agent messages,  actions, and final responses. Traceability makes these artifacts inspectable, while provenance goes further by modeling the typed relations among them.

\noindent \colorbox{covGap}{%
\parbox{0.94\textwidth}{%
\textbf{Execution provenance.}
A structured and typed representation of an agent run, including evidence units (the artifacts, such as \blue{user queries,} retrieved documents, observations, tool outputs, memory items, generated tool parameters, intermediate claims, inter-agent messages, and final outputs); execution units (the occurrences, such as reasoning steps, retrieval calls, tool invocations, memory accesses, environment actions, and delegation steps); and the semantic, derivational, causal, and procedural relations among them.}}
\vspace{1mm}
\noindent \colorbox{covGap}{%
\parbox{0.94\textwidth}{%
\textbf{Evidence tracing.}
A view of execution provenance that records support, contradiction, invalidation, and content derivation for claims, decisions, and actions. It excludes causal and procedural edges and adds \emph{candidate} links for qualifying content-flow paths where no \textsc{Support} or \textsc{Invalidate} relation is recorded between the evidence and the claim. Candidate links indicate possible content contribution, not established support.}}

\blue{We represent execution provenance as a graph whose nodes are evidence and execution units and whose typed edges record their relations. This graph is the \emph{process view} of an agent run; evidence tracing is its \emph{support view}, formalized as a projection in Appendix~\ref{app:projection}. The two views are complementary rather than interchangeable. Different executions can share the same support view, so evidence tracing alone cannot reconstruct how an agent ran. Conversely, recording execution dependencies does not establish evidential support: a tool output can influence a decision without justifying it, and a false memory item can cause an action while supporting nothing. This distinction motivates reviewing evidence attribution alongside tool-use provenance, memory lineage, and failure diagnosis within a common framework.}

\noindent\textbf{Survey scope.}
This survey focuses on works that make LLM-agent behavior traceable beyond final outputs. We include methods, systems, benchmarks, and evaluation protocols when they record, represent, attribute, inspect, constrain, or diagnose the information and execution steps that shape agent behavior. We discuss RAG, memory, graph-based retrieval, tool-use safety, and observability not as standalone research areas, but as settings in which evidence tracing and execution provenance become necessary. The organizing principle is provenance-centered: what should be traced, how trace units are connected, when tracing occurs, how traces are represented, and which trust functions they support.

 \red{Section~\ref{subsec:methodology} states the selection protocol, the inclusion and exclusion criteria, and the resulting corpus. }In summary, this survey makes the following contributions:

\begin{itemize}

\item We formulate evidence tracing and execution provenance as a unified process-level accountability framework for trustworthy LLM agents,  \red{and give a formal definition of the }execution-provenance graph  \red{together with the projection that produces }its evidence-support  \red{view, including what the projection discards and what it adds}.

\item We introduce a six-dimensional taxonomy covering trace sources, evidence and execution units, provenance relations, tracing granularity and timing, representation forms, and trust functions\red{, with two cross-cutting attributes for provenance origin and trace assurance, and we give each of the nine provenance relations an operational definition that separates semantic support from causal and procedural influence}.

\item We review existing methods \blue{by research thread}\red{: provenance representation and evidence attribution (Section~\ref{sec:representation})}, tool-use provenance  \red{and runtime guardrails (Section~\ref{sec:tool-use})}, provenance-bearing memory  \red{(Section~\ref{sec:memory}), and observability }and failure diagnosis\red{, which cut across all three and are assessed in Section~\ref{sec:benchmarks}}.

\item We map benchmarks, datasets, and metrics to provenance-related capabilities, showing how evaluation can move beyond final-answer correctness toward trace availability, evidence support, tool-use safety, memory lineage, observability, and recovery.

\item \blue{We identify three gaps that recur across threads: limited integration of fine-grained provenance with multiple trust functions in the reviewed systems, no benchmark in our catalog that evaluates the complete provenance chain, and little reporting of tamper-evident or attested traces.}

\item We identify open challenges for provenance-aware LLM agents, including unified trace schemas, semantic provenance, \red{memory and multi-agent provenance, }provenance-aware safety, realistic execution-trace benchmarks, recovery-oriented evaluation,  \red{trace assurance and attested provenance, and }privacy-aware audit infrastructure.

\end{itemize}

\subsection{\red{Survey Methodology}}
\label{subsec:methodology}

\red{This is a }\emph{\red{structured curated review}} \red{rather than an exhaustive systematic review}\blue{; the protocol below defines the corpus and the limits of its coverage.}\red{
}

\noindent\textbf{\red{Search and screening.}}
\red{The corpus was assembled by }\emph{\red{seeded snowballing}} \blue{rather than by keyword search over bibliographic databases}\red{. We began from a seed set of foundational and canonical works spanning the areas in scope (classical provenance modeling and information-flow security, retrieval attribution, tool-use safety, memory-augmented agents, agent observability, and trace-based debugging), and expanded it by backward and forward citation following, both from each retained paper and from the benchmarks it evaluates on, iterating until no newly encountered work satisfied the inclusion criteria below. Screening proceeded in two ordered stages, and a candidate had to pass both. }\emph{\red{Stage 1 (title and abstract)}}\red{: retain the candidate if it plausibly satisfies at least one inclusion criterion below; the exclusions for general surveys and for work evaluated only by final-answer accuracy are applied here. }\emph{\red{Stage 2 (full text)}}\red{: retain only if the work demonstrably records, relates, constrains, or evaluates an intermediate execution artifact, rather than merely noting that agents produce traces; the remaining exclusions are applied here. Screening was performed by the authors jointly and disagreements were resolved by discussion; we did not use a second independent coder, which is a limitation of the procedure.
}

\noindent\textbf{\red{Seed set.}}
\red{Because the reach of the closure is fixed by its seeds, we }\blue{list all fifteen, so that the expansion can be re-run and its boundary checked}\red{.\footnote{\red{The seeds, by area. }\emph{\red{Provenance modeling and information-flow security}}\red{: W3C PROV-DM~\citep{w3c2013provdm}, Saltzer and Schroeder~\citep{saltzer1975protection}, the lattice model of secure information flow~\citep{denning1976lattice}, and language-based information-flow security~\citep{sabelfeld2003language}; }\emph{\red{retrieval attribution}}\red{: FEVER~\citep{thorne2018fever}, ALCE~\citep{gao2023alce}, and FActScore~\citep{min2023factscore}; }\emph{\red{tool-use safety}}\red{: ToolEmu~\citep{ruan2024toolemu} and AgentDojo~\citep{debenedetti2024agentdojo}; }\emph{\red{memory-augmented agents}}\red{: MemGPT~\citep{packer2023memgpt} and Generative Agents~\citep{park2023generative}; }\emph{\red{agent observability}}\red{: OpenTelemetry~\citep{opentelemetry2026traces} and AgentOps~\citep{dong2024agentops}; and }\emph{\red{trace-based debugging}}\red{: TRAIL~\citep{deshpande2025trail} and MAST~\citep{cemri2025mast}.}} Expansion ran backward and forward from each seed and from every work subsequently retained.
}

\red{This procedure fixes the boundary of the corpus by the inclusion criteria and the snowballing closure, not by a query. Its reach therefore depends on the seed set, which is the principal reason we describe this review as curated rather than systematic}\blue{; the limitations below follow from this choice.}\red{
}

\noindent\textbf{\red{Inclusion criteria.}} \red{A work is included if it satisfies at least one of the following, each of which is a property of the artifact rather than a judgment about quality:
}\begin{enumerate}
\itemsep2pt
\item \red{it }\emph{\red{records}} \red{intermediate execution artifacts, that is, it defines or releases a trace format, schema, or log covering retrieval, tool calls, memory operations, observations, or inter-agent messages;
}\item \red{it }\emph{\red{relates}} \red{such artifacts, that is, it introduces edges among them, including claim--evidence support, tool-argument dependency, memory lineage, or failure attribution;
}\item \red{it }\emph{\red{constrains}} \red{execution using such artifacts at runtime, including taint tracking, information-flow control, permission enforcement, or policy checking;
}\item \red{it }\emph{\red{evaluates}} \red{such artifacts, that is, it provides a benchmark, dataset, or metric whose ground truth concerns intermediate steps or evidence rather than only final answers; or
}\item \red{it supplies a }\emph{\red{foundation}} \red{that the above build on directly, namely classical provenance modeling, information-flow security, and fact verification.
}\end{enumerate}

\noindent\textbf{\red{Exclusion criteria.}} \red{We exclude work that improves agent capability without exposing process-level structure: prompting and reasoning methods evaluated only by final-answer accuracy; retrieval and re-ranking work that does not attribute claims to sources; agent architectures without released traces or trace schemas; and safety work operating only on model inputs and outputs rather than on execution artifacts. We also exclude general surveys of RAG, agents, or LLM safety except where they define trace or provenance structure, since our unit of analysis is the execution artifact rather than the task.
}

\noindent\textbf{\red{Resulting corpus.}}
\red{Starting from the fifteen seeds, the protocol yields the 71 works that constitute the reviewed corpus: 13 from 2023, 16 from 2024, \blue{13} from 2025, and \blue{24} from 2026, together with five earlier foundations in provenance modeling, information-flow security, and fact verification~\citep{w3c2013provdm,thorne2018fever,saltzer1975protection,denning1976lattice,sabelfeld2003language}. The corpus mixes \blue{51} papers published at peer-reviewed venues, \blue{17} unrefereed preprints, and 3 standards or specifications; the preprint fraction follows directly from the topic's recency. \red{Venue information was verified against the published record rather than taken from first posting}\blue{, and a work is counted as refereed only when its acceptance at a peer-reviewed venue is publicly recorded; a submission under review counts as a preprint.} Twenty-six of the cited works are benchmarks or datasets, cataloged in Table~\ref{tab:benchmarks-provenance} as 25 rows: three rows name two closely related artifacts each, and one of those three, AgenTracer/Aegis, covers two separately cited works. The per-benchmark coding in that catalog is ours, and the family-level coverage ratings of Section~\ref{sec:benchmarks} are computed from it by the rule stated in Figure~\ref{fig:benchmark-coverage-heatmap}.} \blue{Appendix~\ref{app:annotation-protocol} states the annotation protocol behind these per-benchmark ratings and the memory-capability ratings of Table~\ref{tab:memory-provenance}.} \red{By review thread, }\blue{the citation counts are} \red{14 works cited in the representation and attribution review of Section~\ref{sec:representation}, 28 in the tool-use review of Section~\ref{sec:tool-use}, 27 in the memory review of Section~\ref{sec:memory}, and 42 in the benchmark and metric review of Section~\ref{sec:benchmarks}; the totals overlap because many works are cited in more than one thread.
}

\noindent\textbf{\blue{How the main findings were established.}}
\blue{Two of the survey's main findings concern coverage, and they rest on different evidence. The first, that the reviewed systems integrate fine-grained provenance representation with multiple trust functions only to a limited extent, is an observation drawn from Table~\ref{tab:systems-taxonomy} together with the system descriptions of Sections~\ref{sec:representation}--\ref{sec:memory}; because that table records each system's primary trust function only, it supports a statement about limited integration rather than an absolute negative. The second, that no benchmark in our catalog evaluates the complete provenance chain, is computed from Table~\ref{tab:benchmarks-provenance} by a fixed rule and re-derived cell by cell in Appendix~\ref{app:heatmap-derivation}, so each rating can be traced to a specific benchmark and column. Both findings concern the works reachable from the seeds, not the literature as a whole.}\red{
}

\noindent\textbf{\red{Publication status of the corpus.}}
\red{Because the area is young and much of it circulates as preprints, we record }\blue{the publication status of every work.} \red{Every entry in Tables~\ref{tab:systems-taxonomy}, \ref{tab:tool-use-provenance}, \ref{tab:memory-provenance} and~\ref{tab:benchmarks-provenance} that had not been peer reviewed at the time of writing is marked \textsuperscript{\dag} in place, so that confidence can be weighed while reading the row rather than by consulting this section. The exposure is uneven: 6 of 27 rows in Table~\ref{tab:systems-taxonomy}, 8 of the 20 systems named in Table~\ref{tab:tool-use-provenance}, 4 of 11 rows in Table~\ref{tab:memory-provenance}, and \blue{3} of 25 rows in Table~\ref{tab:benchmarks-provenance}. Across the whole corpus, \blue{17} of the 71 works are unrefereed preprints, one of them a non-archival workshop poster~\citep{alsayyad2026agenttrace}, while \blue{51} are peer-reviewed papers and 3 are standards or specifications.
}

\red{Two consequences follow. First, the concentration matters more than the count: the marked rows cluster in runtime enforcement, information-flow auditing, and observability (AgentTrace~\citep{alsayyad2026agenttrace}, Agent-Sentry~\citep{sequeira2026agentsentry}, NeuroTaint~\citep{cai2026neurotaint}, Fides~\citep{costa2025fides}, AgentOps~\citep{dong2024agentops}), which is where the survey identifies its sharpest gaps, so conclusions about those areas deserve more discount than conclusions about retrieval attribution, where the underlying work is refereed. Second, }\emph{\red{no claim in this survey rests on a single unrefereed work}}\red{: each marked system is grouped with refereed work making the same structural point, such as PROV-AGENT~\citep{souza2025provagent} and CaMeL~\citep{debenedetti2025camel}, so discounting a marked row costs an instance rather than a claim. Section~\ref{subsec:trace-assurance}, whose supporting works~\citep{ojewale2026audittrails,nian2026auditableagents,zheng2026acrfence} are also preprints, states a gap rather than a result.
}

\noindent\textbf{\red{Limitations of the protocol.}} \red{Four limitations follow from the above and bound how the survey should be read. Enumerating the seed set makes the closure re-runnable, but not reproducible in the sense a database query would be, and we report no screening yield counts, because screening was concurrent with reading rather than logged prospectively. The corpus is curated rather than exhaustive, so absence from our tables is not evidence that no such work exists. Screening and coding were performed by the authors without an independent second coder, so the coverage judgments carry unquantified subjectivity and we report no inter-coder agreement. Finally, the area is moving quickly enough that a substantial part of the corpus is not yet peer reviewed.
}

\subsection{\red{Relation to Existing Surveys}}
\label{subsec:related-surveys}

\red{Most of the threads reviewed here have been surveyed before, and the contribution of this survey is not to cover works those surveys miss but to organize them by a different unit of analysis. Existing surveys take the }\emph{\red{task}} \red{as the unit (how well does attribution work, how well does memory work) and are therefore organized by method family within one thread. We take the }\emph{\red{execution artifact}} \red{as the unit, which cuts across threads, and the consequence is that questions invisible within a thread become statable across threads.}

\red{The closest work by subject is the survey of attribution, citation, and quotation in evidence-based text generation~\citep{schreieder2025attribution}, which covers claim--source linking in far more depth than Section~\ref{sec:representation} does, but treats the source as given rather than as something an agent produced through a tool call or wrote into memory; in our terms it studies }\textsc{\red{Support}} \red{with the procedural half of the graph out of scope. The survey of memory mechanisms in LLM agents~\citep{zhang2025memorysurvey} is similarly the better reference on memory architectures, and organizes them by read/write/management operation, whereas Section~\ref{sec:memory} asks the orthogonal question of what a memory item's }\emph{\red{lineage}} \red{is once written. Surveys of agent hallucination~\citep{lin2025agenthallucination} classify failure by output symptom; the provenance view instead asks which recorded relation is missing when a symptom appears, which is what makes }\textsc{\red{Derive}} \red{without }\textsc{\red{Support}}\red{, the attributable hallucination, a distinct and separately measurable case. Retrieval-structure surveys such as that on graph-based RAG~\citep{peng2024graphragsurvey} share our use of typed graphs, but the graph is over the }\emph{\red{corpus}}\red{, built before the run, while ours is over the }\emph{\red{execution}}\red{, built by it.}

\red{}\blue{Reading these surveys through a common provenance view brings out three points that none of them shows on its own.} \red{First, the same object appears in several of them under different names: a retrieved passage is evidence in attribution work, context in RAG work, and a taint source in security work. }\blue{A common vocabulary makes it easier to see that they are the same node with different edges attached.} \red{Second, the gaps are }\emph{\red{shared}} \red{rather than thread-specific: }\blue{none of these surveys treats the origin of a recorded relation as a dimension, which is why Section~\ref{subsec:origin-assurance} has no counterpart in them. Third, evaluation is where the fragmentation costs most, since a benchmark built for one thread rarely supplies ground truth for another, as Figure~\ref{fig:benchmark-coverage-heatmap} shows.} \red{We are not aware of a survey of agent observability or of execution-trace infrastructure as such, which is itself consistent with the maturity gap Section~\ref{subsec:trace-assurance} reports. On corpus membership: the attribution and memory surveys are counted among the 71, retained under }\blue{the exclusion criteria's exception for surveys that define trace or provenance structure}\red{; the agent-hallucination and graph-RAG surveys are cited for positioning only, so the counts of Section~\ref{subsec:methodology} are unchanged.}

\noindent\textbf{Structure of the Survey.}
The survey follows a single through-line: turning raw agent traces into accountable provenance. Section~  \ref{sec:taxonomy} first develops a taxonomy of trace sources, evidence and execution units, provenance relations, tracing granularity and timing, representation forms, and trust functions. We then proceed through four steps:  Section~\ref{sec:representation} reviews provenance representations, ranging from structured logs to execution graphs, evidence graphs, claim-support graphs, and runtime provenance structures; Sections~\ref{sec:tool-use} and~\ref{sec:memory} examine two  settings where provenance failures are especially consequential, namely tool use and memory; Section~\ref{sec:benchmarks} maps benchmarks and metrics to provenance-related capabilities; and Section~\ref{sec:open-problems} discusses open challenges and future directions before Section~\ref{sec:conclusion} concludes the survey.

\red{The section allocation follows one criterion: a setting earns a dedicated section when its provenance failure has a }\emph{\red{distinct mechanism}} \red{and there is a reviewable body of released artifacts that act on it. Tool use qualifies on both counts: a contaminated tool argument propagates through execution rather than through content, and Section~\ref{sec:tool-use} reviews 28 works }\blue{that address it}\red{. So does memory, where provenance must survive writes, updates and retrievals across sessions, reviewed over 27 works. Multi-agent and environment provenance are first-class in the taxonomy, as trace sources and unit types. }\blue{We treat them as cross-cutting concerns, discussing them where they intersect with tool use, memory lineage, responsibility attribution, and evaluation; the corpus also contains far fewer works acting specifically on them.} \red{They are accordingly developed where they bear: responsibility attribution in Section~\ref{subsec:memory-multiagent-provenance-eval}, open problems in Section~\ref{subsec:memory-multiagent-open}, and requirements in Appendix~\ref{app:memory-multiagent-checklist}.}

\section{Taxonomy of Evidence Tracing and Execution Provenance}
\label{sec:taxonomy}

\begin{figure*}[!t]
    \centering
    \includegraphics[width=0.9\textwidth]{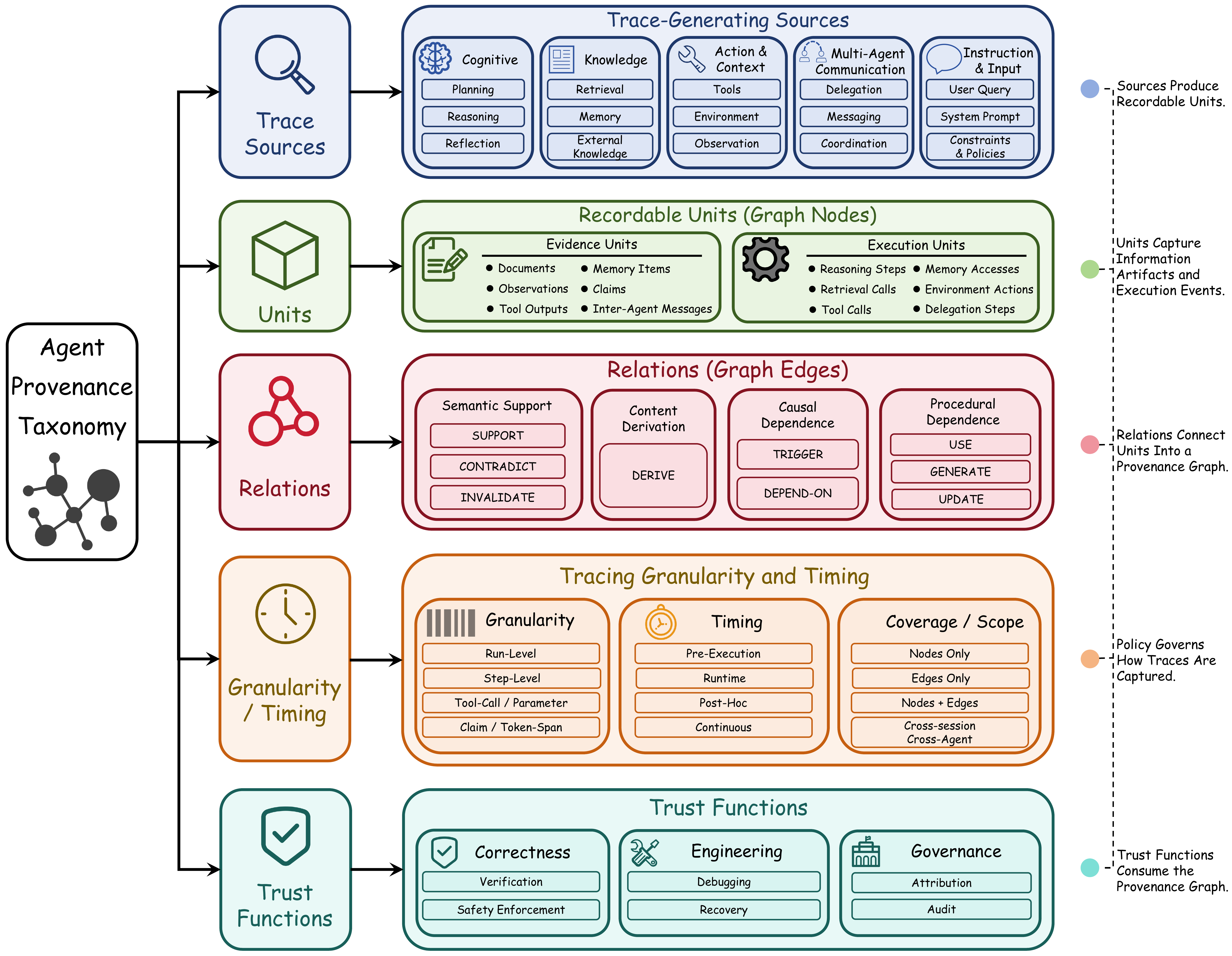}
    \caption{Overview of the agent provenance taxonomy, showing trace sources, evidence and execution units, provenance relations, tracing granularity and timing, and trust functions. Table~\ref{tab:evidence-tracing-taxonomy} summarizes all six dimensions, including representation forms.}
    \label{fig:taxonomy}
\end{figure*}

Having defined agent traces, evidence tracing, and execution provenance, this section introduces the taxonomy used throughout the survey. The goal is not to review every system in detail, but to provide a compact vocabulary for comparing provenance-aware agent work. We organize the space along six dimensions: trace sources, evidence and execution units, provenance relations, tracing granularity and timing, representation forms, and trust functions. Figure~\ref{fig:taxonomy} illustrates the relationships among five of these dimensions. Representation forms describe how provenance is encoded; Table~\ref{tab:evidence-tracing-taxonomy} summarizes all six dimensions and provides a reference for the comparisons throughout the survey.

The taxonomy draws on two existing traditions. Provenance models such as W3C PROV-DM describe how entities, activities, and agents participate in the production of artifacts through relations such as usage, generation, derivation, revision, and invalidation~\citep{w3c2013provdm}. Distributed observability frameworks such as OpenTelemetry model executions as traces and spans, which is useful for describing multi-step agent runs~\citep{opentelemetry2026traces}. LLM agents, however, add semantic and procedural objects that these traditions do not fully capture: retrieved passages, generated claims, tool-call rationales, memory items, natural-language observations, external state changes, and inter-agent messages. Agent-oriented observability systems such as AgentOps and AgentTrace further motivate structured records of both execution artifacts and context~\citep{dong2024agentops,alsayyad2026agenttrace}. We therefore treat classical provenance and observability as foundations, and specialize them to the execution structure of LLM agents.

\begin{table}[!t]
\centering
\footnotesize
\setlength{\tabcolsep}{3.5pt}
\renewcommand{\arraystretch}{1.10}
\caption{\blue{Six-dimensional taxonomy of evidence tracing and execution provenance in LLM agents. Unit types and representative works are illustrative rather than exhaustive.}}
\label{tab:evidence-tracing-taxonomy}
\begin{tabularx}{\textwidth}{>{\raggedright\arraybackslash}p{1.95cm}
                            >{\raggedright\arraybackslash}X
                            >{\raggedright\arraybackslash}X
                            >{\raggedright\arraybackslash\scriptsize}p{4.2cm}}
\toprule
\textbf{Dimension} & \textbf{Categories} & \textbf{Description} & \textbf{Representative Work} \\
\midrule

\rowcolor{covRow}Trace sources
& \blue{Instructions and inputs, reasoning, retrieval, tool use, MCP server/host boundaries, memory, environment, multi-agent communication}
& Components and execution boundaries that generate provenance-relevant artifacts during agent execution.
& ReAct, Reflexion, ToolLLM, WebArena, AutoGen, AgentTrace, MCP-SafetyBench~\citep{yao2023react,shinn2023reflexion,qin2024toolllm,zhou2024webarena,wu2023autogen,alsayyad2026agenttrace,zong2025mcpsafetybench} \\

Evidence and execution units
& \blue{\emph{Evidence units (artifacts):} User queries, system prompts, documents, observations, tool outputs, memory items, generated tool parameters, policies, tool manifests, inter-agent messages, intermediate claims, final answers}
& \blue{Content-bearing artifacts generated or consumed during a run, whose content can be compared across units.}
& ALCE, FActScore, FEVER, RAGChecker~\citep{gao2023alce,min2023factscore,thorne2018fever,ru2024ragchecker} \\
\subrowrule

& \blue{\emph{Execution units (occurrences):} Reasoning steps, retrieval calls, tool calls, memory accesses, environment actions, delegation steps}
& \blue{Occurrences identified by an actor and a time interval, recording what the agent did.}
& AgentOps, AgentTrace, TRAIL, AgenTracer, Aegis~\citep{dong2024agentops,alsayyad2026agenttrace,deshpande2025trail,zhang2025agentracer,kong2025aegis} \\

\rowcolor{covRow}Provenance relations
& \blue{\emph{Semantic support:} \textsc{Support}, \textsc{Contradict}, \textsc{Invalidate}}
& \blue{\textsc{Support}: evidence warrants a claim or action precondition; \textsc{Contradict}: two contents conflict. Neither requires source access; \textsc{Invalidate} also needs authority and temporal context.}
& ALCE, FActScore, SourceCheckup, RAGChecker~\citep{gao2023alce,min2023factscore,wu2024sourcecheckup,ru2024ragchecker} \\
\subrowruleshaded

\rowcolor{covRow}& \blue{\emph{Content derivation:} \textsc{Derive}}
& \blue{The target content was produced by transforming the source content. Derivation establishes content lineage, not the correctness of the transformation or semantic support.}
& W3C PROV-DM, PROV-AGENT~\citep{w3c2013provdm,souza2025provagent} \\
\subrowruleshaded

\rowcolor{covRow}& \blue{\emph{Causal dependence:} \textsc{Trigger}, \textsc{Depend-on}}
& \blue{\textsc{Trigger}: what caused a step; \textsc{Depend-on}: what shaped its arguments or behavior. \textsc{Depend-on} is always inferred; \textsc{Trigger} can be logged directly for event-driven dispatch.}
& AgenTracer, Fides, NeuroTaint~\citep{zhang2025agentracer,costa2025fides,cai2026neurotaint} \\
\subrowruleshaded

\rowcolor{covRow}& \blue{\emph{Procedural dependence:} \textsc{Use}, \textsc{Generate}, \textsc{Update}}
& \blue{Records artifact use, generation, and version succession. \textsc{Update} may be instrumented or inferred; \textsc{Use} over-approximates \textsc{Depend-on} only for dependencies on artifacts.}
& W3C PROV-DM, Agent-Sentry~\citep{w3c2013provdm,sequeira2026agentsentry} \\

Tracing granularity and timing
& \blue{\emph{Granularity:} Run-level, step-level, tool-call-level, parameter-level, claim-level, token/span-level}
& \blue{Determines the level of detail at which units and relations are recorded.}
& FActScore, SourceCheckup, Agent-Sentry~\citep{min2023factscore,wu2024sourcecheckup,sequeira2026agentsentry} \\
\subrowrule

& \blue{\emph{Timing:} Pre-execution, runtime, post-hoc, continuous}
& Determines when trace collection, verification, enforcement, or diagnosis occurs.
& AgentSpec, Agent-Sentry, TRAIL~\citep{wang2025agentspec,sequeira2026agentsentry,deshpande2025trail} \\

\rowcolor{covRow}Representation forms
& \blue{Structured logs, execution graphs, evidence and claim-support graphs, provenance graphs, runtime state}
& \blue{Forms used to encode execution records and relations for inspection, comparison, enforcement, or evaluation.}
& W3C PROV-DM, AgentOps, AgentTrace, PaperTrail~\citep{w3c2013provdm,dong2024agentops,alsayyad2026agenttrace,papertrail2026} \\

Trust functions
& Verification, attribution, debugging, safety enforcement, audit, recovery
& Downstream purposes served by evidence tracing and execution provenance.
& RAGAS, AgentOps, AgentTrace, LADYBUG, AgenTracer, OpenAgentSafety~\citep{es2024ragas,dong2024agentops,alsayyad2026agenttrace,rorseth2025ladybug,zhang2025agentracer,vijayvargiya2026openagentsafety} \\

\bottomrule
\end{tabularx}
\end{table}

\begin{table}[!t]
\centering
\footnotesize
\setlength{\tabcolsep}{3.5pt}
\renewcommand{\arraystretch}{1.12}
\caption{\blue{Two cross-cutting attributes of agent provenance: provenance origin records how a provenance edge came to exist, and trace assurance records how a trace is protected against modification.}}
\label{tab:crosscutting-attributes}
\rowcolors{2}{covRow}{white}
\begin{tabularx}{\textwidth}{>{\raggedright\arraybackslash}p{1.95cm}
                            >{\raggedright\arraybackslash}X
                            >{\raggedright\arraybackslash}X
                            >{\raggedright\arraybackslash\scriptsize}p{4.2cm}}
\toprule
\textbf{\red{Attribute}} & \textbf{\red{Values}} & \textbf{\red{Description}} & \textbf{\red{Motivating Work}} \\
\midrule

\red{Provenance origin
}& \blue{\textsc{Instrumented}, \textsc{Inferred}, \textsc{Declared}, \textsc{Derived}}
& \red{How a provenance edge came to exist: emitted by the runtime, judged post hoc by a model or human, or self-reported by the agent; }\blue{\textsc{Derived}} \red{is reserved for edges computed from a graph rather than observed.
}& \red{AgentOps, PROV-AGENT, AgenTracer, SourceCheckup~\citep{dong2024agentops,souza2025provagent,zhang2025agentracer,wu2024sourcecheckup}}\\

\red{Trace assurance
}& \red{Logged, tamper-evident, attested
}& \red{Whether the record is protected against post-hoc modification.} \blue{The reviewed descriptions do not establish tamper-evident or attested trace guarantees.}\red{
}& \red{Audit trails, auditable agents, checkpoint integrity~\citep{ojewale2026audittrails,nian2026auditableagents,zheng2026acrfence}}\\

\bottomrule
\end{tabularx}
\end{table}

\subsection{Taxonomy Dimensions}
\label{subsec:taxonomy-dimensions}

\paragraph{Trace sources.}\label{subsec:trace-sources}
Trace sources are the components and execution boundaries from which provenance-relevant artifacts are recorded. In LLM agents, they include \blue{instructions and inputs,} reasoning, retrieval, tool use, memory, environment interaction, and multi-agent communication. \blue{Instruction and input traces record user requests, system prompts, constraints, and policies;} reasoning traces record stated plans and rationales for action choices; retrieval traces provide candidate evidence; tool traces record calls, arguments, outputs, permissions, and errors, including interactions across MCP server/host boundaries; memory traces record reads, writes, and updates across sessions; environment traces record observations and state-changing actions; and multi-agent traces capture delegation, messaging, and coordination~\citep{yao2023react,shinn2023reflexion,qin2024toolllm,zhou2024webarena,yao2025taubench,wu2023autogen,li2023camel}. Later sections examine tool and memory sources in depth; here we identify the sources covered by the taxonomy.

\paragraph{Evidence and execution units.}\label{subsec:evidence-units}
Trace sources produce two broad kinds of units. \emph{Evidence units} are  \red{artifacts whose content }can support, contradict, invalidate, or contextualize claims and actions, such as \blue{user queries, system prompts,} documents, passages, observations, tool outputs, memory items,  \red{generated tool parameters, policies, inter-agent messages, intermediate claims, and final answers}. \emph{Execution units} are  \red{occurrences that record }what the agent did, including reasoning steps, retrieval calls, tool invocations,  memory reads or writes, environment actions,  \red{and delegation steps. Membership is decided by a single test, }\blue{so that the same object is always classified the same way}\red{: a unit is an }\emph{\red{evidence unit}} \red{if it is an artifact carrying a content field that can be compared with the content of another unit, and an }\emph{\red{execution unit}} \red{if it is an occurrence carrying an actor and a time interval, so that its identity is }\emph{\red{that it happened}} \red{rather than }\emph{\red{what it says}}\red{. Under this test the two roles are disjoint at graph level, following the entity/activity split of W3C PROV-DM~\citep{w3c2013provdm}. The familiar intuition that a }tool output is  \red{``both'' }the result of an execution step and  evidence for a later claim  \red{reflects the granularity of a flat log, in which one record carries the call, its arguments, and its result; the graph representation splits that record into a tool call (execution unit) and a tool output (evidence unit) joined by a }\textsc{\red{Generate}} \red{edge. What varies across edges is therefore not a unit's role but which part of it an edge reads: a semantic edge reads its content, a procedural edge reads its occurrence. Appendix~\ref{app:node-roles} gives the complete type assignment. }Provenance-aware agents  \red{still }need to connect semantic support with procedural execution rather than recording only one side of the trace~\citep{gao2023alce,min2023factscore,thorne2018fever,ru2024ragchecker,dong2024agentops,alsayyad2026agenttrace}.

\paragraph{Provenance relations.}\label{subsec:provenance-relations}
Relations turn raw trace units into provenance. We retain PROV-compatible base  \red{relations (}\textsc{Use}, \textsc{Generate}, and \textsc{Derive}\red{) and specialize four further relations that PROV expresses in a weaker form: }\textsc{\red{Depend-on}} \red{and }\textsc{\red{Trigger}} \red{refine PROV usage and communication into counterfactual claims, }\textsc{\red{Update}} \red{refines revision, and }\textsc{\red{Invalidate}} \red{refines PROV invalidation from ending an entity's existence to marking a persisting record epistemically unusable. Only }\textsc{Support} \red{and }\textsc{\red{Contradict}} \red{have no PROV counterpart at all, since they compare content rather than record how artifacts were produced.
}

\red{The nine fall into four families, distinguished by }\emph{\red{what kind of evidence establishes them}}\red{: }\emph{\red{semantic support}} \red{(}\textsc{\red{Support}},  \textsc{Contradict}, \textsc{Invalidate}\red{) is decided by reading the contents of both endpoints, and is therefore agent-independent}\blue{, although \textsc{Invalidate} additionally requires authority and temporal context}\red{; }\emph{\red{content derivation}} \red{(}\textsc{\red{Derive}}\red{) asks whether the target's content is a transformation of the source's; }\emph{\red{causal dependence}} \red{(}\textsc{Trigger},  \textsc{Depend-on}\red{) is decided counterfactually, by ablation and replay}\blue{, unless the harness itself dispatches the step on an event, in which case the counterfactual holds by construction and the edge can be logged directly}\red{; it says nothing about whether content is true; and }\emph{\red{procedural dependence}} \red{(}\textsc{\red{Use}}\red{, }\textsc{\red{Generate}}\red{, }\textsc{\red{Update}}\red{) is }\blue{read directly from what the harness emitted, except that \textsc{Update} is inferred when the memory store does not version its keys}\red{. Keeping the families apart is what lets the taxonomy record that a tool output changed a decision without justifying it, or that a false memory item drove an action while supporting nothing.
}

\red{Temporal succession is deliberately }\emph{\red{not}} \red{a relation: ordering is carried by node timestamps and never by an edge, so }\blue{temporal adjacency alone does not establish dependence; without an explicit relation, the graph records no dependency between the steps.} \red{Appendix~\ref{app:relation-semantics} gives each relation a signature over node types, a direction, temporal semantics, a necessary and sufficient condition, and worked examples with the test that separates it from the relation it is most often confused with; Appendix~\ref{app:support-vs-influence} works through the boundary cases on which annotators most often disagree}.

\red{A relation is admitted only if it passes two tests: }\emph{\red{necessity}}\red{, meaning that some trust-function query of this section cannot be answered without it, and }\emph{\red{distinctness}}\red{, meaning that it carries a decision test that no other relation's test subsumes. Appendix~\ref{app:relation-adequacy} applies both tests relation by relation and \blue{states what is lost under three alternative decompositions we considered and rejected, including PROV-DM unextended. It also explains why we prefer explicit relation labels to a single influence edge with a kind attribute, an encoding that carries the same information but makes the distinctions easier to drop.} \blue{The set is minimal only relative to the six trust functions and the corpus surveyed here, and Appendix~\ref{app:relation-adequacy} gives the test a tenth relation would have to pass.}}

\paragraph{\red{Provenance origin and assurance.}}\label{subsec:origin-assurance}
\red{The six dimensions describe }\emph{\red{what}} \red{is traced and }\emph{\red{why}}\red{. Cutting across all of them are two attributes that describe }\emph{\red{how a provenance record came to exist}}\red{, }and \red{that determine how much weight it can bear. We treat them as edge and node attributes rather than as a seventh dimension, because they qualify every relation in the taxonomy rather than partitioning the space.} \blue{Table~\ref{tab:crosscutting-attributes} summarizes the values of each attribute and the work that motivates it; the rest of this subsection explains both.}\red{
}

\emph{\red{Origin}} \red{distinguishes three sources of a provenance edge recorded from an execution; a fourth value, }\textsc{\red{Derived}}\red{, is reserved for edges computed from a graph rather than observed, and is used only by the projection of Appendix~\ref{app:projection}. }\textsc{\red{Instrumented}} \red{edges are emitted by the agent harness or runtime and are reproducible from the log alone; }\textsc{\red{Use}} \red{and }\textsc{\red{Generate}} \red{are always of this kind, and }\textsc{\red{Update}} \red{and }\textsc{\red{Trigger}} \red{are }\blue{Instrumented} \red{when the memory store versions its keys or the harness dispatches on an event. }\textsc{\red{Inferred}} \red{edges are assigned post hoc by a model or a human annotator; }\textsc{\red{Depend-on}} \red{falls here, since counterfactual dependence requires replay, ablation, fault injection, or information-flow tracking rather than reading a log~\citep{zhang2025agentracer,costa2025fides,cai2026neurotaint}; }\textsc{\red{Support}}\red{, }\textsc{Contradict}\red{, and }\textsc{Invalidate} \red{are never directly observable either, and are }\textsc{\red{Inferred}} \red{when an annotator or judge assigns them and }\textsc{\red{Declared}} \red{when the agent asserts them itself. }\textsc{\red{Declared}} \red{edges are agent self-reports, such as a generated citation, and are the weakest: a declared citation need not support the claim it annotates, which is precisely the failure that source-checking work measures~\citep{wu2024sourcecheckup}. A provenance graph is therefore heterogeneous in reliability. Its procedural skeleton can be logged, but its semantic layer is a judgment, and evaluation must treat the two differently rather than reporting a single accuracy number over a mixed graph. \blue{The distinction matters for two downstream uses that appear later in this survey.} A \emph{responsibility} claim (that a particular agent or step is answerable for an outcome, as multi-agent attribution work asserts~\citep{cemri2025mast,zhang2025agentracer,kong2025aegis}) is a claim about counterfactual dependence, so it inherits the origin of \textsc{Depend-on} and \textsc{Trigger} and is \textsc{Inferred}: a log records that a sub-agent produced a message, never that the outcome would have differed without it. And a \emph{reasoning trace} is an artifact the model emitted, not an observation of the computation that produced the output, so an edge read off a stated rationale is \textsc{Declared} \blue{and does not, by itself, establish faithful dependence; its trace assurance is a separate attribute}: a model may cite a passage it did not use, or act on a memory item it never mentions. \blue{Both uses remain legitimate, provided such claims are reported} as inferred or declared rather than as recorded, and validated by replay or annotation rather than by pointing at the log.
}

\emph{\red{Assurance}} \red{distinguishes whether a trace is merely written to a log, or additionally protected against modification after the fact. This is the least developed attribute in the works reviewed here. \blue{The descriptions of the observability and provenance systems in Table~\ref{tab:systems-taxonomy} do not report tamper-evidence, signing, or attestation for their traces}~\citep{dong2024agentops,alsayyad2026agenttrace,souza2025provagent}, so a trace is trustworthy only to the extent that the party holding it is. Recent work has begun to argue for audit-trail infrastructure and auditable agent design~\citep{ojewale2026audittrails,nian2026auditableagents}, and work on checkpoint-restore integrity shows that agent state itself can be attacked~\citep{zheng2026acrfence}, but \blue{the reviewed descriptions do not establish tamper-evident or attested trace guarantees for the systems we survey}. This matters for audit, where the reader of a trace and its producer are often not the same party. We return to it as an open problem in Section~\ref{sec:open-problems}, and Appendix~\ref{app:relation-observability} records the available origins for each relation}.

\paragraph{Tracing granularity and timing.}\label{subsec:granularity-timing}
Granularity specifies how fine-grained the trace is. Run-level traces support coarse audit; step-level traces support debugging; tool-call and parameter-level traces support safety enforcement; claim-level traces support factual attribution; and token- or span-level traces provide the finest but most expensive view. Timing specifies when traces are collected or used: pre-execution checks can block unauthorized actions, runtime tracing supports policy enforcement and information-flow control, post-hoc tracing supports audit and diagnosis, and continuous tracing supports persistent memory and long-horizon workflows~\citep{min2023factscore,wu2024sourcecheckup,costa2025fides,cai2026neurotaint,wang2025agentspec,sequeira2026agentsentry,deshpande2025trail}. The appropriate choice depends on the risk and persistence of the workflow: low-risk RAG may require claim-level post-hoc attribution, whereas high-impact tool use may require runtime parameter-level provenance.

\paragraph{Representation forms.}\label{subsec:representation-forms}
Representation forms determine how traces are encoded. Structured logs preserve chronological events and metadata, but often leave support, contradiction, and influence implicit. Execution graphs make procedural dependencies explicit; evidence and claim-support graphs emphasize semantic grounding; provenance graphs combine evidence and execution units through typed edges; and runtime state representations support online checking and enforcement. We keep this dimension concise here because Section~\ref{sec:representation} reviews representation choices in detail.

\paragraph{Trust functions.}\label{subsec:trust-functions}
The purpose of tracing is not to store more logs, but to support trust functions. \emph{Verification} asks whether claims or actions are supported by evidence; \emph{attribution} identifies which evidence, tool output, memory item, or message contributed to a claim or action; \emph{debugging} localizes failures within an execution; \emph{safety enforcement} prevents untrusted or unauthorized influence from reaching sensitive actions; \emph{audit} reconstructs executions for review and accountability; and \emph{recovery} uses provenance to retry, compensate, roll back, or repair affected states~\citep{gao2023alce,es2024ragas,deshpande2025trail,rorseth2025ladybug,debenedetti2025camel,costa2025fides,sequeira2026agentsentry}. These functions explain why the taxonomy includes both evidence-support links and execution-dependency links.

\subsection{Design Tensions}
\label{sec:core-tensions}

The taxonomy exposes four design tensions that guide the rest of the survey. First, chronological logging and semantic provenance are not the same: a log can record what happened, but it may not show whether evidence supported a claim or whether memory influenced a tool call. Second, component-level tracing does not guarantee end-to-end accountability, because failures may propagate across retrieval, tools, memory, environment states, and other agents. Third, static schemas help standardize trace objects and relations, but runtime safety requires online source tracking, policy checking, and dependency-aware enforcement. Fourth, trace completeness must be balanced against deployability, since fine-grained provenance improves verification, debugging, audit, and recovery but also increases storage cost, privacy exposure, annotation burden, and system complexity.

The taxonomy is used throughout the paper as a compact comparison vocabulary. Sections~\ref{sec:representation}--\ref{sec:benchmarks} return to these tensions when reviewing representation methods, tool-use provenance, memory provenance, and evaluation. Before turning to these topics, Table~\ref{tab:systems-taxonomy} positions representative systems within the taxonomy to expose coverage gaps across trace sources, granularity, timing, representation, and trust functions. Appendix~\ref{app:taxonomy-mappings} provides an explicit mapping to W3C PROV-DM.

\subsection{\red{Two Decisions the Taxonomy Changes}}
\label{subsec:worked-decisions}

\blue{To show what the taxonomy adds beyond enumerating trace artifacts, we work through two decisions} \red{where the two approaches diverge, and in both the divergence comes from the same source: the taxonomy separates }\emph{\red{which relation}} \red{is at stake from }\emph{\red{how a record of it can be obtained}}\red{, so it constrains what a design or an evaluation can deliver before either is built.}

\paragraph{\red{Decision 1: what to instrument in an auditable tool-using agent.}}
\red{Consider an agent with tools and persistent memory, deployed where an auditor must later reconstruct why an action was taken. Enumerating trace artifacts yields the familiar answer: record every artifact in Section~\ref{subsec:trace-sources} at the finest affordable granularity, and if the budget allows, go from step level to token level. The taxonomy gives a different and partly opposite answer, because granularity and origin are independent axes. }\textsc{\red{Use}} \red{and }\textsc{\red{Generate}} \red{are }\textsc{\red{Instrumented}}\red{: they are already reproducible from an ordinary call log, and refining that log buys no new relation. }\textsc{\red{Depend-on}} \red{is }\textsc{\red{Inferred}} \red{only, so }\emph{\red{no}} \red{logging granularity yields it: counterfactual dependence requires replay, ablation, or information-flow tracking, which is a runtime capability rather than a logging setting. And }\textsc{\red{Support}} \red{is never observable at any granularity, so claim-level audit requires an annotation protocol with a fixed warrant threshold and reported agreement, not a finer trace. The recommendation is therefore to spend the budget on }\emph{\red{replay hooks and an annotation protocol}}\red{, and to keep logging at step level, whereas the enumeration answer spends it on token-level logging, which cannot produce either missing layer. The taxonomy also fixes what the audit may claim: without replay the record supports attribution and reconstruction but not counterfactual causal claims, and saying so in advance is more useful than discovering it during an audit.}

\paragraph{\red{Decision 2: how to evaluate a memory-poisoning defense.}}
\red{Now consider evaluating a defense against poisoned memory. Enumeration suggests the obvious protocol: use a memory benchmark and check whether the defense preserves task success. The taxonomy identifies the failure differently. A poisoned item that drives an action while justifying nothing is a }\textsc{\red{Trigger}} \red{edge with no }\textsc{\red{Support}} \red{edge, causal dependence without semantic support, so the quantity to measure is whether that }\textsc{\red{Trigger}} \red{is detected, which is established counterfactually rather than by answer comparison. Two consequences follow that the enumeration protocol misses. First, the memory family of Figure~\ref{fig:benchmark-coverage-heatmap} is rated }\textbf{\red{L}} \red{on safety attacks: }\blue{it supplies no adversarial ground truth}\red{, so a memory benchmark cannot measure the defense's actual target, and the relevant labels (source-to-sink influence, }\blue{that is, causal dependence of the kind} \textsc{Trigger} \blue{and} \textsc{Depend-on} \blue{record}\red{) sit in the taint benchmarks of the provenance/security family instead. Second, }\blue{task success and defense quality \emph{can come apart}. As an illustrative scenario, assume that poisoned items make up a small fraction of the memory store, that they are retrieved only for the queries they target, and that the defense does not also filter benign items. A defense that correctly blocks a poisoned item then removes a \textsc{Trigger} that never carried \textsc{Support}; under these assumptions, answers to unaffected queries need not change, and aggregate accuracy may move only slightly even when the defense is working. The taxonomy does not guarantee this outcome: a defense that also removes benign items, or an attack aimed at frequent queries, can change accuracy appreciably in either direction. What the scenario shows is that answer accuracy is not a reliable proxy for defense quality, so dependency coverage should be reported separately from answer accuracy, and the reason lies in the relation type rather than in a list of artifacts.}

\subsection{Positioning Existing Systems in the Taxonomy}
\label{subsec:systems-taxonomy}

Table~\ref{tab:systems-taxonomy} positions representative agent, attribution, safety, memory, and observability systems within the taxonomy. The table is not intended as an exhaustive benchmark, but as a diagnostic map of where current work concentrates and where coverage remains incomplete. Three patterns emerge. First, \emph{trace-generating systems} such as ReAct, AutoGen, MemGPT, and Generative Agents produce rich execution artifacts, but usually do not implement trust functions such as verification, attribution, enforcement, audit, or recovery on their own. They provide substrates on which provenance mechanisms can be built. Second, \emph{attribution and verification systems} such as ALCE, FActScore, SourceCheckup, and RAGChecker provide claim-level support analysis, but mostly operate over retrieval evidence and final answers rather than tools, memory, environment actions, or multi-agent communication. Third, \emph{safety and enforcement systems} such as CaMeL, Fides, and Agent-Sentry operate at runtime over tool calls, parameters, permissions, or information-flow labels, \red{and NeuroTaint audits the same flows offline}, but often lack semantic claim-evidence attribution.

This mapping shows why evidence tracing and execution provenance should be treated as a unified layer rather than as separate literatures. Existing systems usually cover one or two \blue{of these capabilities} well: they may record traces, verify claims, enforce tool-use policies, audit executions, or diagnose failures.  \blue{Within the corpus surveyed here, however, these capabilities are integrated only to a limited extent: few lines of work combine diverse trace sources, fine-grained units, runtime timing, explicit provenance representation, and multiple trust functions.} This gap motivates the rest of the survey, which reviews representation, attribution, tool-use provenance, memory provenance, observability, safety, and evaluation as connected parts of the same provenance problem.

\begin{table}[!t]
\centering
\scriptsize
\setlength{\tabcolsep}{2.0pt}
\renewcommand{\arraystretch}{1.05}
\caption{\blue{Representative systems mapped to the taxonomy. Rows are grouped by each system's primary trust function; other functions a system supports are not recorded.} \red{\textsuperscript{\dag}\,marks a work that had not been peer reviewed at the time of writing; for an entry naming two works, at least one of the two.}}
\label{tab:systems-taxonomy}
\begingroup
\hypersetup{hidelinks}
\newcommand{\TrustGroupRule}{\arrayrulecolor{covStrong!55}\specialrule{0.35pt}{1.6pt}{1.6pt}\arrayrulecolor{black}}
 \begin{tabularx}{\textwidth}{>{\raggedright\arraybackslash\bfseries}p{1.75cm}
                            >{\centering\arraybackslash\scriptsize}X
                            >{\raggedright\arraybackslash}p{1.95cm}
                            >{\raggedright\arraybackslash}p{1.62cm}
                            >{\raggedright\arraybackslash}p{1.52cm}
                            >{\raggedright\arraybackslash}p{1.92cm}}
\toprule
\rowcolor{covHeader}
\textbf{Primary trust} & \textbf{System} & \textbf{Trace source(s)} & \textbf{Granularity} & \textbf{Timing} & \textbf{ \red{Repre\-sentation}} \\
\midrule

\multirow[c]{7}{=}{Substrate}
& ReAct~\citep{yao2023react} & Reasoning, tool, env & Step & Runtime & Reasoning--action log \\
& Generative Agents~\citep{park2023generative} & Memory, env & Step & Continuous & Memory stream \\
& Toolformer~\citep{schick2023toolformer} & Tool & Tool-call & Pre-exec. & --- \\
& ToolLLM~\citep{qin2024toolllm} & Tool & Tool-call, param. & Runtime & API-call path \\
& AutoGen~\citep{wu2023autogen} & Multi-agent, tool & Step & Runtime & Message log \\
& MemGPT\red{\textsuperscript{\dag}}~\citep{packer2023memgpt} & Memory & Step & Continuous & Tiered memory \\
& A-MEM~\citep{xu2025amem} & Memory & Step & Continuous & Linked memory \\

\TrustGroupRule
\multirow[c]{5}{=}{Attribution / Verification}
& ALCE~\citep{gao2023alce} & Retrieval & Claim & Post-hoc & Citation links \\
& FActScore~\citep{min2023factscore} & Retrieval & Atomic claim & Post-hoc & Claim--evidence \\
& SourceCheckup~\citep{wu2024sourcecheckup} & Retrieval & Claim & Post-hoc & Source--support \\
& RAGChecker~\citep{ru2024ragchecker} & Retrieval & Claim & Post-hoc & Claim--evidence \\
& \red{RAGAS~\citep{es2024ragas}} & \red{Retrieval} & \red{Claim} & \red{Post-hoc} & \red{Claim--context} \\

\TrustGroupRule
\multirow[c]{6}{=}{Safety}
& CaMeL~\citep{debenedetti2025camel} & Tool & Parameter & Runtime & Control/data flow \\
& Fides\red{\textsuperscript{\dag}}~\citep{costa2025fides} & Tool, memory & Parameter & Runtime & IFC labels \\
& NeuroTaint\red{\textsuperscript{\dag}}~\citep{cai2026neurotaint} & Tool, memory & Param. \slash token & \red{Post-hoc} & Taint labels \\
& AgentSpec~\citep{wang2025agentspec} & Tool, env & Tool-call & Pre \slash Runtime & Policy rules \\
& AgentBound~\citep{buhler2025agentbound} & Tool & Tool-call & Pre \slash Runtime & Permissions \\
& Agent-Sentry\red{\textsuperscript{\dag}}~\citep{sequeira2026agentsentry} & Tool & Parameter & Runtime & Provenance graph \\

\TrustGroupRule
\multirow[c]{4}{=}{Safety (eval)}
& ToolEmu~\citep{ruan2024toolemu} & Tool & Tool-call & Runtime & Action log \\
& InjecAgent~\citep{zhan2024injecagent} & Tool, env & Tool-call & Runtime & --- \\
& AgentDojo~\citep{debenedetti2024agentdojo} & Tool, env & Tool-call & Runtime & --- \\
& $\tau$-bench~\citep{yao2025taubench} & Tool, env & Tool-call & Post-hoc & State log \\

\TrustGroupRule
Recovery
& Reflexion~\citep{shinn2023reflexion} & Reasoning, memory & Step & Continuous & Episodic memory \\

\TrustGroupRule
\multirow[c]{4}{=}{Debugging / Audit}
& TRAIL\red{\textsuperscript{\dag}}~\citep{deshpande2025trail} & All sources & Step & Post-hoc & Annotated trace \\
& MAST~\citep{cemri2025mast} & Multi-agent & Step & Post-hoc & Failure taxonomy \\
& WebArena~\citep{zhou2024webarena} & Env, tool & Step & Post-hoc & Action/state log \\
& AgentOps / AgentTrace\red{\textsuperscript{\dag}}~\citep{dong2024agentops,alsayyad2026agenttrace} & All sources & Step & Post-hoc/Cont. & Structured log \\

\bottomrule
\end{tabularx}
\endgroup
\end{table}

\section{Provenance Representation and Evidence Attribution}
\label{sec:representation}

\begin{figure}[!t] \centering \includegraphics[width=0.96\textwidth]{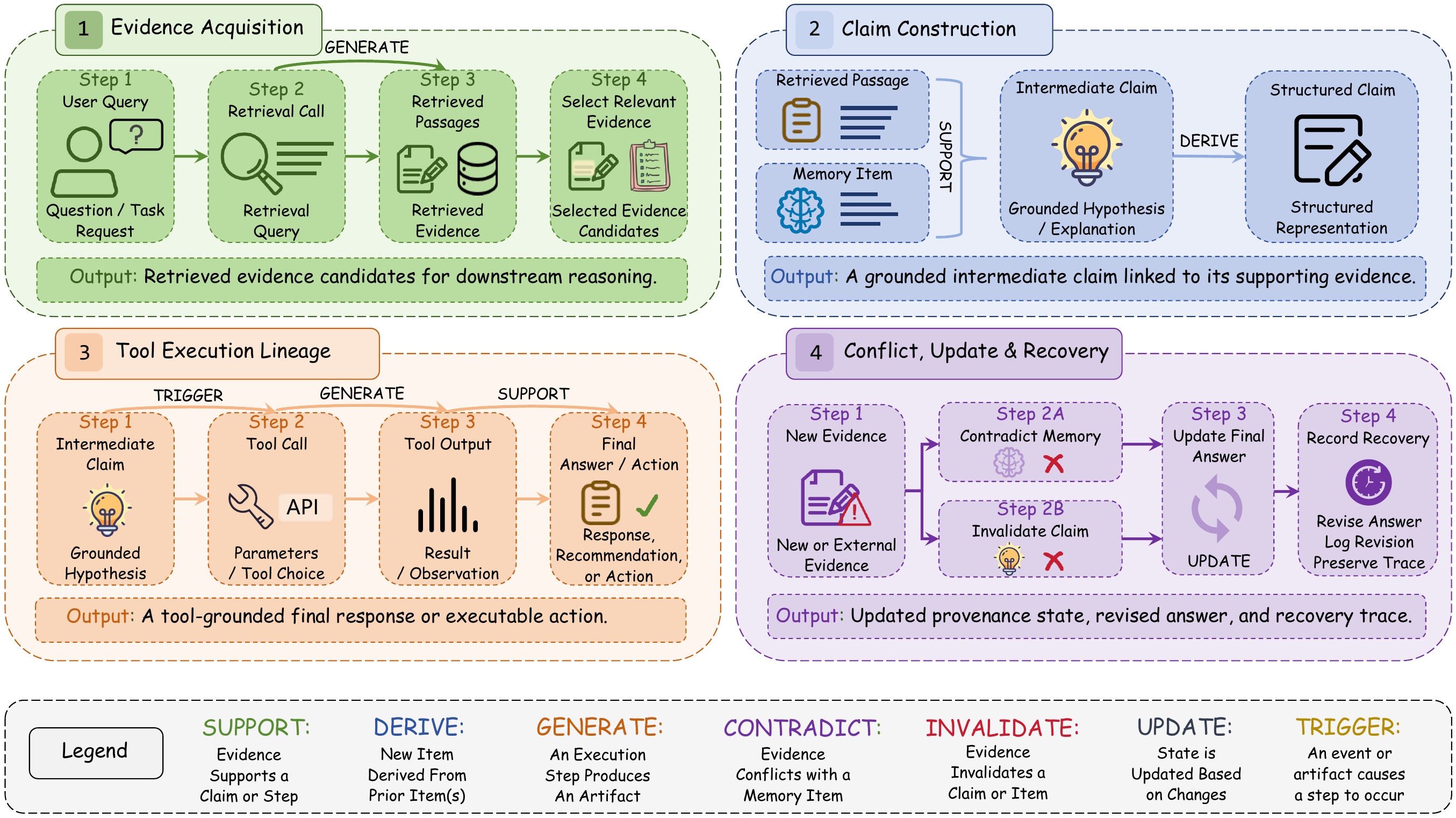} \caption{\blue{Running example of a provenance graph for an LLM-agent execution, linking user queries, retrieved passages, intermediate claims, tool calls and outputs, memory items, new evidence, and final actions through typed relations. For legibility, the panels show seven of the nine relations and omit \textsc{Use} and \textsc{Depend-on}; Table~\ref{tab:app-projection-example} lists every edge of the example.}} \label{fig:example-provenance-graph} \end{figure}

The previous section identifies what should be traced; this section asks how traces should be represented. We organize provenance representation into three layers. \emph{Structured logs} record typed execution events. \emph{Execution graphs} connect instructions, retrievals, tool calls, memory operations, observations, claims, and actions through dependency relations. \emph{Evidence and claim-support graphs} further specify whether evidence units support, contradict, or fail to support generated claims. This layered view is central to the rest of the survey: logs make executions reconstructable, graphs make dependencies inspectable, and claim-support links make grounding and accountability measurable. Figure~\ref{fig:example-provenance-graph} provides a running example. It shows how evidence acquisition, claim construction, tool execution, memory update, and recovery can be represented as a typed provenance graph rather than a flat chronological transcript.

\subsection{From Logs to Execution Graphs}
\label{subsec:logs-to-execution-graphs}

A natural starting point is the structured execution log. Instead of storing free-form transcripts alone, logs record typed events such as user requests, model outputs, retrieval calls, tool invocations, tool outputs, memory reads and writes, environment observations, inter-agent messages, and final responses. W3C PROV-DM provides a general vocabulary of entities, activities, agents, and derivation relations~\citep{w3c2013provdm}. PROV-AGENT adapts this vocabulary to agentic workflows by modeling prompts, responses, decisions, tool interactions, and workflow context~\citep{souza2025provagent}. AgentOps emphasizes relationships among artifacts across the agent lifecycle~\citep{dong2024agentops}, while AgentTrace organizes execution records into operational, cognitive, and contextual logs~\citep{alsayyad2026agenttrace}.

Logs provide chronological observability, but dependency analysis requires graph structure. A log can show that a retrieval, tool call, and memory write occurred; it may not show which retrieved passage supported a claim, which tool output changed the plan, or which memory item influenced an action. Execution graphs address this gap by representing execution artifacts as nodes and their influence relations as typed edges. TRAIL demonstrates how annotated trajectories can localize failures to particular steps, components, or interactions~\citep{deshpande2025trail}. AgenTracer further uses counterfactual replay and fault injection to attribute failures to responsible agents and steps~\citep{zhang2025agentracer}, whereas Aegis uses positive--negative trajectory pairs to identify faulty agents and error modes~\citep{kong2025aegis}. \blue{These systems differ in method, but share one contrast:} logs answer \emph{what happened}; execution graphs answer \emph{what depended on what}.

\subsection{Evidence and Claim-Support Attribution}
\label{subsec:evidence-claim-support-graphs}

Evidence attribution adds a semantic layer on top of execution dependencies. The key issue is whether a generated claim is actually supported by the evidence it cites or depends on. Existing work motivates this distinction at different granularities: citation-based evaluation links generated answers to sources~\citep{gao2023alce}; atomic-fact evaluation checks fine-grained factual support~\citep{min2023factscore}; source-checking work shows that a cited source may be relevant without supporting the specific claim~\citep{wu2024sourcecheckup}; and claim--evidence mapping explicitly distinguishes supported claims, unsupported claims, and omitted evidence~\citep{papertrail2026}.

For provenance representation, these correspond to distinct \red{relationships }rather than a single citation relation. A source may be \textsc{Cited}\red{, asserted by the agent as supporting, without being }\textsc{\red{Supported}}\red{, that is, without an adjudicated support link}; a relevant passage may be available but \textsc{Omitted} from the answer; a tool output may \textsc{Contradict} a memory item; and new evidence may \textsc{Invalidate} an earlier belief. \textsc{\red{Cited}} \red{and }\textsc{\red{Omitted}} \red{are not additional taxonomy relations: they are predicates derived from the nine, and Appendix~\ref{app:derived-predicates} defines them in those terms. }Evidence graphs therefore distinguish evidence availability from evidence use. In LLM agents, evidence-bearing units include not only retrieved documents, but also tool outputs, memory items, environment observations, policies, and inter-agent messages.

\subsection{Representation Trade-offs}
\label{subsec:representation-tradeoffs}

Static schemas and runtime provenance \blue{play complementary roles in} provenance representation. A static schema defines object types, relation types, trust labels, and required fields, including evidence units, tool calls, memory items, claims, actions, and relations such as \textsc{Support}, \textsc{Depend-on}, \textsc{Contradict}, \textsc{Update}, and \textsc{Invalidate}. Runtime provenance instantiates this schema during a concrete execution, recording the actual dependencies among sources, tool calls, memory operations, claims, and actions. Observability systems emphasize runtime trace capture and inspection~\citep{dong2024agentops,alsayyad2026agenttrace}. Safety-oriented systems additionally use runtime provenance for enforcement~\citep{sequeira2026agentsentry}, information-flow control~\citep{costa2025fides}, and source-to-sink influence analysis~\citep{cai2026neurotaint}.

The central trade-off concerns granularity. Fine-grained provenance enables claim verification, failure localization, rollback, audit, and policy enforcement, but increases storage cost, logging complexity, privacy exposure, and annotation burden. Coarse-grained provenance is easier to collect, but may miss the evidence that supported a claim, the memory item that caused an error, or the tool argument that crossed a trust boundary. The appropriate granularity therefore depends on the trust function: claim-level links may suffice for low-risk RAG, tool-using agents require parameter-level provenance, and long-term personalized agents require memory lineage, invalidation, and recovery support. Table~\ref{tab:provenance-representation} summarizes the main representation forms discussed in this section.

\begin{table}[t]
\centering
\footnotesize
\setlength{\tabcolsep}{3.5pt}
\renewcommand{\arraystretch}{1.12}
\caption{Main provenance representation forms for LLM agents.}
\label{tab:provenance-representation}
\rowcolors{2}{covRow}{white}
\begin{tabularx}{\linewidth}{p{2.2cm} X X}
\toprule
\textbf{Form} & \textbf{What it captures} & \textbf{Main role} \\
\midrule

Structured logs
& Typed execution events such as requests, retrieval calls, tool calls, memory operations, observations, and outputs.
& Execution observability and reconstruction. \\

Execution graphs
& Dependencies among instructions, evidence, tool calls, memory items, claims, and actions.
& Influence analysis for debugging, audit, and recovery. \\

Evidence graphs
& Links among evidence units, citations, atomic claims, and generated answers.
& Separation of citation presence, relevance, support, contradiction, and omission. \\

Static schemas
& Object types, relation types, trust labels, and required provenance fields.
& A reusable vocabulary for trace representation and comparison. \\

Runtime provenance
& Concrete source-to-sink, tool, memory, and action dependencies during execution.
& Runtime monitoring, enforcement, recovery, and information-flow analysis. \\

\bottomrule
\end{tabularx}
\end{table}

\section{Execution Provenance in Tool-Using Agents}
\label{sec:tool-use}

Tool use is the \emph{outward} axis of agent provenance: it is where an LLM agent moves from generating text to interacting with APIs, databases, code interpreters, browsers, filesystems, enterprise systems, and other external environments. A tool call may retrieve private data, modify external state, send messages, execute code, or trigger irreversible actions. Trustworthy tool use therefore requires tracing why a tool was selected, where its arguments came from, whether its output was reliable, and how the result shaped later decisions. This section reviews five \blue{questions central to tool-use provenance}: what to trace in tool execution, how external content contaminates tool use, how unsafe influence can be constrained, how action boundaries are enforced, and how traces support verification and recovery.

\subsection{Tool-Call, Argument, and Output Provenance}
\label{subsec:tool-call-argument-output-provenance}

Tool-call provenance records the path from tool selection to downstream influence: which tool was selected, which arguments were passed, what output was returned, and how that output affected later reasoning or actions. Toolformer~\citep{schick2023toolformer} and ToolLLM~\citep{qin2024toolllm} show that external tools can expand LLM capabilities through API calls, calculation, search, and other operations. Beyond execution success, provenance asks whether the call is justified, whether its arguments are derived from trusted or authorized sources, and whether its output is used appropriately.

A tool call contains several \blue{objects that provenance must record}: the selected tool, schema, argument values, execution result, and downstream consumers. Argument provenance is especially important because legitimate tools can become unsafe when their parameters are derived from untrusted or incorrectly generated values. Agent-Sentry tracks the sources of values flowing into sensitive tool arguments, \blue{which motivates} parameter-level lineage beyond tool-level permission checks~\citep{sequeira2026agentsentry}. Tool outputs are also provenance-bearing units: they may support claims, update memory, trigger additional tools, modify external state, or conflict with other evidence. WebArena highlights the importance of correctly interpreting observations and executing state-changing actions in realistic web environments~\citep{zhou2024webarena}, while $\tau$-bench evaluates tool-agent-user interactions under domain-specific policies~\citep{yao2025taubench}. Tool-use provenance should therefore connect sources, arguments, outputs, and downstream consumers into a single dependency chain.

\subsection{Indirect Prompt Injection and Tool-Use Contamination}
\label{subsec:indirect-prompt-injection}

Tool-using agents are vulnerable to indirect prompt injection because they consume external content that may contain adversarial instructions. Unlike direct prompt injection, where the malicious instruction is placed in the user prompt, indirect injection embeds the instruction in webpages, documents, emails, database entries, tool outputs, or third-party API responses. The attack exploits a boundary failure: untrusted data may be interpreted as executable guidance rather than passive evidence~\citep{greshake2023not,liu2023promptinjection}.

InjecAgent benchmarks indirect prompt injection in tool-integrated agent scenarios~\citep{zhan2024injecagent}, while AgentDojo evaluates attacks and defenses in realistic task environments~\citep{debenedetti2024agentdojo}. Related benchmarks broaden the tool-risk surface. ToolEmu evaluates risky tool execution in an emulated sandbox~\citep{ruan2024toolemu}; OpenAgentSafety covers real tools such as browsers, code execution, filesystems, shells, and messaging platforms~\citep{vijayvargiya2026openagentsafety}; and MCP-SafetyBench targets real MCP servers and multi-server workflows~\citep{zong2025mcpsafetybench}. Together, these benchmarks show that contaminated tool contexts can cause unauthorized tool use, privacy leakage, ignored user intent, and unintended state changes.

From a provenance perspective, indirect prompt injection is an information-flow and influence-tracking problem. The key question is whether malicious content influenced a later tool call, argument value, memory update, or external action. A flat log may show that a webpage was retrieved and an email tool was later called, but not whether the recipient, subject, or body came from the webpage, the user instruction, or the model's reasoning. Relations such as \textsc{Derive}, \textsc{Depend-on}, \textsc{Trigger}, and \textsc{Use} make this influence explicit. Tool-use provenance should therefore distinguish trusted intent from untrusted content and support enforcement against unsafe flows into privileged execution channels.

\subsection{Information Flow, Taint Tracking, and Provenance Enforcement}
\label{subsec:information-flow-taint}

Recent defenses increasingly treat tool-use safety as an information-flow problem. Instead of relying only on prompting or output filtering, they track where information originates, how it propagates, and whether it is allowed to influence sensitive actions. This view extends classical information-flow control, where security policies regulate flows among subjects, objects, and security classes~\citep{denning1976lattice,sabelfeld2003language}. For LLM agents, the analogous goal is to prevent low-integrity or untrusted data from exerting unauthorized influence over privileged tool calls, memory updates, or state-changing actions.

Defenses instantiate this principle at different layers. Instruction Hierarchy establishes precedence among instruction sources~\citep{wallace2024instructionhierarchy}, whereas StruQ separates instructions from untrusted data through structured channels~\citep{chen2024struq}. CaMeL makes this separation explicit at execution time by isolating control flow from data flow and restricting how external data can influence agent behavior~\citep{debenedetti2025camel}. Fides formalizes agent-level information-flow control through confidentiality and integrity labels enforced during execution~\citep{costa2025fides}. At finer granularity, NeuroTaint propagates taint across neural and symbolic components~\citep{cai2026neurotaint}, while Agent-Sentry tracks whether sensitive tool arguments are influenced by untrusted sources~\citep{sequeira2026agentsentry}.

The key lesson is that unsafe behavior can arise from influence, not content alone. A webpage may be safe to summarize but unsafe if an embedded instruction determines an email recipient, database parameter, or authorization signal. Runtime provenance makes this distinction explicit by connecting sources, transformations, and sinks, allowing enforcement policies to account for a value's origin and use as well as its content.

\subsection{Access Control and Execution Boundaries}
\label{subsec:access-control-execution-boundaries}

Information-flow tracking must be paired with access control and execution boundaries. Even if an agent can reason about many tools, it should not be allowed to invoke every tool in every context. Classical principles such as least privilege and complete mediation require sensitive actions to be explicitly authorized and access decisions to be checked at the point of use~\citep{saltzer1975protection}. In tool-using agents, these principles translate into tool permissions, argument constraints, user confirmations, sandboxing, and policy-mediated execution.

Recent systems operationalize these principles at different layers. AgentSpec enforces user-defined runtime constraints over agent actions and tool use~\citep{wang2025agentspec}. AgentBound provides capability-constrained access control for MCP servers, limiting the resources available to each server~\citep{buhler2025agentbound}. Agent-Sentry instead uses execution provenance, including per-argument data flow, to bound tool calls according to learned behavior and user intent~\citep{sequeira2026agentsentry}. Together, these systems show that tool-use safety requires controlling which tools and resources can be accessed, under what context, with which arguments, and with what external effects.

Provenance makes these controls context-sensitive. Agent-Sentry, for example, uses per-argument data flow and user-intent alignment to assess proposed tool calls~\citep{sequeira2026agentsentry}. A read-only search tool may be permitted to consume untrusted web content, whereas an email, database-update, or code-execution tool may require arguments derived from authorized user intent, verified tool outputs, or explicit approval. Access control and provenance are therefore complementary: capability policies delimit the tools and resources available to an agent~\citep{buhler2025agentbound}, while provenance and runtime rules determine whether a particular action should proceed, be restricted, require confirmation, or be blocked~\citep{wang2025agentspec,sequeira2026agentsentry}.

\subsection{Runtime Guardrails and Pre-/Post-Execution Verification}
\label{subsec:runtime-guardrails-verification}

Runtime guardrails check agent behavior before, during, or after execution. Pre-execution guardrails decide whether a proposed tool call should proceed; runtime monitors track information flow and state changes; post-execution verification inspects traces, outputs, and side effects for debugging, audit, and recovery. General guardrail systems illustrate how safety mechanisms can be placed around LLM applications: NeMo Guardrails provides programmable conversational guardrails~\citep{rebedea2023nemo}, while Llama Guard provides model-based input/output safety classification~\citep{inan2023llamaguard}. Tool-using agents, however, require checks over execution state, tool arguments, and downstream effects, not only input--output text.

For high-impact tools, pre-execution verification is critical. A provenance-aware guardrail can check whether the tool is authorized, whether user intent is present, whether arguments are well formed, whether sensitive fields come from trusted sources, and whether the action violates policy. AgentSpec formalizes runtime constraints over agent actions~\citep{wang2025agentspec}. Agent-Sentry uses execution provenance to verify whether sensitive tool arguments are influenced by untrusted sources~\citep{sequeira2026agentsentry}. ToolEmu complements these defenses by evaluating risky tool executions in an emulated sandbox~\citep{ruan2024toolemu}.

Post-execution verification supports failure localization and repair. When an agent gives a wrong answer or performs an unsafe action, traces can help identify whether the failure arose from evidence, tool use, memory, or downstream reasoning. AgentOps records agent artifacts and their lifecycle relationships~\citep{dong2024agentops}, while AgentTrace structures execution records into operational, cognitive, and contextual traces~\citep{alsayyad2026agenttrace}. TRAIL localizes failures using annotated trajectories~\citep{deshpande2025trail}, and LADYBUG uses trace-based debugging to support failure localization and repair~\citep{rorseth2025ladybug}.

Effective guardrails therefore require both semantic and procedural provenance: semantic provenance links claims to supporting or contradicting evidence, while procedural provenance links actions to authorized paths and trusted inputs. Together, they help detect unsupported claims, contaminated arguments, unauthorized tool calls, and invalid state updates.

Table~\ref{tab:tool-use-provenance} summarizes representative systems according to \blue{the objects they track, the mechanisms they use, and their role in tool-use provenance}.

\begin{table}[!t]
\centering
\scriptsize
\setlength{\tabcolsep}{2.3pt}
\renewcommand{\arraystretch}{0.98}
\caption{Representative work on tool-use execution provenance. The table groups systems by the objects they track, the mechanisms they use, and the provenance role they support. \red{\textsuperscript{\dag}\,marks a work that had not been peer reviewed at the time of writing; for an entry naming two works, at least one of the two.}}
\label{tab:tool-use-provenance}
\rowcolors{2}{covRow}{white}
\begingroup
\hypersetup{hidelinks}
\begin{tabularx}{\textwidth}{>{\raggedright\arraybackslash}p{1.95cm}
                            >{\raggedright\arraybackslash\scriptsize}p{4.15cm}
                            >{\raggedright\arraybackslash}p{2.55cm}
                            >{\raggedright\arraybackslash}X
                            >{\raggedright\arraybackslash}p{2.45cm}}
\toprule
\textbf{Line of Work} & \textbf{Representative Systems} & \textbf{Tracked Objects} & \textbf{Core mechanism (how)} & \textbf{Provenance Role} \\
\midrule
Tool-use learning & Toolformer~\citep{schick2023toolformer}; ToolLLM~\citep{qin2024toolllm} & Tool calls, API arguments, tool outputs & Learn when and how to call tools from (self-)supervised tool-use trajectories & Tool-call trace construction \\
\addlinespace[0.5pt]
\blue{Prompt-injection and tool-risk benchmarks} & InjecAgent~\citep{zhan2024injecagent}; AgentDojo~\citep{debenedetti2024agentdojo}; ToolEmu~\citep{ruan2024toolemu} & External content, injected instructions, private data & Embed adversarial instructions in external content and measure unsafe actions in (emulated) tool execution & Contamination and risk evaluation \\
\addlinespace[0.5pt]
Instruction--data separation & Instruction Hierarchy\red{\textsuperscript{\dag}}~\citep{wallace2024instructionhierarchy}; StruQ~\citep{chen2024struq}; CaMeL~\citep{debenedetti2025camel} & Privileged instructions, untrusted data, control/data flow & Mark privileged instructions vs.\ untrusted data and separate control flow from data flow via training or structured channels & Data--instruction boundary control \\
\addlinespace[0.5pt]
Information-flow tracking & Fides\red{\textsuperscript{\dag}}~\citep{costa2025fides}; NeuroTaint\red{\textsuperscript{\dag}}~\citep{cai2026neurotaint} & Security labels, tainted values, influence paths & Attach confidentiality/integrity (or taint) labels and propagate them source-to-sink, including through semantic transformations & Source-to-sink enforcement \\
\addlinespace[0.5pt]
Argument provenance & Agent-Sentry\red{\textsuperscript{\dag}}~\citep{sequeira2026agentsentry} & Tool arguments, sources, trust labels & Build an execution-provenance graph and check whether sensitive tool arguments derive from untrusted sources & Sensitive-argument verification \\
\addlinespace[0.5pt]
Execution boundaries
& AgentSpec~\citep{wang2025agentspec}; AgentBound~\citep{buhler2025agentbound}; MCP-SafetyBench~\citep{zong2025mcpsafetybench}
& Agent actions, permissions, policy constraints, and tool interfaces
& Specify, enforce, or evaluate constraints over agent actions and tool access
& Runtime action constraints and boundary-aware provenance \\
\addlinespace[0.5pt]
Guardrails & NeMo Guardrails~\citep{rebedea2023nemo}; Llama Guard\red{\textsuperscript{\dag}}~\citep{inan2023llamaguard} & Inputs, outputs, dialogue states, safety policies & External rule/classifier layer that filters inputs, outputs, and dialogue states against safety policies & Application-level safety filtering \\
\addlinespace[0.5pt]
Trace-based debugging & AgentOps\red{\textsuperscript{\dag}}~\citep{dong2024agentops}; AgentTrace\red{\textsuperscript{\dag}}~\citep{alsayyad2026agenttrace}; TRAIL\red{\textsuperscript{\dag}}~\citep{deshpande2025trail}; LADYBUG~\citep{rorseth2025ladybug} & Structured traces, artifacts, failure locations & Record structured execution traces and localize failures post hoc for audit and repair & Audit, localization, and recovery \\
\bottomrule
\end{tabularx}
\endgroup
\end{table}

\subsection{Summary}
\label{subsec:tool-use-summary}

Tool-using agents make execution provenance necessary because tool calls can consume untrusted information, expose private data, modify external state, and trigger downstream consequences. The core questions are where tool arguments come from, whether tool outputs are trustworthy, how external content influences actions, and whether execution remains within authorized boundaries. Recent work on prompt-injection benchmarks, information-flow control, taint tracking, execution bounding, runtime guardrails, observability, and trace-based debugging marks a shift from output-only safety toward provenance-aware execution control.

\noindent\textbf{Synthesis: from tool logging to influence control.}
Tool-use provenance is moving from recording events to governing influence. Tool-level traces show which API was called, but not whether the call was necessary, justified, or consistent with the user's goal. Parameter lineage can detect untrusted values in sensitive arguments, but may miss invalid interpretation of tool outputs. Information-flow and taint-based methods constrain unsafe influence, but remain difficult under summarization, paraphrasing, memory consolidation, and multi-step reasoning. The next step is to unify tool-call lineage, semantic evidence support, source trust, state changes, policy compliance, and recovery dependencies across the full agent workflow.

\section{Memory as Provenance-Bearing Evidence}
\label{sec:memory}

Memory is the \emph{inward} counterpart of tool use. \blue{Whereas tool calls act on external systems, memory retains evidence and transforms, retrieves, and reuses it across turns, tasks, sessions, and environments.} This makes memory a persistent part of the provenance chain: a memory item may later support a claim, shape a plan, fill a tool argument, update another memory, or justify a final answer. Memory therefore sharpens the trace-completeness tension: agents benefit from compact, persistent, and adaptive memory, but accountability requires source lineage, retrieval context, temporal validity, conflict status, and downstream influence.

This section does not survey memory architectures in general. Existing work has already reviewed memory mechanisms and evaluation protocols for LLM agents~\citep{zhang2025memorysurvey,du2026memoryagents}, as well as graph-based memory systems~\citep{yang2026graphmemory}. We instead treat memory items as evidence-bearing objects. Under this view, a memory is not only something stored and retrieved; it is an artifact with an origin, transformations, revisions, validity conditions, and later effects. Without these links, long-term memory becomes an opaque evidence source rather than an auditable component of agent execution.

\begin{figure*}[t]
    \centering
    \includegraphics[width=0.95\textwidth]{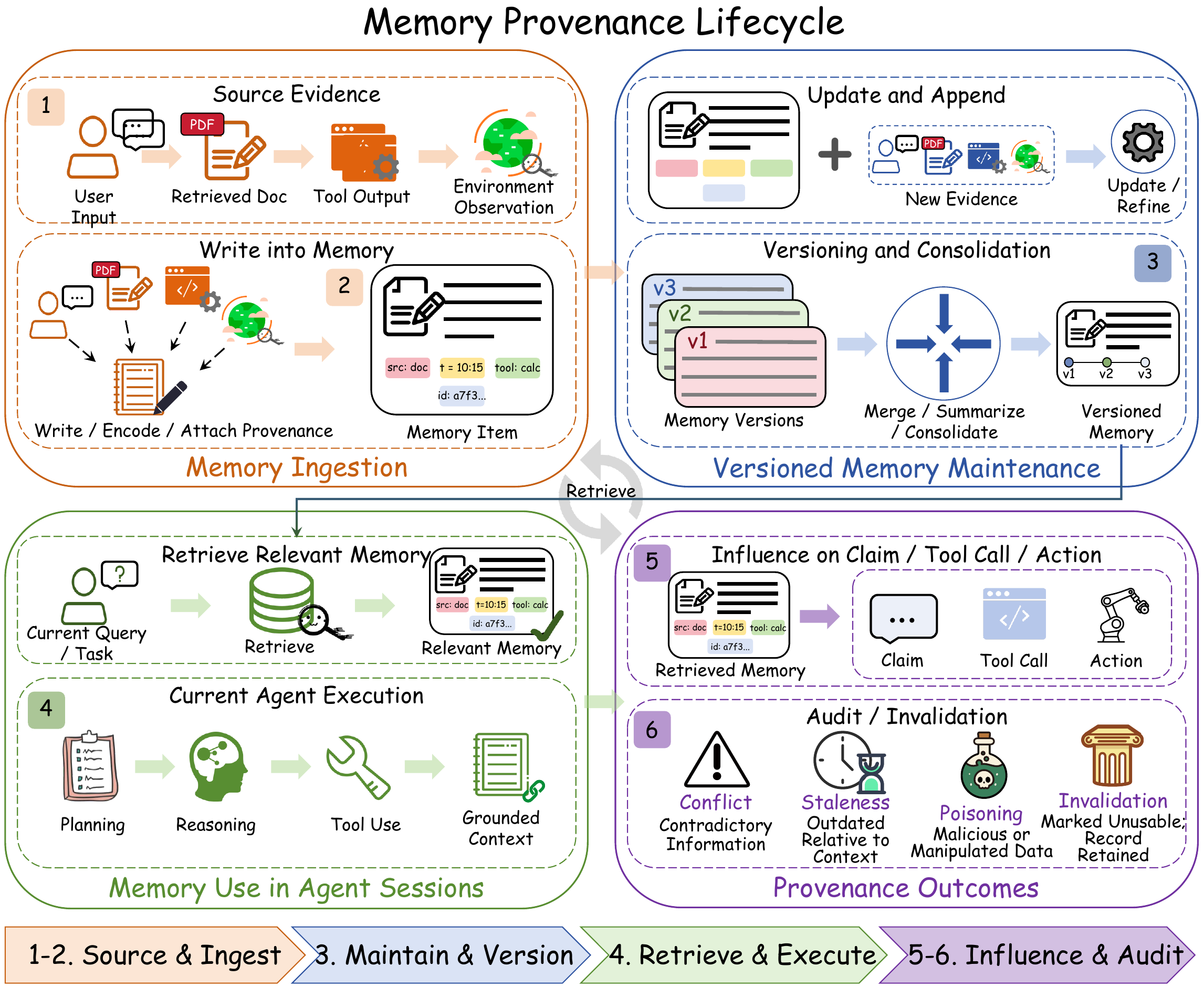}
    \caption{Memory as provenance-bearing evidence across agent sessions. Memory items are written from source evidence, maintained through updates and consolidation, retrieved into later agent executions, used to influence claims and actions, and audited for conflicts, staleness, poisoning, and invalidation.}
    \label{fig:memory-provenance-lifecycle}
\end{figure*}

Figure~\ref{fig:memory-provenance-lifecycle} summarizes this lifecycle. Unlike storage-centric views of memory, it treats memory items as provenance-bearing evidence objects whose origins, updates, retrievals, validity, and downstream influence should remain inspectable.

\subsection{Memory Writes, Source Attribution, and Lineage}
\label{subsec:memory-write-lineage}

Memory provenance begins when a memory item is written. A memory write may originate from user input, retrieved documents, environment observations, tool outputs, feedback, reflections, or messages from other agents. Existing memory systems illustrate different write paths: Generative Agents store observations for later reflection and planning~\citep{park2023generative}; Reflexion writes verbal feedback into episodic memory after failures~\citep{shinn2023reflexion}; MemoryBank updates long-term user memories through interaction~\citep{zhong2024memorybank}; MemGPT manages movement across memory tiers through explicit read and write operations~\citep{packer2023memgpt}; and A-MEM links new memories into an evolving memory structure~\citep{xu2025amem}.

From a provenance perspective, the important unit is not the memory record alone, but the relation between the record and the evidence that produced it. Each memory write should carry source and lineage metadata, including source type, timestamp, authoring agent, supporting evidence, transformation operation, confidence, and update history. This follows the broader provenance principle that generated artifacts should be linked to the entities, activities, and agents that produced them~\citep{w3c2013provdm,souza2025provagent}, and aligns with tracing frameworks that record execution artifacts and operational context~\citep{alsayyad2026agenttrace} \blue{and with provenance graphs that track the sources of tool-call arguments~\citep{sequeira2026agentsentry}}. Such lineage is crucial because memory writes often transform information: documents are summarized, observations are merged, preferences are inferred, and failures are abstracted into reflections. A provenance-bearing memory should therefore distinguish raw observations, extracted facts, inferred memories, and revised memories.

\textbf{Synthesis.} Memory accountability starts at write time. If a memory is stored without source and transformation metadata, later verification can inspect only its surface content, not its evidential basis.

\subsection{Memory Retrieval, Temporal Validity, and Downstream Influence}
\label{subsec:memory-retrieval-validity}

Retrieval determines which stored information re-enters the current execution. Retrieved memories may affect task decomposition, answer generation, tool choice, argument construction, user modeling, and future memory updates. Existing memory systems motivate this retrieval-centered view: MemoryBank and Mem0 emphasize personalization and multi-session continuity~\citep{zhong2024memorybank,chhikara2025mem0}; MemGPT and LongMem manage information beyond the current context window~\citep{packer2023memgpt,wang2023longmem}; and A-MEM improves retrieval through dynamic memory linking~\citep{xu2025amem}.

The provenance risk is that retrieved memory may be only partially relevant, outdated, over-generalized, derived from hallucinated content, poisoned by adversarial interaction, or invalid under the current context. Memory poisoning and injection attacks show that persistent memory can steer later responses and actions across sessions~\citep{chen2024agentpoison,dong2025memoryinjection,tian2026injecmem,pulipaka2026hiddenmemory}. Temporal validity is therefore central: a memory item should record when it was created or updated, what evidence supported it, whether that evidence remains valid, and whether later observations superseded it.

A provenance-aware retrieval trace should record the triggering query or context, retrieved memory items, original sources, relevance signals, validity status, and downstream links to claims, tool calls, actions, final answers, or later memory updates. This would allow systems to identify whether a final output was influenced by a particular memory and whether that influence was justified. It also enables selective invalidation: when a memory is stale, unsupported, private, or contaminated, affected claims, actions, and derived memories can be located instead of treating the execution history as opaque~\citep{wang2025memoryprivacy,wei2025amemguard,sequeira2026agentsentry}.

\textbf{Synthesis.} Retrieval is not only a relevance problem. It is an influence problem: once memory enters context, provenance must track what it shaped and whether that influence remains defensible.

\subsection{Conflicting, Contradictory, and Poisoned Memory}
\label{subsec:conflicting-poisoned-memory}

Persistent memory can accumulate incompatible, invalidated, or contaminated information. A stored memory may conflict with newer evidence, encode an adversarial instruction, or preserve a weak inference as if it were a stable fact. Recent attacks make this risk concrete. AgentPoison shows that poisoned long-term memory or RAG knowledge bases can act as backdoor substrates for agents~\citep{chen2024agentpoison}. Memory-injection attacks show that malicious records can be introduced through ordinary interactions~\citep{dong2025memoryinjection}. Environment-injected and sleeper memory poisoning studies further show that untrusted webpages, documents, or repositories can enter memory, persist silently, and later steer behavior~\citep{zou2026poison,pulipaka2026hiddenmemory}.

This connects memory provenance to tool safety and information-flow control. Fides tracks confidentiality and integrity labels to restrict unsafe information flows~\citep{costa2025fides}, while NeuroTaint emphasizes that unsafe influence can survive semantic transformation and memory reuse rather than appearing as exact text reuse~\citep{cai2026neurotaint}. Indirect prompt-injection benchmarks such as InjecAgent and AgentDojo show how untrusted external content can manipulate tool-using agents~\citep{zhan2024injecagent,debenedetti2024agentdojo}. When such content is written into memory, a single-run attack becomes a long-horizon provenance failure.

\textbf{Synthesis.} Memory quality cannot be measured only by recall, personalization, or coherence. It also depends on source trust, write-path legitimacy, conflict status, contamination risk, and whether a memory is later activated in consequential decisions.

\subsection{Memory-Grounded Verification and Long-Term Audit}
\label{subsec:memory-verification-audit}

Memory-grounded verification extends evidence tracing from retrieved documents and current tool outputs to persistent memory. When a claim or tool action relies on memory, the system should link that memory item to the observation, document, tool call, user statement, or reflection that created it, as well as to later revisions, conflicts, invalidations, and downstream uses. Claim--evidence interfaces such as PaperTrail motivate this granularity by decomposing answers and sources into discrete claims and evidence units~\citep{papertrail2026}. AgentOps and AgentTrace provide foundations for structured execution logs and inspection~\citep{dong2024agentops,alsayyad2026agenttrace}, while Agent-Sentry shows how provenance graphs can support source-aware enforcement over tool-call arguments~\citep{sequeira2026agentsentry}. Long-term memory requires an analogous structure across sessions.

Despite progress in systems such as MemGPT, A-MEM, and Mem0~\citep{packer2023memgpt,xu2025amem,chhikara2025mem0}, \blue{the memory systems reviewed here mainly optimize} for recall, personalization, efficiency, or coherence, with limited support for tracing origins, temporal validity, conflicts, contamination, privacy exposure, and downstream influence. This gap is critical because memory itself can become an attack surface: poisoned, injected, stale, or private memories may later shape claims and actions without appearing as external evidence~\citep{chen2024agentpoison,tian2026injecmem,pulipaka2026hiddenmemory,wang2025memoryprivacy}. Defense work such as A-MemGuard suggests that memory should support checking and correction, not only input-time filtering~\citep{wei2025amemguard}.

Table~\ref{tab:memory-provenance} summarizes this gap. Across representative memory systems and security frameworks, write and retrieval support is common, but \blue{conflict handling, staleness detection, and evidence-aware verification} remain much less developed.

\begin{table}[!t]
\centering
\scriptsize
\setlength{\tabcolsep}{2.0pt}
\renewcommand{\arraystretch}{1.04}
\caption{\blue{Agent memory through a provenance lens. \MemYes{} explicit, documented mechanism; \MemPartial{} achieved indirectly or only for some memory types; \MemLimited{} information exposed but not implemented; \MemNo{} absent; \MemAttack{} capability studied adversarially rather than provided.} \red{\textsuperscript{\dag}\,marks a work that had not been peer reviewed at the time of writing; for an entry naming two works, at least one of the two.}}
\label{tab:memory-provenance}
\begingroup
\hypersetup{hidelinks}
\rowcolors{2}{covRow}{white}
\begin{tabularx}{\textwidth}{>{\raggedright\arraybackslash\scriptsize}p{3.55cm}
                            >{\raggedright\arraybackslash}X
                            >{\raggedright\arraybackslash}p{2.45cm}
                            C{0.80cm}
                            C{0.78cm}
                            C{0.72cm}
                            C{0.72cm}
                            C{0.84cm}}
\toprule
\rowcolor{covHeader}
\textbf{Paper / System}
& \textbf{Memory Type}
& \textbf{Update / Evolution}
& \textbf{Write}
& \textbf{Retr.}
& \textbf{Conf.}
& \textbf{Stale}
& \textbf{Verify} \\
\midrule

Generative Agents~\citep{park2023generative}
& Memory stream / reflection
& Reflection-based
& \MemPartial
& \MemPartial
& \MemNo
& \MemNo
& \MemNo \\

Reflexion~\citep{shinn2023reflexion}
& Episodic memory / feedback
& Feedback-driven
& \MemPartial
& \MemPartial
& \MemNo
& \MemNo
& \MemPartial \\

MemoryBank~\citep{zhong2024memorybank}
& Long-term interaction memory
& Continuous update
& \MemPartial
& \MemYes
& \MemLimited
& \MemLimited
& \MemNo \\

MemGPT\red{\textsuperscript{\dag}}~\citep{packer2023memgpt}
& Hierarchical virtual memory
& Memory-tier management
& \MemYes
& \MemYes
& \MemLimited
& \MemLimited
& \MemNo \\

LongMem~\citep{wang2023longmem}
& Long-term memory for LMs
& Cache and update
& \MemPartial
& \MemYes
& \MemNo
& \MemPartial
& \MemNo \\

Mem0~\citep{chhikara2025mem0}
& Production long-term memory
& Extraction and consolidation
& \MemYes
& \MemYes
& \MemLimited
& \MemPartial
& \MemPartial \\

A-MEM~\citep{xu2025amem}
& Agentic interconnected memory
& Dynamic linking and evolution
& \MemYes
& \MemYes
& \MemLimited
& \MemPartial
& \MemNo \\

AgentPoison / Memory Injection~\citep{chen2024agentpoison,dong2025memoryinjection}
& Poisoned or injected memory
& Cross-session persistence
& \MemAttack
& \MemAttack
& \MemNo
& \MemNo
& \MemNo \\

Fides / NeuroTaint\red{\textsuperscript{\dag}}~\citep{costa2025fides,cai2026neurotaint}
& Information-flow and taint tracking
& Tracks influence
& \MemYes
& \MemYes
& \MemYes
& \MemPartial
& \MemYes \\

AgentOps / AgentTrace\red{\textsuperscript{\dag}}~\citep{dong2024agentops,alsayyad2026agenttrace}
& Execution logs and traces
& Trace-level history
& \MemPartial
& \MemPartial
& \MemNo
& \MemNo
& \MemPartial \\

Agent-Sentry\red{\textsuperscript{\dag}}~\citep{sequeira2026agentsentry}
& Execution provenance graph
& Runtime provenance
& \MemYes
& \MemYes
& \MemPartial
& \MemPartial
& \MemYes \\

\bottomrule
\end{tabularx}
\endgroup
\end{table}

\noindent\textbf{Synthesis: memory usefulness versus memory accountability.}
Compression, consolidation, and persistence make memory useful; lineage, validity, conflict tracking, and influence links make it governable. \blue{The columns of Table~\ref{tab:memory-provenance} divide by literature.} \red{Write and retrieval support is present in every memory system reviewed here, and not one of them reaches full support on conflict handling or on evidence-aware verification; the only rows anywhere above partial in those two columns are Fides/NeuroTaint, which reaches full support on both, and Agent-Sentry, which reaches it on verification and stays partial on conflict. Both are information-flow and execution-provenance systems that observe memory from outside rather than memory systems that record their own lineage. The division has a consequence. Those systems establish conflict and support by tracking influence during a run, so what they recover is the current execution's use of a memory item rather than the item's history, which is why staleness remains partial even for them.} \blue{Among the systems reviewed here, memory accountability is therefore supplied mainly from outside memory systems, and it does not persist beyond the session in which it was computed.}

\section{Benchmarks, Datasets, and Metrics}
\label{sec:benchmarks}

Evaluation is central to evidence tracing and execution provenance. Beyond final-answer correctness, provenance-aware agent evaluation must assess evidence support, tool-call justification, unsafe influence, trace completeness, failure localization, and recovery. Existing benchmarks provide partial views across RAG attribution, citation evaluation, agent execution, tool-use safety, memory, and multi-agent analysis, but \red{no entry in the catalog of Table~\ref{tab:benchmarks-provenance} evaluates} complete provenance across evidence, tools, memory, traces, safety, and recovery. This section summarizes representative benchmark families, metrics, and limitations.

To provide a high-level view of these gaps, Figure~\ref{fig:benchmark-coverage-heatmap} summarizes the coverage of major benchmark families across provenance-related evaluation capabilities.

\begin{figure*}[t]
\centering
\small
\setlength{\tabcolsep}{4pt}
\renewcommand{\arraystretch}{1.35}

\noindent\resizebox{\textwidth}{!}{%
\begin{tabular}{L{3.6cm} *{5}{C{1.55cm}} C{1.9cm} C{1.55cm}}
\toprule
\rowcolor{covHeader}
\makecell[c]{\textbf{Benchmark}\\\textbf{Family}}
& \makecell[c]{\textbf{Evidence}\\\textbf{Labels}}
& \makecell[c]{\textbf{Tool}\\\textbf{Calls}}
& \makecell[c]{\textbf{Memory}}
& \makecell[c]{\textbf{Multi-}\\\textbf{agent}}
& \makecell[c]{\textbf{Safety}\\\textbf{Attacks}}
& \makecell[c]{\textbf{Provenance}\\\textbf{Relations}}
& \makecell[c]{\textbf{Recovery}} \\
\midrule

\rowcolor{covRow}
RAG benchmarks
& \Strong & \Limited & \Limited & \Limited & \Limited & \Partial & \Limited \\

Tool-use benchmarks
& \Limited & \Strong & \Partial & \Limited & \Partial & \Limited & \Limited \\

\rowcolor{covRow}
Memory benchmarks
& \Partial & \Partial & \Strong & \Limited & \Limited & \Partial & \Partial \\

Multi-agent benchmarks
& \Partial & \Partial & \Partial & \Strong & \Limited & \Partial & \Limited \\

\rowcolor{covRow}
Trace-debugging benchmarks
& \Partial & \Partial & \Partial & \Strong & \Limited & \Partial & \Partial \\

Provenance/security benchmarks
& \Partial & \Strong & \Strong & \Partial & \Strong & \Partial & \Partial \\

\bottomrule
\end{tabular}%
}
\vspace{2mm}
\noindent\begin{minipage}{0.96\textwidth}
\footnotesize
\textbf{Legend:}
\colorbox{covStrong}{\textcolor{white}{\strut\textbf{S}}} Strong coverage \quad
\colorbox{covPartial}{\strut\textbf{P}} Partial coverage \quad
\colorbox{covLimited}{\strut\textbf{L}} Limited or absent coverage.
\vspace{0.6mm}
\textit{\blue{Coding rule.}} \blue{Each cell reports the highest capability rating among the benchmarks in that family. Ratings follow the column-specific criteria detailed in Appendix~\ref{app:heatmap-derivation}, which derives every cell. Family-level ratings therefore indicate the strongest available coverage, not coverage shared by all members.}
\vspace{0.6mm}
\textit{\red{Family membership.}} \textbf{\red{RAG}}\red{: ALCE, RAGAS, ARES, RAGChecker, RAGTruth, FActScore, SourceCheckup. }\textbf{\red{Tool-use}}\red{: AgentBench, WebArena, ToolBench/ToolLLM, $\tau$-bench. }\textbf{\red{Memory}}\red{: LoCoMo, MemoryArena. }\textbf{\red{Multi-agent}}\red{: MAST, AgenTracer/Aegis. }\textbf{\red{Trace-debugging}}\red{: TRAIL, MAST. }\textbf{\red{Provenance/security}}\red{: ToolEmu, InjecAgent, AgentDojo, OpenAgentSafety, MCP-SafetyBench, Agent-Sentry Bench, NeuroTaint/TaintBench, AgentSpec, AgentBound.
}

\end{minipage}

\caption{Benchmark coverage heatmap across provenance-related evaluation capabilities. The heatmap summarizes benchmark families at a high level rather than listing individual datasets. It highlights that existing benchmarks provide partial coverage of isolated components, while \blue{end-to-end} evaluation of evidence tracing and execution provenance remains underdeveloped.}
\label{fig:benchmark-coverage-heatmap}
\end{figure*}

\blue{The heatmap shows that strong coverage is concentrated in the capabilities each family was built around, such as evidence grounding, tool use, memory, multi-agent interaction, or safety attacks, and that no family reaches it on provenance relations or recovery.} This motivates the more detailed discussion of benchmark families below.

\subsection{Benchmark Families for Evidence Tracing}
\label{subsec:rag-attribution-benchmarks}

RAG and attribution benchmarks provide the closest foundation for evidence tracing because they evaluate whether generated outputs are grounded in retrieved or cited evidence~\citep{gao2023alce,es2024ragas,saadfalcon2024ares,min2023factscore,niu2024ragtruth,ru2024ragchecker,wu2024sourcecheckup}. These benchmarks generally focus on the relationship between a final answer and its supporting sources, rather than the full execution trajectory of an agent. Nevertheless, they provide important metrics and datasets for evaluating evidence recall, citation quality, faithfulness, and claim-level support.

At the answer level, ALCE treats citations as explicit links between answer spans and external sources and checks whether those citations support the generated content, while RAGAS and ARES score reference-free faithfulness, context relevance, and answer relevance using lightweight model judges~\citep{gao2023alce,es2024ragas,saadfalcon2024ares}. These frameworks share one assumption, that an answer should be supported by retrieved evidence. \red{RAGAS does decompose the answer, extracting a set of statements and scoring each for support, but the check is made against the retrieved context as a whole; what }\blue{none of these frameworks reports} \red{is a link from a claim to the particular passage that supports it.}

Finer-grained benchmarks move toward claim-level provenance, differing mainly in the unit they score: retrieval-versus-generation error separation (RAGChecker), word-level hallucination annotation (RAGTruth), atomic-fact support (FActScore), source-supportiveness of cited references (SourceCheckup), and supported/refuted/unverifiable claim classification (FEVER)~\citep{ru2024ragchecker,niu2024ragtruth,min2023factscore,wu2024sourcecheckup,thorne2018fever}. The trend is a shift from answer-level to claim-level grounding; SourceCheckup sharpens the point by showing that citation presence does not imply evidence support.

These benchmarks are important for evidence tracing, but they remain limited from an agent-provenance perspective. Their primary evaluation targets are answer-source grounding, citation support, faithfulness, hallucination annotation, or claim-level factual support~\citep{gao2023alce,es2024ragas,saadfalcon2024ares,min2023factscore,niu2024ragtruth,ru2024ragchecker,wu2024sourcecheckup}. As a result, they provide only partial coverage of agent provenance: they help evaluate whether claims are supported by evidence, but they do not directly evaluate how tool calls, memory operations, environment observations, intermediate decisions, or multi-agent communication shape the final answer or action. Thus, they should be viewed as foundations for evidence attribution rather than complete benchmarks for execution provenance.

\subsection{Benchmarks for Agent Execution Traces and Tool-Use Safety}
\label{subsec:agent-trace-tool-safety-benchmarks}

Agent benchmarks extend evaluation from static answer generation to interactive, multi-step execution, exposing trajectories, tool calls, environment states, and intermediate decisions. They differ mainly in setting and in what they score: multi-environment capability (AgentBench), long-horizon web interaction (WebArena), API selection and execution paths (ToolBench/ToolLLM), and tool-agent-user workflows that grade policy following and final database state rather than only the final response ($\tau$-bench)~\citep{liu2024agentbench,zhou2024webarena,qin2024toolllm,yao2025taubench}. The common trend is a move from scoring the answer toward scoring the process and its end state.

Trace-oriented benchmarks connect execution provenance to failure diagnosis. AgentTrace provides structured operational, cognitive, and contextual logs for agent observability and downstream audit~\citep{alsayyad2026agenttrace}, while TRAIL uses human-annotated traces to evaluate whether systems can localize failures to specific steps, components, or interactions~\citep{deshpande2025trail}. Recent multi-agent work moves further from trace inspection to explicit failure attribution: AgenTracer constructs TracerTraj through counterfactual replay and fault injection to identify responsible agents and steps~\citep{zhang2025agentracer}, and Aegis generates positive--negative trajectory pairs with faulty-agent and error-mode labels~\citep{kong2025aegis}. Together, these works show that provenance can be evaluated \blue{by its ability to support failure localization and attribution, not only by whether a trace was recorded}.

Tool-use safety benchmarks examine whether agents remain safe when tools introduce external risks. ToolEmu evaluates high-risk tool use in an LM-emulated sandbox~\citep{ruan2024toolemu}; InjecAgent and AgentDojo study indirect prompt injection and adversarial tool interactions in tool-integrated settings~\citep{zhan2024injecagent,debenedetti2024agentdojo}; and OpenAgentSafety extends evaluation to real tools and multi-turn, multi-user tasks across multiple risk categories~\citep{vijayvargiya2026openagentsafety}. MCP-SafetyBench further studies safety risks in real MCP servers and multi-server workflows~\citep{zong2025mcpsafetybench}. These benchmarks matter for provenance because unsafe behavior often depends on hidden flows from untrusted sources into tool arguments, execution boundaries, or state-changing actions.

Recent work evaluates more explicit execution-provenance and information-flow mechanisms. Agent-Sentry introduces execution provenance graphs to bound LLM-agent behavior and detect out-of-bounds tool calls~\citep{sequeira2026agentsentry}. NeuroTaint and TaintBench evaluate semantic taint and source-to-sink influence in LLM agents, emphasizing that unsafe influence may be transformed through summarization, inference, or memory rather than copied verbatim~\citep{cai2026neurotaint}. Fides applies information-flow control to secure AI agents by tracking confidentiality and integrity labels~\citep{costa2025fides}. AgentSpec evaluates runtime policy enforcement through customizable constraints~\citep{wang2025agentspec}. AgentBound extends this line of work to execution boundaries and access control in MCP-style agent-tool ecosystems~\citep{buhler2025agentbound}. Together, these benchmarks and systems shift evaluation from ``can the agent use tools?'' to ``can the agent use tools safely, with traceable and enforceable provenance?''

\subsection{Memory and Multi-Agent Provenance Evaluation}
\label{subsec:memory-multiagent-provenance-eval}

Memory and multi-agent settings introduce additional provenance challenges because information may persist across sessions, be updated over time, or propagate across agents. Existing memory benchmarks and systems mainly evaluate recall, personalization, long-term consistency, memory updating, retrieval, or multi-session task performance~\citep{zhong2024memorybank,packer2023memgpt,xu2025amem,chhikara2025mem0,maharana2024locomo,he2026memoryarena}. MemoryBank evaluates long-term memory recall, continuous updating, and personalization~\citep{zhong2024memorybank}; MemGPT makes memory read and write operations explicit through hierarchical memory management~\citep{packer2023memgpt}; A-MEM organizes agent memory through dynamically evolving interconnected memory structures~\citep{xu2025amem}; Mem0 evaluates memory extraction, consolidation, and retrieval in multi-session dialogue settings~\citep{chhikara2025mem0}; LoCoMo evaluates long-term conversational memory~\citep{maharana2024locomo}; and MemoryArena evaluates memory use in interdependent multi-session agentic tasks~\citep{he2026memoryarena}. Existing memory and graph-memory surveys further summarize progress in memory mechanisms, evaluation protocols, and graph-based memory systems~\citep{zhang2025memorysurvey,du2026memoryagents,yang2026graphmemory}.

From a provenance perspective, these benchmarks provide useful but incomplete proxies for memory provenance. They often test whether an agent remembers useful information, but \red{neither of the two memory benchmarks in Table~\ref{tab:benchmarks-provenance} evaluates} whether a memory item is traceable to its source, temporally valid, consistent with newer evidence, protected from contamination, or responsible for a downstream action. Dedicated memory-provenance evaluation should therefore assess memory source attribution, update history, conflict resolution, stale-memory invalidation, contamination detection, and downstream influence, which remain underdeveloped in current memory evaluation protocols~\citep{maharana2024locomo,he2026memoryarena,zhang2025memorysurvey,du2026memoryagents}.

Multi-agent systems introduce another layer of provenance complexity because evidence, assumptions, messages, and errors can propagate across agent boundaries. AutoGen and CAMEL provide representative frameworks for multi-agent communication, role assignment, and collaborative task solving~\citep{wu2023autogen,li2023camel}. Evaluating provenance in such settings requires more than measuring final task success: it requires identifying which agent contributed which evidence, which communication step introduced an error, and how responsibility is distributed across agents. MAST analyzes failures from specification errors, inter-agent misalignment, and weak task verification~\citep{cemri2025mast}, while TRAIL evaluates issue localization from agent traces~\citep{deshpande2025trail}. AgenTracer and Aegis further construct labeled multi-agent trajectories for identifying faulty agents, responsible steps, and error modes~\citep{zhang2025agentracer,kong2025aegis}. Together, these works suggest that multi-agent provenance benchmarks should evaluate evidence propagation, communication-level error introduction, cross-agent dependency, and responsibility attribution, rather than only final coordination success.

\subsection{Metrics for Evidence Tracing and Execution Provenance}
\label{subsec:provenance-metrics}

Existing benchmarks use diverse metrics for retrieval grounding, citation quality, tool use, safety, memory, and agent execution~\citep{gao2023alce,min2023factscore,ru2024ragchecker,wu2024sourcecheckup,ruan2024toolemu,debenedetti2024agentdojo,maharana2024locomo,deshpande2025trail,cemri2025mast}. From the perspective of evidence tracing and execution provenance, these metrics can be reorganized into four evaluation aspects: evidence attribution, execution provenance, safety and robustness, and debugging and recovery. Table~\ref{tab:provenance-metrics} summarizes representative metric dimensions and their related foundations. Some metrics are already used in RAG, attribution, tool-use, and agent benchmarks, while others are natural extensions for provenance-aware evaluation.

\begin{table}[!t]
\centering
\footnotesize
\setlength{\tabcolsep}{3pt}
\renewcommand{\arraystretch}{1.08}
\caption{Representative metric dimensions for evaluating evidence tracing and execution provenance in LLM agents. The \textbf{Status} column flags maturity: \emph{Established} metrics are operationalized in existing RAG, attribution, tool-use, or safety benchmarks; \emph{Partly established} metrics exist for some sub-tasks (e.g., failure localization in TRAIL/MAST) but not for recovery; \emph{Proposed} metrics (trace completeness, provenance accuracy, dependency coverage, temporal consistency) are desiderata that are not yet operationalized with agreed definitions. \ding{51}: Established; \textbullet: Partly established; \ding{72}: Proposed.}
\label{tab:provenance-metrics}
\resizebox{\textwidth}{!}{%
\rowcolors{2}{covRow}{white}
\begin{tabular}{p{2.1cm} p{3.2cm} p{4.0cm} >{\centering\arraybackslash}p{1.2cm} p{4.0cm}}
\toprule
\textbf{Evaluation Aspect} & \textbf{Core Question} & \textbf{Representative Metrics} & \textbf{Status} & \textbf{Related Basis} \\
\midrule

Evidence attribution
& Are the agent's claims supported by reliable evidence?
& Evidence recall, citation precision, source supportiveness, faithfulness, claim support accuracy.
& \ding{51}
& RAG and attribution evaluation~\citep{gao2023alce,min2023factscore,niu2024ragtruth,ru2024ragchecker,wu2024sourcecheckup}. \\

\midrule

Execution provenance
& Is the agent's execution process completely and correctly traceable?
& Trace completeness, provenance accuracy, dependency coverage, temporal consistency.
& \ding{72}
& Provenance standards and agent observability~\citep{w3c2013provdm,dong2024agentops,alsayyad2026agenttrace,souza2025provagent}. \\

\midrule

Safety and robustness
& Can the system detect unsafe or untrusted influence during execution?
& Unsafe influence detection, attack success rate, policy violation rate\blue{, false-positive rate}.
& \ding{51}
& Tool-use safety, prompt injection, and information-flow control~\citep{ruan2024toolemu,zhan2024injecagent,debenedetti2024agentdojo,debenedetti2025camel,costa2025fides,zong2025mcpsafetybench}. \\

\midrule

Debugging and recovery
& Can traces help identify and repair failures?
& Failure localization accuracy, diagnosis correctness, auditability, recovery success.
& \textbullet
& Trace-based debugging, multi-agent failure analysis, and failure-attribution datasets~\citep{deshpande2025trail,cemri2025mast,alsayyad2026agenttrace,zhang2025agentracer,kong2025aegis}. \\

\bottomrule
\end{tabular}%
}
\end{table}

\begin{table}[!t]
\centering
\scriptsize
\setlength{\tabcolsep}{2.2pt}
\renewcommand{\arraystretch}{1.08}
\caption{\blue{Benchmark catalog for evidence tracing and execution provenance (25 entries). \BenchYes{} per-instance ground-truth labels released with a metric over them; \BenchPartial{} present in the released data but unlabeled or only partly covered; \BenchNo{} neither released; \BenchTask{} coverage varies across the benchmark's task suite; \BenchState{} realized as external environment state rather than as an agent-side artifact.} \red{\textsuperscript{\dag}\,marks a work that had not been peer reviewed at the time of writing; for an entry naming two works, at least one of the two.}}
\label{tab:benchmarks-provenance}
\resizebox{\textwidth}{!}{%
\rowcolors{2}{covRow}{white}
\begin{tabular}{l p{2.7cm} C{0.9cm} C{0.9cm} C{0.8cm} C{0.9cm} C{0.9cm} C{0.95cm} p{3.7cm}}
\toprule
\textbf{Benchmark} & \textbf{Task Type} & \makecell{\textbf{Trace}\\\textbf{Avail.}} & \makecell{\textbf{Evid.}\\\textbf{Labels}} & \makecell{\textbf{Tool}\\\textbf{Calls}} & \textbf{Mem.} & \makecell{\textbf{Multi}\\\textbf{Agent}} & \makecell{\textbf{Safety}\\\textbf{Attack}} & \textbf{Main Metrics} \\
\midrule

ALCE~\citep{gao2023alce}
& Citation generation
& \BenchPartial{}
& \BenchYes{}
& \BenchNo{}
& \BenchNo{}
& \BenchNo{}
& \BenchNo{}
& Citation quality, correctness, fluency \\

RAGAS~\citep{es2024ragas}
& RAG evaluation
& \BenchNo{}
& \BenchPartial{}
& \BenchNo{}
& \BenchNo{}
& \BenchNo{}
& \BenchNo{}
& Faithfulness, context relevance, answer relevance \\

ARES~\citep{saadfalcon2024ares}
& RAG evaluation
& \BenchNo{}
& \BenchPartial{}
& \BenchNo{}
& \BenchNo{}
& \BenchNo{}
& \BenchNo{}
& Context relevance, answer faithfulness, answer relevance \\

RAGChecker~\citep{ru2024ragchecker}
& Fine-grained RAG diagnosis
& \BenchPartial{}
& \BenchPartial{}
& \BenchNo{}
& \BenchNo{}
& \BenchNo{}
& \BenchNo{}
& Retrieval and generation diagnosis \\

RAGTruth~\citep{niu2024ragtruth}
& RAG hallucination annotation
& \BenchPartial{}
& \BenchYes{}
& \BenchNo{}
& \BenchNo{}
& \BenchNo{}
& \BenchNo{}
& Word-level hallucination labels \\

FActScore~\citep{min2023factscore}
& Atomic factuality
& \BenchNo{}
& \BenchYes{}
& \BenchNo{}
& \BenchNo{}
& \BenchNo{}
& \BenchNo{}
& Atomic fact support \\

SourceCheckup~\citep{wu2024sourcecheckup}
& Source supportiveness
& \BenchNo{}
& \BenchYes{}
& \BenchNo{}
& \BenchNo{}
& \BenchNo{}
& \BenchNo{}
& Source relevance and supportiveness \\

AgentBench~\citep{liu2024agentbench}
& Agent capability evaluation
& \BenchPartial{}
& \BenchNo{}
& \BenchPartial{}
& \BenchTask{}
& \BenchNo{}
& \BenchNo{}
& Task success, environment-specific scores \\

WebArena~\citep{zhou2024webarena}
& Web-agent execution
& \BenchYes{}
& \BenchNo{}
& \BenchYes{}
& \BenchNo{}
& \BenchNo{}
& \BenchNo{}
& Task success, web interaction outcome \\

ToolBench / ToolLLM~\citep{qin2024toolllm}
& API tool use
& \BenchYes{}
& \BenchNo{}
& \BenchYes{}
& \BenchNo{}
& \BenchNo{}
& \BenchNo{}
& Tool selection, argument correctness, execution success \\

$\tau$-bench~\citep{yao2025taubench}
& Tool-agent-user interaction
& \BenchYes{}
& \BenchNo{}
& \BenchYes{}
& \BenchState{}
& \BenchNo{}
& \BenchPartial{}
& Task success, policy following, final state \\

TRAIL\red{\textsuperscript{\dag}}~\citep{deshpande2025trail}
& Trace debugging
& \BenchYes{}
& \BenchPartial{}
& \BenchPartial{}
& \BenchPartial{}
& \BenchPartial{}
& \BenchNo{}
& Issue localization, error type diagnosis \\

MAST~\citep{cemri2025mast}
& Multi-agent failure analysis
& \BenchYes{}
& \BenchNo{}
& \BenchTask{}
& \BenchTask{}
& \BenchYes{}
& \BenchNo{}
& Failure taxonomy, error attribution \\

ToolEmu~\citep{ruan2024toolemu}
& Tool-use risk evaluation
& \BenchYes{}
& \BenchNo{}
& \BenchYes{}
& \BenchNo{}
& \BenchNo{}
& \BenchYes{}
& Risk detection, unsafe action rate \\

InjecAgent~\citep{zhan2024injecagent}
& Indirect prompt injection
& \BenchYes{}
& \BenchNo{}
& \BenchYes{}
& \BenchNo{}
& \BenchNo{}
& \BenchYes{}
& Attack success rate, defense performance \\

AgentDojo~\citep{debenedetti2024agentdojo}
& Prompt injection attacks and defenses
& \BenchYes{}
& \BenchNo{}
& \BenchYes{}
& \BenchNo{}
& \BenchNo{}
& \BenchYes{}
& Attack success, utility, defense robustness \\

Agent-Sentry Bench\red{\textsuperscript{\dag}}~\citep{sequeira2026agentsentry}
& Out-of-bounds tool execution
& \BenchYes{}
& \BenchPartial{}
& \BenchYes{}
& \BenchNo{}
& \BenchNo{}
& \BenchYes{}
& Out-of-bounds detection, false positives \\

NeuroTaint / TaintBench\red{\textsuperscript{\dag}}~\citep{cai2026neurotaint}
& Semantic taint tracking
& \BenchYes{}
& \BenchPartial{}
& \BenchYes{}
& \BenchYes{}
& \BenchPartial{}
& \BenchYes{}
& Unsafe influence detection, taint propagation \\

AgentBound~\citep{buhler2025agentbound}
& MCP execution boundary
& \BenchYes{}
& \BenchNo{}
& \BenchYes{}
& \BenchNo{}
& \BenchNo{}
& \BenchYes{}
& Access-control violations, policy enforcement \\

\red{AgenTracer / Aegis~\citep{zhang2025agentracer,kong2025aegis}}
& \red{Multi-agent failure attribution}
& \BenchYes{}
& \BenchPartial{}
& \BenchTask{}
& \BenchTask{}
& \BenchYes{}
& \BenchNo{}
& \red{Responsible-agent and error-mode attribution} \\

\red{OpenAgentSafety~\citep{vijayvargiya2026openagentsafety}}
& \red{Real-tool safety evaluation}
& \BenchYes{}
& \BenchPartial{}
& \BenchYes{}
& \BenchTask{}
& \BenchPartial{}
& \BenchYes{}
& \red{Unsafe action rate, task utility, risk categories} \\

\red{MCP-SafetyBench~\citep{zong2025mcpsafetybench}}
& \red{MCP workflow safety}
& \BenchYes{}
& \BenchNo{}
& \BenchYes{}
& \BenchNo{}
& \BenchPartial{}
& \BenchYes{}
& \red{Attack success, policy violation, workflow risk} \\

\red{AgentSpec~\citep{wang2025agentspec}}
& \red{Runtime policy enforcement}
& \BenchYes{}
& \BenchNo{}
& \BenchYes{}
& \BenchNo{}
& \BenchPartial{}
& \BenchYes{}
& \red{Policy violation rate, enforcement accuracy} \\

LoCoMo~\citep{maharana2024locomo}
& Long-term conversational memory
& \BenchPartial{}
& \BenchPartial{}
& \BenchNo{}
& \BenchYes{}
& \BenchNo{}
& \BenchNo{}
& Long-range QA, event summarization, memory recall \\

MemoryArena~\citep{he2026memoryarena}
& Multi-session agent memory
& \BenchYes{}
& \BenchPartial{}
& \BenchTask{}
& \BenchYes{}
& \BenchNo{}
& \BenchNo{}
& Multi-session task success, memory use \\

\bottomrule
\end{tabular}%
}
\end{table}

\paragraph{\blue{Research agenda for provenance metrics.}}
\red{The execution-provenance row of Table~\ref{tab:provenance-metrics} is the only row marked }\ding{72}\red{: trace completeness, provenance accuracy, dependency coverage, and temporal consistency are }\emph{\red{not}} \blue{yet} \red{available as measures. None has an agreed definition in the work we surveyed, none has an established ground-truth construction there, and }\blue{they are not offered as an evaluation framework ready for adoption. We describe each with the ground truth it would require, so that the gap can be acted on.}\red{
}

\emph{\red{Trace completeness}} \red{would ask what fraction of the units and relations that actually occurred are present in the recorded trace. Its ground truth is the hardest of the four to obtain from observation, because the missing part of a trace is by definition unrecorded; the practical route is synthetic or instrumented execution, where a reference harness emits an exhaustive trace and the system under evaluation is scored against it. }\emph{\red{Dependency coverage}} \red{would ask what fraction of the }\textsc{\red{Depend-on}} \red{and }\textsc{\red{Trigger}} \red{edges implied by an execution are recovered. Its ground truth is obtainable by ablation and replay: perturb each candidate dependency and observe whether the dependent step changes~\citep{zhang2025agentracer}. This makes it the most tractable of the \blue{three that need external ground truth}, at a cost of one replay per candidate edge. }\emph{\red{Temporal consistency}} \red{would ask whether the recorded ordering and validity intervals are internally coherent, for example whether an item was used after it was invalidated. This is the only one of the four that needs no external ground truth, since it is a self-consistency check on the graph, and it is best reported as a violation count rather than an accuracy. }\emph{\red{Provenance accuracy}} \red{would ask whether the recorded relations are the correct ones, and it is the metric whose definition is genuinely open, for the reason given next.
}

\paragraph{\red{Why provenance accuracy resists a single gold graph.}}
\red{The same execution frequently admits several defensible dependency graphs. Annotators may disagree on granularity, on whether a passage that was present but unnecessary is a dependency, and on whether a step was triggered by an observation or by the plan that anticipated it. Scoring exact-match against one designated gold graph therefore measures annotator agreement as much as system quality. We see three defensible responses}\blue{; which one the field should adopt remains open.} \red{The first is to score against an }\emph{\red{admissible set}}\red{: annotate the set of graphs consistent with the trace under stated conventions and count a system correct if its output lies in that set. The second is to score edges rather than graphs, reporting per-relation precision and recall against a union of annotations, weighted by how many annotators asserted each edge, so that contested edges contribute less than unanimous ones. The third is to define a canonical form and score against it, for instance the }\emph{\red{minimal sufficient graph}}\red{, }\blue{a subgraph in which every edge is necessary} \red{to explain the observed outputs under replay. This sidesteps disagreement about redundant edges at the cost of discarding genuine over-determination. In all three cases the measurement is meaningful only if inter-annotator agreement is reported alongside it, since that agreement is the ceiling on any achievable score. This is compounded by the origin heterogeneity of Section~\ref{subsec:origin-assurance}: instrumented edges can be checked against the log and inferred edges cannot, so a single accuracy number over a mixed graph conflates a reproducible check with a judgment call, and the two should be reported separately.
}

The table highlights that provenance-aware evaluation should go beyond final-answer correctness and component-level task success. Building on prior work in attribution evaluation, provenance modeling, agent observability, tool-use safety, and trace-based debugging~\citep{gao2023alce,min2023factscore,ru2024ragchecker,w3c2013provdm,dong2024agentops,alsayyad2026agenttrace,souza2025provagent,ruan2024toolemu,debenedetti2024agentdojo,deshpande2025trail,cemri2025mast}, we organize these metrics by their evaluation function. Evidence attribution metrics assess claim grounding; execution-provenance metrics\blue{, once operationalized, would} assess traceability; safety and robustness metrics assess unsafe influence and policy violations; and debugging and recovery metrics assess whether traces support failure localization, audit, and repair. Together, these dimensions provide a broader evaluation view for agents whose outputs depend on retrieval, tools, memory, communication, and external state changes.

\subsection{Limitations of Current Evaluation Protocols}
\label{subsec:evaluation-limitations}

Existing benchmarks provide useful foundations for evaluating retrieval grounding, tool use, memory, web interaction, safety, and multi-agent behavior, but they remain insufficient for full evidence tracing and execution provenance. RAG and attribution benchmarks mainly focus on answer-source grounding or citation quality~\citep{gao2023alce,ru2024ragchecker,wu2024sourcecheckup}. Tool-use and agent benchmarks emphasize task success, unsafe tool behavior, prompt injection, environment interaction, real-world tool safety, or MCP-specific attacks~\citep{liu2024agentbench,zhou2024webarena,ruan2024toolemu,debenedetti2024agentdojo,yao2025taubench,vijayvargiya2026openagentsafety,zong2025mcpsafetybench}. Memory and multi-agent benchmarks evaluate recall, \blue{multi-session memory use}, coordination, failure modes, or error attribution~\citep{maharana2024locomo,he2026memoryarena,deshpande2025trail,cemri2025mast,zhang2025agentracer,kong2025aegis}. Although recent work has improved realistic tool-safety evaluation and multi-agent failure attribution, \red{no benchmark in the catalog of Table~\ref{tab:benchmarks-provenance} evaluates} the complete provenance chain from evidence and tool outputs to memory updates, intermediate claims, final answers, external state changes, and recovery actions.

A key limitation is the lack of unified execution traces and provenance-relation annotations. Trace-oriented benchmarks and observability frameworks have made useful progress, but they still differ in what artifacts they record and how provenance is represented~\citep{dong2024agentops,alsayyad2026agenttrace,souza2025provagent,deshpande2025trail,cemri2025mast}. \red{No entry in Table~\ref{tab:benchmarks-provenance} jointly captures} user requests, retrieved evidence, tool calls and outputs, memory reads and writes, inter-agent messages, intermediate claims, final responses, and state changes. \blue{None annotates the typed relations of such a joint trace as ground truth; existing labels cover single relations in isolation, such as claim-level} \textsc{Support} \blue{in attribution benchmarks or source-to-sink influence in taint benchmarks}, despite the availability of general provenance concepts in standards such as W3C PROV-DM~\citep{w3c2013provdm}. As a result, current evaluation often measures correctness, faithfulness, or attack success indirectly, rather than directly assessing trace completeness, provenance accuracy, claim support, memory validity, or cross-agent responsibility.

Recovery and governance are also under-evaluated. \red{The benchmarks cataloged in Table~\ref{tab:benchmarks-provenance} commonly test} whether an agent succeeds, fails, hallucinates, or takes an unsafe action, \red{but none scores} whether provenance can help the system repair execution, such as by invalidating stale memory, quarantining contaminated evidence, retrying a tool call, requesting human approval, or rolling back an unsafe state change~\citep{ruan2024toolemu,debenedetti2024agentdojo,deshpande2025trail,cemri2025mast}. Realistic evaluation is further complicated by long-horizon and enterprise traces involving private tools, changing data, user-specific memory, organizational policies, credentials, and confidential intermediate states~\citep{maharana2024locomo,he2026memoryarena,dong2024agentops,alsayyad2026agenttrace}. Future provenance benchmarks should therefore evaluate not only whether agents succeed, but also whether their behavior is traceable, auditable, safe, privacy-aware, and recoverable~\citep{costa2025fides}.

Table~\ref{tab:benchmarks-provenance} \blue{catalogs all 25 benchmark entries of the corpus by} their task type, trace availability, evidence labels, tool calls, memory, multi-agent setting, safety attacks, and main metrics.

\red{Read across the catalog, the shortfall is structural rather than incidental. The two columns no family reaches the top rating on, provenance relations and recovery, are also the two that adding tasks }\blue{of the current kind does not fill}\red{: a relation counts as ground truth only once it is released as a per-instance label, and recovery becomes measurable only once a benchmark scores a repair rather than a prevention. Every benchmark here is built by writing tasks, }\blue{so adding more such benchmarks would change neither column.} \red{What the field lacks is therefore not more agent tasks but released relation annotations and replayable executions, which is the requirement Section~\ref{sec:open-problems} takes up. Appendix~\ref{app:heatmap-derivation} reports the per-family counts behind this reading and the size effect that qualifies them.}

\section{Open Problems and Future Directions}
\label{sec:open-problems}

Despite progress in RAG attribution, agent observability, tool-use safety, memory systems, graph-based representation, debugging, and multi-agent evaluation, evidence tracing and execution provenance for LLM agents remain fragmented. Many existing systems record execution artifacts, but \red{no system in Table~\ref{tab:systems-taxonomy} explains} how evidence, tools, memory, observations, messages, and actions jointly shape an agent's final answer or external behavior. This section outlines open problems toward provenance-aware agents that support verification, audit, attribution, runtime safety, recovery, and governance.

\subsection{Toward Unified and Interoperable Trace Schemas}
\label{subsec:unified-trace-schema}

A central open problem is the lack of unified trace schemas for LLM agents. Existing frameworks and benchmarks record different artifacts, including prompts, model responses, tool calls, retrieved documents, memory operations, environment observations, execution metadata, and error events~\citep{dong2024agentops,alsayyad2026agenttrace,souza2025provagent,deshpande2025trail,cemri2025mast}. However, few schemas jointly represent reasoning steps, retrieval, tool calls and outputs, memory reads and writes, inter-agent messages, intermediate claims, final responses, and external state changes.

Existing standards provide useful building blocks but do not fully solve the agent-provenance problem. W3C PROV-DM offers a general vocabulary for entities, activities, agents, and derivation relations~\citep{w3c2013provdm}; OpenTelemetry and OpenLineage provide abstractions for distributed traces and data workflow lineage~\citep{opentelemetry2026traces,openlineageSpec2026}; and PROV-AGENT adapts provenance modeling to agentic workflows~\citep{souza2025provagent}. Future schemas should connect these operational trace structures with semantic evidence relations. They should represent claims, evidence units, tool observations, memory items, environment states, and agent messages, together with relations such as \textsc{Support}, \textsc{Contradict}, \textsc{Depend-on}, \textsc{Update}, and \textsc{Invalidate}. Such schemas would make agent traces easier to compare, replay, annotate, audit, and exchange across frameworks.

\subsection{Claim-Level and Semantic Provenance}
\label{subsec:claim-semantic-provenance}

Current attribution methods often operate at coarse granularity, such as answer-level citation, context-level faithfulness, or step-level logging~\citep{gao2023alce,ru2024ragchecker,wu2024sourcecheckup,schreieder2025attribution}. However, agent outputs frequently contain multiple claims, intermediate assumptions, and action-relevant inferences. A final response may be partially correct: one claim may be supported by retrieved evidence, another may be inferred beyond the source, and another may be contradicted by a tool output or memory item. This motivates claim-level provenance, building on fine-grained factuality and attribution work such as ALCE, FActScore, RAGTruth, RAGChecker, and SourceCheckup~\citep{gao2023alce,min2023factscore,niu2024ragtruth,ru2024ragchecker,wu2024sourcecheckup}.

For agents, the harder problem is semantic provenance across transformations. Evidence may be paraphrased, summarized, aggregated, compressed into memory, passed through tools, or reused by another agent before influencing a claim or action. String-level matching and citation presence are therefore insufficient. Future work should trace semantic dependencies across retrieval, tool outputs, memory updates, reflection, and inter-agent communication, including cases where evidence is indirectly used or transformed across multiple steps. Work on semantic taint and causal influence points toward this direction, but robust claim-level provenance for long agent executions remains open~\citep{cai2026neurotaint}.

\subsection{Memory and Multi-Agent Provenance}
\label{subsec:memory-multiagent-open}

Memory provenance remains underdeveloped. Long-term memory systems allow agents to store, retrieve, update, and reuse information across sessions, but they often provide limited support for tracing where memory items came from, whether they remain valid, whether they conflict with newer evidence, and how they influence later claims or actions. Existing memory and graph-memory surveys cover architectures, evaluation methods, and relational memory structures~\citep{zhang2025memorysurvey,du2026memoryagents,yang2026graphmemory}, but memory as provenance-bearing evidence remains a distinct open problem.

Future memory systems should attach provenance metadata to memory items, including source attribution, write time, update history, retrieval context, downstream usage, conflict status, and invalidation events. Memory should not be treated as an unqualified context reservoir: a memory item derived from user input, retrieved evidence, a tool output, an environment observation, a reflection, or another agent message should carry different trust implications. This is especially important when memory affects tool calls, personalization, or long-term decisions. Memory poisoning and memory-injection attacks further show that compromised memory can influence future behavior without modifying model weights~\citep{chen2024agentpoison,dong2025memoryinjection}.

Multi-agent provenance introduces additional complexity because agents may communicate, delegate tasks, verify one another's outputs, and jointly produce final decisions. Frameworks such as AutoGen and CAMEL demonstrate these distributed workflows~\citep{wu2023autogen,li2023camel}, while MAST shows that failures can arise from specification errors, inter-agent misalignment, weak verification, and coordination breakdowns~\citep{cemri2025mast}. AgenTracer and Aegis further show that responsible-agent and responsible-step attribution can be operationalized through labeled, counterfactual, or contrastive multi-agent trajectories~\citep{zhang2025agentracer,kong2025aegis}. Future provenance models should therefore track how claims, evidence, messages, and errors propagate across agents: which agent introduced a claim, which agent verified or reused it, which message carried it, which step induced failure, and where responsibility lies. The remaining challenge is to integrate responsibility attribution with claim-level evidence provenance, memory lineage, and recovery actions, rather than treating failure attribution as a separate diagnostic task.

\subsection{Runtime Safety, Enforcement, and Recovery}
\label{subsec:runtime-safety-recovery}

Runtime safety for LLM agents requires more than output filtering. In tool-using agents, dangerous behavior may arise when untrusted content influences tool arguments, when private information flows into public outputs, when stale memory affects a privileged action, or when an agent invokes a tool outside its intended boundary. Guardrails that inspect only the final answer or the current tool call may miss these source-to-sink dependencies. Provenance-aware guardrails should reason over the evidence, memory items, tool outputs, observations, and external content that influenced a proposed action~\citep{zhan2024injecagent,debenedetti2024agentdojo,debenedetti2025camel,costa2025fides,sequeira2026agentsentry,buhler2025agentbound}.

Recent work points toward provenance-aware enforcement through prompt-injection benchmarks, control/data-flow separation, information-flow control, semantic taint tracking, runtime constraints, execution provenance graphs, and access boundaries~\citep{ruan2024toolemu,zhan2024injecagent,debenedetti2024agentdojo,debenedetti2025camel,costa2025fides,cai2026neurotaint,wang2025agentspec,sequeira2026agentsentry,buhler2025agentbound}. The remaining challenge is recovery. Most systems focus on detecting or blocking unsafe behavior, but trustworthy agents also need to repair unsupported or unsafe executions. Future systems should use provenance to invalidate stale memory, quarantine contaminated evidence, retry tool calls with corrected parameters, request human approval, roll back unsafe state changes, or replay execution from safe checkpoints. This requires provenance representations that support intervention and recovery, not only explanation.

\subsection{Realistic Benchmarks, Privacy, and Governance}
\label{subsec:benchmarks-privacy-governance}

Current benchmarks still lack realistic full execution traces. \red{Read across Table~\ref{tab:benchmarks-provenance}, the pattern is consistent: }RAG benchmarks provide evidence labels but \red{none releases} tools \red{or} memory; \blue{provenance and security benchmarks} provide attacks and tool calls but \red{none releases} claim-level evidence labels; memory benchmarks evaluate \blue{long-range recall and multi-session memory use} but \red{neither evaluates} lineage, conflict handling, or downstream influence; and multi-agent benchmarks analyze failures but \red{neither releases} complete provenance annotations. Benchmarks such as AgentBench, WebArena, $\tau$-bench, TRAIL, MAST, LoCoMo, and MemoryArena provide important foundations for agent evaluation, trace analysis, multi-agent failure diagnosis, and long-term memory assessment~\citep{liu2024agentbench,zhou2024webarena,yao2025taubench,deshpande2025trail,cemri2025mast,maharana2024locomo,he2026memoryarena}. However,  \red{none of the benchmark families derived from Table~\ref{tab:benchmarks-provenance} in Appendix~\ref{app:heatmap-derivation} }fully evaluates evidence tracing and execution provenance across retrieval, tool use, memory, multi-agent communication, safety attacks, and recovery.

Future benchmarks should move from isolated answer-, task-, or attack-level evaluation toward trace-level evaluation. Provenance standards such as W3C PROV-DM provide a useful vocabulary for representing entities, activities, agents, and derivation relations~\citep{w3c2013provdm}. Agent observability frameworks such as AgentOps and AgentTrace show how agent executions can be collected and inspected as structured traces~\citep{dong2024agentops,alsayyad2026agenttrace}. PROV-AGENT further adapts provenance modeling to agentic workflows~\citep{souza2025provagent}. Building on these foundations, future benchmarks should capture complete execution traces across agent workflows and annotate the provenance relations needed to evaluate evidence support, unsafe influence, responsibility attribution, and recovery.

Existing work highlights different parts of this agenda. TRAIL shows how execution traces can support failure analysis in agent workflows~\citep{deshpande2025trail}. MAST focuses on diagnosing failure modes in multi-agent systems~\citep{cemri2025mast}. \blue{AgenTracer attributes failures to responsible agents and steps through counterfactual replay and fault injection}~\citep{zhang2025agentracer}. \blue{Aegis generates multi-agent trajectories with injected errors labeled by faulty agent and error mode}~\citep{kong2025aegis}. OpenAgentSafety emphasizes realistic tool-use safety risks in open agent settings~\citep{vijayvargiya2026openagentsafety}. MCP-SafetyBench extends safety evaluation to protocol-level agent-tool environments~\citep{zong2025mcpsafetybench}. The remaining challenge is to integrate these dimensions into benchmarks that jointly test claim-level evidence support, memory validity, unsafe influence detection, auditability, and recovery.

Privacy and governance introduce a separate but equally important challenge. Provenance traces may contain sensitive user data, confidential documents, credentials, tool outputs, API arguments, internal reasoning artifacts, and long-term memory items. Observability frameworks such as AgentOps and AgentTrace provide useful foundations for collecting and inspecting traces~\citep{dong2024agentops,alsayyad2026agenttrace}, while security mechanisms such as Fides highlight the need to track confidentiality and integrity constraints in agent workflows~\citep{costa2025fides}. Future provenance infrastructure should therefore support trace minimization, access control, anonymization, retention policies, encryption, and selective disclosure. Without privacy-aware governance, provenance infrastructure itself may become a source of compliance and security risk.

\subsection{\red{Trace Assurance and Attested Provenance}}
\label{subsec:trace-assurance}

\red{A dimension that current work leaves almost entirely open is }\emph{\red{assurance}}\red{: whether a trace can be trusted by a party other than the one that produced it. Section~\ref{subsec:origin-assurance} introduces assurance as a cross-cutting attribute, and \blue{the reviewed descriptions of the systems in Table~\ref{tab:systems-taxonomy} do not establish tamper-evident or attested trace guarantees. As described, they write traces to a log that could be edited, truncated, or reordered after the fact by whoever operates the agent.} This is adequate for debugging, where the reader and the producer are the same party, and inadequate for audit, where they are not. Work on audit-trail infrastructure and auditable agent design has begun to argue for }\blue{auditable trace infrastructure}\red{~\citep{ojewale2026audittrails,nian2026auditableagents}, and work on checkpoint-restore integrity shows that agent state itself is an attack surface~\citep{zheng2026acrfence}, but no assurance tier has been standardized across the agent provenance systems we survey.
}

\red{Three questions follow. First, what should be attested? Signing an entire trace is }\blue{costly and may expose sensitive content}\red{, whereas committing to a trace's structure (its nodes, edges, and ordering) may be enough to detect deletion and reordering without disclosing tool arguments or memory contents. Second, how should assurance interact with the privacy requirements above? Trace minimization and tamper-evidence pull in opposite directions, since selective disclosure }\blue{can remove} \red{exactly the material a verifier would check. Third, how should assurance interact with }\emph{\red{origin}}\red{? An attested trace }\blue{can show} \red{that recorded procedural edges were not altered after the fact, but }\blue{does not establish} \red{whether an inferred }\textsc{\red{Support}} \red{edge is correct. Assurance and semantic correctness are independent axes}\blue{: an attested trace can still contain an incorrect inferred edge, so attestation should not be read as evidence of correctness.}\red{
}

Overall, provenance-aware agents require more than richer logs. They require interoperable schemas, semantic claim-level attribution, memory and multi-agent lineage, runtime enforcement with recovery, \blue{assured traces,} and benchmarks that evaluate traceability, auditability, safety, privacy, and recoverability as first-class capabilities.

 \section{Conclusion}
\label{sec:conclusion}

LLM agents are moving beyond isolated text generation toward systems that retrieve evidence, plan intermediate steps, invoke tools, update memory, interact with environments, and coordinate with other agents. This shift makes final-answer correctness insufficient for trustworthiness: reliable agent systems must also expose where answers come from, which evidence supports them, how tools and intermediate actions contribute to them, and where failures or unsafe influences enter the execution process. This survey positions evidence tracing and execution provenance as a unified perspective for studying these questions across the agent lifecycle. We reviewed key provenance representations, including structured logs, execution graphs, evidence graphs, tool dependency graphs, memory provenance, and multi-agent communication traces, and discussed their roles in attribution, debugging, verification, safety enforcement, evaluation, and recovery.

Our review shows that the field has made important progress, but remains fragmented. Existing systems often record sources or tool calls, \blue{but the capabilities that build on these records remain fragmented across the reviewed systems, with limited integration of claim-level support, influence tracking, memory lineage, cross-agent provenance, and adversarially robust tracing. No benchmark in our catalog evaluates the complete provenance chain, and the reviewed descriptions do not establish tamper-evident or attested trace guarantees.} Looking forward, evidence tracing and execution provenance should become a first-class infrastructure layer for reliable agent systems. We hope this survey helps consolidate this emerging research space and motivates the development of traceable, auditable, and recoverable LLM agents.

\bibliographystyle{unsrtnat}
\bibliography{references}

\appendix


\section{Additional Taxonomy Mappings}
\label{app:taxonomy-mappings}

This appendix provides supplementary material for the taxonomy in Section~\ref{sec:taxonomy}. The main text keeps the taxonomy compact by using Figure~\ref{fig:taxonomy} and Table~\ref{tab:evidence-tracing-taxonomy} as the backbone, while the mapping below makes explicit how the proposed agent-provenance vocabulary relates to standard provenance modeling. The concrete provenance-graph example is discussed in Section~\ref{sec:representation}, where it serves as a running example for provenance representation rather than as appendix-only material.

\subsection{Mapping to W3C PROV-DM}
\label{app:prov-mapping}

Table~\ref{tab:app-prov-mapping} makes the relationship to the W3C PROV data model explicit. \red{Node types align with PROV entities, activities, and agents, so much of agent execution can reuse a standard provenance vocabulary. The relations are then grouped into three tiers. }\textsc{\red{Use}}\red{, }\textsc{\red{Generate}} \red{and }\textsc{\red{Derive}} \red{are }\emph{\red{inherited}}\red{: their PROV semantics are retained unchanged. }\textsc{\red{Depend-on}}\red{, }\textsc{\red{Trigger}}\red{, }\textsc{\red{Invalidate}} \red{and }\textsc{\red{Update}} \red{are }\emph{\red{specialized}}\red{: PROV expresses each in a weaker form, and agent execution requires the stronger reading given in the table. Only the third tier is }\emph{\red{novel}}\red{, and it is what PROV does not directly express:} agent execution introduces semantic relations such as \textsc{Support} and \textsc{Contradict}, which depend on comparing the content of evidence and claims rather than only on bookkeeping about how artifacts were produced. In this sense, agent provenance extends, rather than merely instantiates, classical provenance models.

\begin{table}[!t]
\centering
\footnotesize
\setlength{\tabcolsep}{3.5pt}
\renewcommand{\arraystretch}{1.12}
\caption{Correspondence between the W3C PROV data model~\citep{w3c2013provdm} and the agent-provenance model of this survey\blue{, grouped into inherited, specialized, and novel relations}.}
\label{tab:app-prov-mapping}
\begingroup
\rowcolors{2}{covRow}{white}
\begin{tabularx}{\textwidth}{>{\raggedright\arraybackslash}p{3.05cm}
                            >{\raggedright\arraybackslash}p{3.15cm}
                            >{\raggedright\arraybackslash}X}
\toprule
\textbf{W3C PROV construct} & \textbf{Agent-provenance counterpart} & \textbf{Notes and agent-specific extension} \\
\midrule
\rowcolor{covHeader}\multicolumn{3}{l}{\emph{Node types (PROV-aligned)}} \\
\addlinespace[1pt]
Entity & Evidence / execution artifact & Documents, passages, tool outputs, memory items, generated claims, messages. \\
Activity & Execution unit & Reasoning steps, retrieval and tool calls, memory operations, environment actions. \\
Agent & Agent & LLM, tool/API, user, or sub-agent; multi-agent roles and delegation become first-class. \\
\midrule
\rowcolor{covHeader}\multicolumn{3}{l}{\emph{Tier 1: Inherited relations (PROV semantics retained)}} \\
\addlinespace[1pt]
\texttt{used} & \textsc{Use} & A step consumed an artifact. Declarative and instrumentable, exactly as in PROV; agent provenance adds trust and taint labels on the consumed value. \\
\texttt{wasGeneratedBy} & \textsc{Generate} & A tool output is generated by a tool call; a claim by a reasoning step. \\
\texttt{wasDerivedFrom} & \textsc{Derive} & Memory derived from a document, or a claim from a table; captures summarization and transformation, not only copying. \\
\midrule
\rowcolor{covHeader}\multicolumn{3}{l}{\emph{Tier 2: Specialized relations (PROV-compatible notions refined for agent execution)}} \\
\addlinespace[1pt]
\texttt{used} (strengthened) & \textsc{Depend-on} & PROV \texttt{used} records consumption; agent \textsc{Depend-on} additionally asserts \emph{counterfactual} dependence, established by replay or ablation. The two are deliberately distinct relations here, since a step may consume ten artifacts and depend on one. Step-on-step dependence is outside \texttt{used} altogether, since \texttt{used} relates an activity to an entity; the nearest PROV construct is \texttt{wasInformedBy}, which records that one activity was informed by another without asserting counterfactual dependence. \\
\texttt{wasInformedBy} & \textsc{Trigger} & An observation triggers a tool call; an error triggers replanning. \\
\texttt{wasInvalidatedBy} & \textsc{Invalidate} & PROV invalidation ends an entity's existence; agent \textsc{Invalidate} marks a claim, plan, or memory item as epistemically invalid while the record persists. \\
\texttt{wasRevisionOf} & \textsc{Update} & Memory, external-state, or delivered-output updates across turns or sessions; agent \textsc{Update} identifies the superseded version by \emph{location} rather than by content similarity. \\
\midrule
\rowcolor{covHeader}\multicolumn{3}{l}{\emph{Tier 3: Novel relations and units (no PROV counterpart)}} \\
\addlinespace[1pt]
--- & \textsc{Support} & Evidence supports a claim, decision, or action; central to attribution, but PROV has no support/justification semantics. \\
--- & \textsc{Contradict} & A tool output contradicts a memory item or retrieved passage; requires semantic comparison, not provenance bookkeeping. \\
--- & Semantic units & Generated claims, tool-call rationales, natural-language observations, inter-agent messages, and trust labels, which PROV treats as opaque entities. \\
\bottomrule
\end{tabularx}
\endgroup
\end{table}


\section{\red{Operational Semantics of Provenance Relations}}
\label{app:relation-semantics}

Section~\ref{sec:taxonomy} introduces nine provenance relations. This appendix makes them operational, so that two independent annotators reading the same execution trace should produce the same typed graph. For each relation we give (i) a \emph{signature} constraining which node types it may connect, (ii) its \emph{direction}, (iii) whether it carries a \emph{temporal} requirement, (iv) the \emph{necessary and sufficient condition} under which it holds, and (v) a positive example, a negative or boundary example, and the relation it is most often confused with together with the test that separates them. Section~\ref{app:support-vs-influence} then works through the two boundary cases on which annotators most often disagree. Section~\ref{app:projection} then states formally what it means for evidence tracing to be a projection of execution provenance.

\subsection{Node Types and Unit Roles}
\label{app:node-roles}

The distinction between evidence units and execution units in Section~\ref{subsec:evidence-units} becomes sharp once artifacts are separated from occurrences, following the entity/activity split of W3C PROV-DM~\citep{w3c2013provdm}. An \emph{evidence unit} is an artifact: it carries a content field $\rho(v)$ that can be compared with the content of another node. An \emph{execution unit} is an occurrence: it carries an actor and a time interval, and its identity is \emph{that it happened}, not \emph{what it says}. Table~\ref{tab:app-node-roles} assigns every node type in the taxonomy to exactly one role.

\begin{table}[!t]
\centering
\footnotesize
\setlength{\tabcolsep}{3.5pt}
\renewcommand{\arraystretch}{1.12}
\caption{\blue{Node types and the unit role assigned to each.}}
\label{tab:app-node-roles}
\begingroup
\rowcolors{2}{covRow}{white}
\begin{tabularx}{\textwidth}{>{\raggedright\arraybackslash}p{2.05cm}
                            >{\raggedright\arraybackslash}p{5.4cm}
                            >{\centering\arraybackslash}p{1.3cm}
                            >{\centering\arraybackslash}p{1.5cm}
                            >{\raggedright\arraybackslash}X}
\toprule
\textbf{Role} & \textbf{Node types} & \textbf{Content $\rho$} & \textbf{Actor + interval} & \textbf{Eligible as} \\
\midrule
Evidence unit
& User query or instruction, retrieved document or passage, environment observation, tool output, memory item, generated tool parameter, policy, tool manifest, inter-agent message, intermediate claim, final answer
& \ding{51} & \ding{55}
& Endpoint of every semantic and derivational relation, and of \textsc{Update}; source of \textsc{Trigger}; target of \textsc{Use}, \textsc{Generate}, \textsc{Depend-on}. Which \emph{end} is permitted in each case is fixed by the signatures in Table~\ref{tab:app-relation-signatures}: only assertive units may be the target of \textsc{Support}. \\
Execution unit
& Reasoning step, retrieval call, tool call, memory read or write, environment action, delegation step
& \ding{55} & \ding{51}
& Source of \textsc{Use}, \textsc{Generate}, \textsc{Depend-on}; source and target of \textsc{Trigger}; target of \textsc{Depend-on}; target of \textsc{Support} and \textsc{Invalidate} via its precondition (Section~\ref{app:relation-signatures}). \\
Actor
& LLM, tool or API, user, sub-agent
& \ding{55} & \ding{55}
& Attribution of responsibility only; never an endpoint of a semantic relation. \\
\bottomrule
\end{tabularx}
\endgroup
\end{table}

Two consequences follow, and they answer the question of when an object is an evidence unit, an execution unit, or both.

\begin{enumerate}
\item \textbf{At graph level the roles are disjoint.} A tool output is an evidence unit; the tool call that produced it is an execution unit; the two are distinct nodes joined by a \textsc{Generate} edge. The intuition that a tool output is ``both'' reflects the granularity of a flat log, in which a single record carries the call, its arguments, and its result. Converting a log to a provenance graph therefore always involves splitting such records.
\item \textbf{Roles are stable, but the \emph{part of the node an edge reads} is not.} A semantic edge incident on a node reads its content $\rho(v)$; a procedural edge reads its occurrence. This is why an intermediate claim can be the target of \textsc{Support} (a statement about its content) and, in the same graph, the source of \textsc{Trigger} to a tool call (a statement about its occurrence) without any ambiguity of role.
\end{enumerate}

\subsection{Relation Signatures, Directionality, and Temporal Semantics}
\label{app:relation-signatures}

We write $\mathcal{E}$ for the set of evidence units, $\mathcal{X}$ for execution units, and $\mathcal{C} \subset \mathcal{E}$ for the \emph{assertive} evidence units, namely intermediate claims and final answers. A relation is \emph{semantic} if whether it holds is determined by the contents of its endpoints, and \emph{procedural} if whether it holds is determined by the execution, independently of whether the contents are true. Table~\ref{tab:app-relation-signatures} records the signature of each relation; Table~\ref{tab:app-relation-conditions} records the condition under which it holds together with worked examples.

\blue{The ``Temporal'' column of Table~\ref{tab:app-relation-signatures} states the ordering each relation requires between its endpoints. We write $t(v)$ for the timestamp of an evidence unit and, for an execution unit, for the \emph{start} of its interval. An artifact produced by an activity is timestamped at that activity's completion, so \textsc{Generate} admits equality; \textsc{Use} also admits equality, since a step may begin at the instant its input was produced. \textsc{Invalidate} is stated against $t_{\mathrm{est}}(t)$, the time at which the target was \emph{established} (created for an artifact, scheduled for a step), because a step may be invalidated before it starts. The required order is not always source-before-target: for \textsc{Invalidate}, \textsc{Depend-on}, \textsc{Update}, and \textsc{Use}, the source is the later node, because the invalidator, the dependent step, the new version, and the consuming activity all come after what they act on.}

\begin{table}[!t]
\centering
\footnotesize
\setlength{\tabcolsep}{3.5pt}
\renewcommand{\arraystretch}{1.12}
\caption{\blue{Signature, direction, and temporal ordering of the nine provenance relations.}}
\label{tab:app-relation-signatures}
\begingroup
\rowcolors{2}{covRow}{white}
\begin{tabularx}{\textwidth}{>{\raggedright\arraybackslash}p{1.85cm}
                            >{\raggedright\arraybackslash}p{4.05cm}
                            >{\raggedright\arraybackslash}p{2.35cm}
                            >{\centering\arraybackslash}p{1.45cm}
                            >{\raggedright\arraybackslash}X}
\toprule
\textbf{Relation} & \textbf{Signature (source $\rightarrow$ target)} & \textbf{Direction} & \textbf{Temporal} & \textbf{Family and algebra} \\
\midrule
\textsc{Support}
& $\mathcal{E} \rightarrow \mathcal{C} \cup \mathcal{X}$
& Supporter $\rightarrow$ supported
& None
& Semantic support. Asymmetric; \emph{not} transitive. \\
\textsc{Contradict}
& $\mathcal{E} \leftrightarrow \mathcal{E}$
& Symmetric (unordered pair)
& None
& Semantic support. Symmetric; not transitive. \\
\textsc{Invalidate}
& $\mathcal{E} \rightarrow \mathcal{C} \cup \{\text{memory item}\} \cup \mathcal{X}$
& Invalidator $\rightarrow$ invalidated
& $t(s) > t_{\mathrm{est}}(t)$
& Semantic support (epistemic status). Asymmetric. \\
\textsc{Derive}
& $\mathcal{E} \rightarrow \mathcal{E}$
& Origin $\rightarrow$ derived
& $t(s) < t(t)$
& Content transformation. Asymmetric; transitive. \\
\textsc{Depend-on}
& $\mathcal{X} \rightarrow \mathcal{E} \cup \mathcal{X}$
& Dependent step $\rightarrow$ dependency
& $t(s) > t(t)$
& \red{Causal} (counterfactual: parameterization). Asymmetric; \blue{not transitive in general}. \\
\textsc{Trigger}
& $\mathcal{E} \cup \mathcal{X} \rightarrow \mathcal{X}$
& Cause $\rightarrow$ activated step
& $t(s) < t(t)$
& \red{Causal} (counterfactual: occurrence). Asymmetric; not transitive. \\
\textsc{Update}
& $\mathcal{E} \rightarrow \mathcal{E}$ (same identified location)
& New version $\rightarrow$ superseded version
& $t(s) > t(t)$
& Procedural, state succession (instrumented). Asymmetric. \\
\textsc{Use}
& $\mathcal{X} \rightarrow \mathcal{E}$
& Activity $\rightarrow$ consumed artifact
& $t(s) \geq t(t)$
& Procedural, declarative (instrumented; PROV \texttt{used}). \\
\textsc{Generate}
& $\mathcal{X} \rightarrow \mathcal{E}$
& Activity $\rightarrow$ produced artifact
& $t(s) \leq t(t)$
& Procedural, declarative (instrumented; PROV \texttt{wasGeneratedBy}). Functional: each artifact has at most one generator within a run. \\
\bottomrule
\end{tabularx}
\endgroup
\end{table}

\noindent
One further label appears in provenance graphs but is not one of the nine: \textsc{Support-candidate}, generated by the projection of Appendix~\ref{app:projection}. Its signature is $\mathcal{E} \rightarrow \mathcal{C}$, it is asymmetric, its temporal constraint is \blue{$t(s) \le t(t)$, inherited from the edges of its witness path and strict when the path contains a \textsc{Derive} edge}, and its origin is always \textsc{Derived}, the value reserved for edges computed rather than observed. It asserts availability of content, never support, and no instrumented or annotated graph contains one.

\FloatBarrier
\begingroup
\setlength{\tabcolsep}{3pt}
\setlength{\LTcapwidth}{\textwidth}
\renewcommand{\arraystretch}{1.1}
\begin{xltabular}{\textwidth}{>{\raggedright\arraybackslash\scriptsize}p{1.5cm}
                            >{\raggedright\arraybackslash\scriptsize}X
                            >{\raggedright\arraybackslash\scriptsize}p{4.0cm}
                            >{\raggedright\arraybackslash\scriptsize}p{4.0cm}}
\caption{Necessary and sufficient conditions, worked examples, and boundary cases. The final column names the relation each is most often confused with, together with the test that separates them. $s$ denotes the source node and $t$ the target; $\rho(\cdot)$ denotes node content.}
\label{tab:app-relation-conditions}\\
\toprule
\textbf{Relation} & \textbf{Holds if and only if} & \textbf{Positive / negative example} & \textbf{Confusable with, and the test} \\
\midrule
\endfirsthead
\multicolumn{4}{l}{\scriptsize\itshape Table~\ref{tab:app-relation-conditions} (continued)}\\
\toprule
\textbf{Relation} & \textbf{Holds if and only if} & \textbf{Positive / negative example} & \textbf{Confusable with, and the test} \\
\midrule
\endhead
\midrule
\multicolumn{4}{r}{\scriptsize\itshape continued on next page}\\
\endfoot
\bottomrule
\endlastfoot

\rowcolor{covRow}\textsc{Support}
& Taking $\rho(s)$ as given raises the warrant for $\rho(t)$, or for the precondition of step $t$, above a threshold $\theta$ fixed in advance by the annotation protocol. The relation is a property of the two contents, not of the execution; given $\theta$ it is adjudicable by a third party who reads only $\rho(s)$ and $\rho(t)$. Because $\theta$ is task-specific and not derivable from the graph, a protocol that leaves it unstated will not reproduce, which is why inter-annotator agreement bounds any metric built on this relation (Section~\ref{subsec:provenance-metrics}). \emph{Not necessary}: that the agent read $s$. \emph{Not sufficient}: that the agent cited $s$, that $s$ was in context, or that $s$ changed the output.
& \textbf{+} Passage ``Q3 revenue was \$2.1B'' supports the claim ``Q3 revenue exceeded \$2B''. \newline \textbf{--} A passage on the same topic that does not entail the claim: relevant and possibly cited, but not supporting~\citep{wu2024sourcecheckup}.
& \textsc{Depend-on}. Ablate $s$: if the claim would still be warranted, \textsc{Support} is a property of content and survives; if the \emph{step} would have been parameterized differently, that is \textsc{Depend-on}. Both may hold; they are not alternatives. \\

\textsc{Contradict}
& $\rho(s)$ and $\rho(t)$ cannot both hold under the task's interpretation, at the same time index.
& \textbf{+} Tool output ``balance = 0'' contradicts memory item ``account is funded''. \newline \textbf{--} Two figures at different granularity or different time indices that are jointly satisfiable: \blue{no contradiction, although the older figure may be stale}.
& \textsc{Invalidate}. \textsc{Contradict} is a symmetric statement about content and asserts nothing about which endpoint survives. \\

\rowcolor{covRow}\textsc{Invalidate}
& (i) $s$ contradicts $t$ or establishes that a precondition of $t$ no longer holds; (ii) $s$ is at least as authoritative as the basis of $t$, and $t(s) > t_{\mathrm{est}}(t)$; (iii) after $s$, $t$ may no longer be used as a warrant. The record of $t$ persists.
& \textbf{+} A 404 response invalidates the planned tool call to that endpoint, whose precondition no longer holds. \newline \textbf{--} A conflicting but less authoritative source: record \textsc{Contradict} only, and leave adjudication to a later step.
& \textsc{Contradict} (see above) and \textsc{Update}. \textsc{Invalidate} marks the old item unusable; \textsc{Update} installs the new one. They commonly co-occur as two distinct edges. \\

\textsc{Derive}
& \blue{$\rho(t)$ was produced by transforming $\rho(s)$}: copying, extraction, summarization, translation, or aggregation. \blue{Derivation establishes content lineage, not the correctness of the transformation or semantic support.} \emph{Not sufficient}: topical similarity, or co-presence in the context window.
& \textbf{+} A memory item derived from a conversation turn; a claim derived from a table row. \newline \textbf{--} A claim produced from parametric knowledge while a document happened to be in context: no \textsc{Derive} edge.
& \textsc{Support}. \textsc{Derive} says where content came from; \textsc{Support} says whether content warrants. A claim mis-summarized from a passage is \textsc{Derive} \emph{without} \textsc{Support}, which is precisely the attributable-hallucination case~\citep{min2023factscore,ru2024ragchecker}. \\

\rowcolor{covRow}\textsc{Depend-on}
& Had $t$ been absent or different, step $s$ would have been issued with different arguments or would otherwise have behaved differently. Whether the step occurred at all is \textsc{Trigger}, not \textsc{Depend-on}. Established by replay, ablation, or fault injection~\citep{zhang2025agentracer}, not by reading content.
& \textbf{+} A tool call whose argument string was extracted from a retrieved page. \newline \textbf{--} A document present in the context window that provably does not change the arguments under replay: \textsc{Use} but not \textsc{Depend-on}.
& \textsc{Use}. On dependencies whose target is an artifact, \textsc{Use} is the observable over-approximation that the harness emits and \textsc{Depend-on} is the counterfactual subset inside it. Dependencies on another \emph{step} have no \textsc{Use} counterpart at all, and can be established only by replay. \\

\textsc{Trigger}
& The occurrence of $s$ is what caused step $t$ to be scheduled at all.
& \textbf{+} A tool error triggers replanning; a new observation triggers a diagnostic call. \newline \textbf{--} An input that only changed the arguments of an already-scheduled step.
& \textsc{Depend-on}. The test is \emph{whether} versus \emph{how}: \textsc{Trigger} governs whether the step happened, \textsc{Depend-on} governs how it was parameterized. Boundary case: a step that would have occurred regardless but was brought forward or delayed by $s$ is \emph{not} triggered by it, since scheduling time is an attribute rather than an occurrence; record the re-timing in the timestamps and leave the edge out. \\

\rowcolor{covRow}\textsc{Update}
& $s$ and $t$ occupy the same identified location (a store slot, memory key, external state variable, or an externally identified output slot such as the response delivered for a given user turn), and $s$ supersedes $t$ from its timestamp onward. Two artifacts occupy the same location iff they are addressed by the same identifier within the same scope, so a revised answer to one turn updates that turn's response while an answer to a later turn is a different location. Identity of location, not similarity of content, is what the relation asserts.
& \textbf{+} A user-preference memory revised in a later session. \newline \textbf{--} A second, independent memory item about the same topic that does not overwrite the first: two items, no \textsc{Update}.
& \textsc{Derive}. \textsc{Update} requires identity continuity of the slot; \textsc{Derive} does not. A summary written to a \emph{new} key is \textsc{Derive}; overwriting the old key is both. \\

\textsc{Use}
& The harness supplied artifact $t$ to activity $s$. A recorded fact about the execution; asserts no influence and no support.
& \textbf{+} A reasoning step uses all ten passages returned by a retrieval call. \newline \textbf{--} A passage retrieved but filtered out before the step: not used.
& \textsc{Depend-on} (see above). Reporting \textsc{Use} as if it were \textsc{Support} is a \blue{common} over-claim in citation-style evaluation. \\

\rowcolor{covRow}\textsc{Generate}
& Activity $s$ produced artifact $t$, whose timestamp equals the completion of $s$.
& \textbf{+} A tool call generates a tool output; a reasoning step generates a claim. \newline \textbf{--} A corpus document that pre-existed the run: it enters the graph with an external-origin attribute and no generator.
& \textsc{Derive}. \textsc{Generate} connects an activity to an artifact; \textsc{Derive} connects two artifacts. Both are typically present: a reasoning step \textsc{Generate}s a claim that \textsc{Derive}s from a passage. \\

\end{xltabular}
\endgroup

\subsection{Separating Semantic Support from Procedural Influence}
\label{app:support-vs-influence}

The four notions that Table~\ref{tab:app-relation-signatures} keeps apart are frequently conflated in trace-based evaluation, so we state the discriminating tests directly.

\begin{itemize}
\item \textbf{Semantic support} (\textsc{Support}, \textsc{Contradict}, and the first clause of \textsc{Invalidate}) is decided by reading the contents of the two endpoints. It is agent-independent: a passage supports a claim whether or not any agent ever retrieved it. When the target of \textsc{Support} is an execution unit, which carries no content of its own, the content read is the step's \emph{precondition}: the proposition that must hold for the step to be justified, which the annotation protocol is required to make explicit before the edge can be adjudicated.
\item \textbf{Content derivation} (\textsc{Derive}) is decided by asking whether the target's content is a transformation of the source's. It is agent-dependent and content-level at once, which is why it can hold while \textsc{Support} fails.
\item \textbf{Causal dependence} (\textsc{Trigger}, \textsc{Depend-on}) is decided counterfactually, by removing or perturbing the source and re-running\blue{, unless the harness dispatched the step on an event, in which case the edge is logged directly}. It says nothing about whether the content is true.
\item \textbf{Procedural dependence} (\textsc{Use}, \textsc{Generate}, \textsc{Update}) is \blue{read directly from what the harness emitted, except that \textsc{Update} is inferred when the store does not version its keys (Table~\ref{tab:app-relation-observability})}. It is the observable layer beneath causal dependence.
\item \textbf{Temporal succession} is deliberately \emph{not} a relation. Adjacency in a log is an artifact of ordering, not a dependency, so ordering is carried by node timestamps and never by an edge. \blue{Temporal adjacency alone does not establish dependence: without an explicit relation, the provenance graph records no dependency between two steps, which means that none has been established, not that none exists.}
\end{itemize}

\noindent
Two boundary cases deserve explicit treatment because they are the ones on which annotators most often disagree.

\vspace{1mm}
\noindent\colorbox{covGap}{%
\parbox{0.94\textwidth}{%
\textbf{Case 1: a tool output changes the final decision without justifying the final claim.}
Record \textsc{Use}(step, output) because the harness passed it; record \textsc{Depend-on}(action, output) if ablating the output changes the action; record \textsc{Derive} only if the answer's content is a transformation of the output's content; record \textsc{Support} only if a third party reading the two contents judges that the output warrants the claim. The correct annotation is frequently $\{\textsc{Use}, \textsc{Depend-on}\}$ with neither \textsc{Derive} nor \textsc{Support}, and the resulting graph says exactly the right thing: the decision was influenced by something that did not justify it.

\vspace{1mm}
\textbf{Case 2: an incorrect memory item causally triggers an action.}
Record \textsc{Trigger}(memory item, action), because the relation is \blue{causal} and holds regardless of the item's truth. \blue{Do \emph{not} record \textsc{Support} merely because the item drove the action: \textsc{Support} requires a third party to judge that the item's content, taken as given, warrants the action's precondition, and an item that only prompted the step need not do so.} When the error is discovered, add \textsc{Contradict}(new evidence, memory item) and \textsc{Invalidate}(new evidence, dependent claim). The graph then simultaneously records that the item drove the execution and that it \blue{was later shown to be false}, which is the distinction the taxonomy exists to preserve.}}

\subsection{Which Relations Are Observable, and Which Must Be Inferred}
\label{app:relation-observability}

Provenance relations differ not only in meaning but in how they can be obtained. We therefore annotate every edge with an \emph{origin} attribute. Three values apply to edges of a recorded execution: \textsc{Instrumented} (emitted by the agent harness or runtime), \textsc{Inferred} (assigned post hoc by a model or a human annotator), and \textsc{Declared} (asserted by the agent itself, for example a self-reported citation). A fourth, \textsc{Derived}, is reserved for edges computed from a graph by a defined procedure rather than obtained from the execution at all; the only label carrying it is \textsc{Support-candidate}, produced by the projection of Appendix~\ref{app:projection}. Table~\ref{tab:app-relation-observability} states which relations admit which origins. The practical consequence is that a provenance graph is heterogeneous in reliability: its procedural skeleton can be logged, but its semantic layer is a judgment, and evaluation must treat the two differently.

\begin{table}[!t]
\centering
\footnotesize
\setlength{\tabcolsep}{3.5pt}
\renewcommand{\arraystretch}{1.12}
\caption{Origin of each of the nine provenance relations\blue{. \textsc{Support-candidate} is omitted, since its origin is always \textsc{Derived}.}}
\label{tab:app-relation-observability}
\begingroup
\rowcolors{2}{covRow}{white}
\begin{tabularx}{\textwidth}{>{\raggedright\arraybackslash}p{2.0cm}
                            >{\centering\arraybackslash}p{2.35cm}
                            >{\raggedright\arraybackslash}X}
\toprule
\textbf{Relation} & \textbf{Available origins} & \textbf{How it is obtained in practice} \\
\midrule
\textsc{Use}, \textsc{Generate}
& Instrumented
& Emitted directly by the agent loop, tool router, or memory store; reproducible from logs alone~\citep{dong2024agentops,alsayyad2026agenttrace,souza2025provagent}. \\
\textsc{Update}
& Instrumented, Inferred
& Instrumented when the store versions its keys; otherwise inferred by matching successive items to the same slot. \\
\textsc{Trigger}
& Instrumented, Inferred
& Instrumented when the harness dispatches on an event, such as retry-on-error; inferred when the model itself chose to act on an observation. \\
\textsc{Depend-on}
& Inferred
& Not directly observable. Obtained by replay, ablation, fault injection, or information-flow tracking; \textsc{Use} over-approximates it only for dependencies on artifacts~\citep{zhang2025agentracer,costa2025fides,cai2026neurotaint}. \\
\textsc{Support}, \textsc{Contradict}, \textsc{Invalidate}
& Inferred, Declared
& Never directly observable. Require semantic judgment by a human annotator or model judge, and therefore require an annotation protocol and reported inter-annotator agreement~\citep{gao2023alce,min2023factscore,thorne2018fever,ru2024ragchecker}. \\
\textsc{Derive}
& Inferred, Declared
& Inferred by content comparison, or declared by an agent that reports its own sources for a summary or memory write. \\
\bottomrule
\end{tabularx}
\endgroup
\end{table}

Orthogonally to origin, an edge may carry an \emph{assurance} attribute recording whether the trace it came from is merely logged, or additionally protected against post-hoc modification. Among the systems surveyed in Table~\ref{tab:systems-taxonomy}, none reports tamper-evidence, signing, or attestation for its traces, so provenance is trustworthy only to the extent that the party holding the log is~\citep{ojewale2026audittrails,nian2026auditableagents,zheng2026acrfence}. Section~\ref{sec:open-problems} treats this as an open problem rather than a solved dimension.

\subsection{Derived Predicates: Cited and Omitted}
\label{app:derived-predicates}

Section~\ref{subsec:evidence-claim-support-graphs} refers to \textsc{Cited} and \textsc{Omitted}. These are not additional taxonomy relations; they are predicates derived from the nine above, and defining them requires the \emph{origin} attribute of Section~\ref{app:relation-observability}, because a citation is an agent's \emph{assertion} that support holds, which is a different thing from support holding.

Write $\mathrm{avail}(d)$ for ``artifact $d$ was available to the run'', that is, $d$ was generated within the run or supplied to some step: $\mathrm{avail}(d) \equiv \exists x.\, \textsc{Generate}(x,d) \vee \textsc{Use}(x,d) \vee \mathrm{external}(d)$, where $\mathrm{external}(d)$ marks an artifact that pre-existed the run and entered the graph without a generator, such as a corpus document. Distinguish two kinds of \textsc{Support} edge by origin:
\[
\textsc{Cited}(d, c) \;\equiv\; \exists\, e.\; \lambda(e) = \textsc{Support},\ e : d \rightarrow c,\ \mathrm{origin}(e) = \textsc{Declared},
\]
\[
\textsc{Supported}(d, c) \;\equiv\; \exists\, e.\; \lambda(e) = \textsc{Support},\ e : d \rightarrow c,\ \mathrm{origin}(e) = \textsc{Inferred}.
\]
\textsc{Cited} records that the agent asserted the link; \textsc{Supported} records that an annotator or judge adjudicated it. The two are independent, and their disagreement is exactly what citation-quality evaluation measures:
\[
\text{citation error} \;\equiv\; \textsc{Cited}(d,c) \wedge \neg\textsc{Supported}(d,c),
\qquad
\text{attribution gap} \;\equiv\; \textsc{Supported}(d,c) \wedge \neg\textsc{Cited}(d,c),
\]
\[
\textsc{Omitted}(d, a) \;\equiv\; \mathrm{avail}(d) \;\wedge\; \neg\exists\, c \in \mathrm{claims}(a).\; \textsc{Supported}(d, c),
\]
where $\mathrm{claims}(a)$ denotes the decomposition of answer $a$ into assertive units, which any claim-level protocol must fix. A source that is cited but does not support is therefore a citation error and not a contradiction in terms~\citep{gao2023alce,wu2024sourcecheckup,papertrail2026}; \textsc{Omitted} covers a retrieved-but-unused passage as well as a used one, which is the case Section~\ref{subsec:evidence-claim-support-graphs} describes. Note that this is the one place where the origin attribute is load-bearing rather than descriptive: without it, an agent's self-report and an adjudicated judgment are the same edge, and citation quality cannot be expressed at all.

\subsection{\red{Derivation and Adequacy of the Relation Set}}
\label{app:relation-adequacy}

\blue{Each of the nine relations is admitted only if it passes two tests. The tests also keep the set open to revision: a relation that fails them, or a query that none of them answers, would change it.}

\begin{itemize}
\itemsep2pt
\item \red{\textbf{Necessity.} Some trust-function query of Section~\ref{subsec:trust-functions} (verification, attribution, debugging, safety enforcement, audit, or recovery) becomes unanswerable if the relation is removed or merged into another.}
\item \red{\textbf{Distinctness.} The relation carries a decision test that no other relation's test subsumes. Because the four families of Section~\ref{app:support-vs-influence} are decided by different kinds of evidence (content comparison, content transformation, counterfactual ablation, and }\blue{harness records}\red{), two relations in different families can never be merged without discarding one of those tests.}
\end{itemize}

\red{Table~\ref{tab:app-relation-adequacy} applies both tests relation by relation. The third column is the operative one: it names what a reader would no longer be able to express, and the fourth names surveyed work that already draws the distinction operationally, so that necessity rests on existing practice.}

\begin{table}[!t]
\centering
\footnotesize
\setlength{\tabcolsep}{3.5pt}
\renewcommand{\arraystretch}{1.12}
\caption{\red{Derivation of the relation set. Each relation is admitted because removing or merging it makes a trust-function query unanswerable, and because at least one surveyed system or benchmark already acts on the distinction it draws.}}
\label{tab:app-relation-adequacy}
\begingroup
\rowcolors{2}{covRow}{white}
\begin{tabularx}{\textwidth}{>{\raggedright\arraybackslash}p{1.75cm}
                            >{\raggedright\arraybackslash}p{4.05cm}
                            >{\raggedright\arraybackslash}X
                            >{\raggedright\arraybackslash}p{3.15cm}}
\toprule
\textbf{\red{Relation}} & \textbf{\red{Query it answers}} & \textbf{\red{What is lost if dropped or merged}} & \textbf{\red{Already operational in}} \\
\midrule
\red{\textsc{Support}} & \red{Is this claim warranted by the evidence available? (verification)} & \red{Claim-level verification collapses into retrieval logging. No PROV relation compares content, so this cannot be recovered from the base model.} & \red{ALCE, FActScore, SourceCheckup, RAGChecker~\citep{gao2023alce,min2023factscore,wu2024sourcecheckup,ru2024ragchecker}} \\
\red{\textsc{Contradict}} & \red{Which two artifacts cannot both be relied on? (verification, safety)} & \red{Treated as negative \textsc{Support}, conflict between two non-assertive artifacts, such as two memory items or two retrieved passages, becomes unrepresentable, since \textsc{Support} requires an assertive target.} & \red{FEVER, RAGTruth, memory-conflict work~\citep{thorne2018fever,niu2024ragtruth,zhong2024memorybank}} \\
\red{\textsc{Invalidate}} & \red{Which recorded item ceased to be usable, and from when? (recovery, }\blue{audit}\red{)} & \red{A stale or poisoned item can be deleted but not \emph{marked}: recovery loses its audit trail. }\blue{Distinguished from \textsc{Contradict}, which is symmetric, by being directed and time-stamped: it marks which item ceased to be usable, and from when, while leaving that item in the graph.} & \red{A-MemGuard, memory-poisoning defenses~\citep{wei2025amemguard,chen2024agentpoison}} \\
\red{\textsc{Derive}} & \red{Is this content a transformation of that content? (attribution)} & \red{Merged into \textsc{Support}, the attributable hallucination, a claim mis-summarized from a real passage and hence derived but unsupported, becomes inexpressible, and it is precisely the case attribution benchmarks target.} & \red{ALCE, RAGTruth, NeuroTaint~\citep{gao2023alce,niu2024ragtruth,cai2026neurotaint}} \\
\red{\textsc{Depend-on}} & \red{Would this step have been parameterized differently without that artifact? (debugging, safety)} & \red{Merged into \textsc{Use}, availability is mistaken for dependence: every artifact the harness supplied counts as a cause, and ablation-verified influence becomes indistinguishable from mere presence.} & \red{NeuroTaint, Fides, AgenTracer~\citep{cai2026neurotaint,costa2025fides,zhang2025agentracer}} \\
\red{\textsc{Trigger}} & \red{Would this step have occurred at all? (debugging, safety)} & \red{Merged into \textsc{Depend-on}, \emph{whether} a step happened collapses into \emph{how} it was parameterized, but a guardrail that blocks a call acts on the former and a taint check on the latter.} & \red{AgentSpec, Agent-Sentry, CaMeL~\citep{wang2025agentspec,sequeira2026agentsentry,debenedetti2025camel}} \\
\red{\textsc{Use}} & \red{What did the harness supply to this step? (audit)} & \red{The loggable skeleton of the graph disappears, and every procedural edge becomes inferred rather than instrumented, which is the distinction Section~\ref{app:relation-observability} rests on.} & \red{OpenTelemetry, AgentOps, AgentTrace, PROV-AGENT~\citep{opentelemetry2026traces,dong2024agentops,alsayyad2026agenttrace,souza2025provagent}} \\
\red{\textsc{Generate}} & \red{Which step produced this artifact? (audit, attribution)} & \red{Artifacts lose their origin, so no chain runs from a tool call to a downstream claim, and evidence cannot be traced back past the point at which it entered the run.} & \red{W3C PROV-DM, AgentOps, PROV-AGENT~\citep{w3c2013provdm,dong2024agentops,souza2025provagent}} \\
\red{\textsc{Update}} & \red{Which version of this location superseded which? (memory, recovery)} & \red{Merged into \textsc{Derive}, the same-location constraint is dropped, so memory lineage cannot be separated from ordinary content copying and versioned stores lose their history.} & \red{MemGPT, A-MEM, Mem0~\citep{packer2023memgpt,xu2025amem,chhikara2025mem0}} \\
\bottomrule
\end{tabularx}
\endgroup
\end{table}

\paragraph{\red{Alternative decompositions we considered and rejected.}}
\red{Four alternatives are natural.} \blue{Three of them lose a distinction the trust functions need or add cost without letting the graph answer a new query. The second, an attribute encoding of relation kind, loses no information, and we decline it as a matter of schema design rather than expressiveness.}

\begin{enumerate}
\itemsep2pt
\item \red{\textbf{W3C PROV-DM unextended.} Its relations record how artifacts were produced and never compare content, so \textsc{Support} and \textsc{Contradict} have no counterpart at all (Section~\ref{subsec:provenance-relations}). Claim--evidence grounding, the concern of an entire literature reviewed in Section~\ref{sec:representation}, would be outside the model.}
\item \red{\textbf{A single attributed \textsc{Influence} edge}, with kind recorded as an attribute.} \blue{Provided the attribute is mandatory and ranges over the same nine kinds, this encoding carries the same information as explicit labels: the projection $\pi$ of Section~\ref{app:projection} can be restated as a filter on the attribute value rather than on $\lambda$, and P3 (that procedural influence has \emph{no} image under $\pi$) holds in the same form. The choice is therefore one of schema design, and we prefer explicit relation labels for three practical reasons. First, the signature constraints of Table~\ref{tab:app-relation-signatures} attach directly to a label, whereas under the attribute encoding they become conditional constraints on attribute values that a validator must enforce separately. Second, a label cannot be omitted, whereas an attribute can be dropped or defaulted by a consumer that does not recognize it; when traces pass between tools, that is the most likely way for ``does this justify the claim'' and ``did this cause the step'' to be silently conflated. Third, labeled edges correspond directly to edge types in property-graph stores, and the PROV-compatible ones to the typed relations of W3C PROV-DM (Table~\ref{tab:app-prov-mapping}), so $\pi$ reduces to selection by edge type in standard query languages. An implementation that adopts a single edge type with a mandatory, validated kind attribute is equivalent for the purposes of this survey.}
\item \red{\textbf{A finer decomposition}, splitting \textsc{Support} into entailment, paraphrase, and partial support, or \textsc{Use} into reading and parameterizing. Each split is a \emph{refinement} of a single decision test rather than a new one, and the resulting distinctions are already carried by edge attributes and by the threshold $\theta$ of Table~\ref{tab:app-relation-conditions}. Splitting them raises annotation burden and lowers inter-annotator agreement without letting the graph answer a query it could not answer before.}
\item \red{\textbf{Merging \textsc{Trigger} into \textsc{Depend-on}}, on the view that both are counterfactual. They are, but over different questions (occurrence versus parameterization), and runtime enforcement acts on the first while information-flow tracking acts on the second, so the merge would erase the boundary between the two guardrail families surveyed in Section~\ref{sec:tool-use}.}
\end{enumerate}

\paragraph{\red{Candidate relations considered and excluded.}}
\red{Six further candidates were considered while constructing the set and excluded, each on one of the two tests. }\textsc{\red{Follows}}\red{, a temporal-succession edge, fails distinctness: order is already carried by node timestamps, and adjacency establishes no semantic, causal, or procedural dependence (Section~\ref{app:support-vs-influence}). }\textsc{\red{Cited}} \red{and }\textsc{\red{Omitted}} \red{fail necessity: both are predicates derived from the nine, and Section~\ref{app:derived-predicates} defines them in those terms rather than as edges of their own. }\textsc{\red{Retrieved-from}} \red{fails distinctness: a retrieval is an execution unit, so the edge it would add is already }\textsc{\red{Use}} \red{on the query and }\textsc{\red{Generate}} \red{on the returned passage. }\textsc{\red{Responsible-for}} \red{and }\textsc{\red{Authorized-by}} \red{fail necessity only under the current corpus, in which delegation and policy are expressed indirectly; they are the two candidates we regard as open, and the next paragraph states the test they would have to pass.}

\paragraph{\blue{Scope of the relation set.}}
\red{The set is not proved minimal in any formal sense, and minimality is not a property a relation vocabulary can have absolutely. It is minimal \emph{relative} to two things we state: the six trust functions of Section~\ref{subsec:trust-functions}, and the corpus of Section~\ref{subsec:methodology}. A different set of trust functions would admit a different set: provenance for cost accounting or for licensing compliance would need relations we do not have. \blue{The vocabulary therefore stays open}: a tenth relation is warranted when it answers a trust-function query that no combination of the nine answers, \emph{and} its decision test is not a refinement of one already listed in Table~\ref{tab:app-relation-conditions}. Two candidates we regard as open are an explicit authorization relation linking an action to the policy that permitted it, and a responsibility relation for multi-agent delegation; both are currently expressed indirectly, and Section~\ref{sec:open-problems} treats them as such.}

\section{\red{The Evidence-Tracing Projection}}
\label{app:projection}

Section~\ref{sec:introduction} describes evidence tracing as the projection of execution provenance onto its evidence-support relations, together with the content-derivation links that connect them. \blue{This section states precisely what that means.} One consequence is worth flagging at the outset: $\pi(G)$ is not a sub\-graph of $G$. It retains a subset of $G$'s edges, but it also \emph{adds} a layer of explicitly labeled \textsc{Support-candidate} edges recording content flow that no semantic edge has yet adjudicated. Evidence tracing is a sub-problem of execution provenance in the sense that it answers a strictly weaker question, not in the sense that its graph is contained in the larger one.

\subsection{Definition}

An \textbf{execution provenance graph} of a single agent run is a typed attributed directed multigraph
\[
G \;=\; (V,\; E,\; \tau,\; \lambda,\; \alpha),
\]
where $V$ is the set of nodes; $\tau : V \rightarrow \mathcal{T}$ assigns each node a type from the vocabulary of Table~\ref{tab:app-node-roles}, inducing the partition $V = \mathcal{E} \uplus \mathcal{X} \uplus \mathcal{A}$ into evidence units, execution units, and actors; $E$ is an abstract set of edges with endpoint maps $\mathrm{src}, \mathrm{tgt} : E \rightarrow V$, so that one ordered pair of nodes may carry several edges (a passage can both \textsc{Derive} and \textsc{Support} the same claim, a tool call can both \textsc{Use} and \textsc{Depend-on} the same parameter, and, crucially, two edges may share a label and differ only in \emph{origin}, which is how a citation the agent \textsc{Declared} and an annotator subsequently \textsc{Inferred} is recorded as two \textsc{Support} edges rather than one); $\lambda : E \rightarrow \mathcal{R}^{+}$ labels each edge, where $\mathcal{R}^{+} = \mathcal{R} \cup \{\textsc{Support-candidate}\}$ extends the nine relations with the derived label introduced below, subject to the signature constraints of Table~\ref{tab:app-relation-signatures}. An execution provenance graph produced by instrumentation or annotation carries no \textsc{Support-candidate} edge; the label exists so that $\pi$'s output lies in the same class as its input, and it has signature $\mathcal{E} \rightarrow \mathcal{C}$, is asymmetric, and carries a temporal requirement inherited from the \textsc{Derive}, \textsc{Use} and \textsc{Generate} edges of its witness path; and $\alpha$ assigns attributes to nodes and edges, including timestamps, actor identity, content or content hash, trust and taint labels, and the \emph{origin} and \emph{assurance} values of Section~\ref{app:relation-observability}.

Partition the relation vocabulary into
\[
\begin{aligned}
\mathcal{R}_{\mathrm{sem}} &= \{\textsc{Support}, \textsc{Contradict}, \textsc{Invalidate}\},
\qquad
\mathcal{R}_{\mathrm{der}} = \{\textsc{Derive}\}, \\[2pt]
\mathcal{R}_{\mathrm{proc}} &= \{\textsc{Use}, \textsc{Generate}, \textsc{Depend-on}, \textsc{Trigger}, \textsc{Update}\}.
\end{aligned}
\]

\red{Here $\mathcal{R}_{\mathrm{proc}}$ groups the }\emph{\red{causal}} \red{and }\emph{\red{procedural}} \red{families of Appendix~\ref{app:support-vs-influence} together, because the projection treats them alike; the two remain distinct relation families everywhere else in the paper.}

The \textbf{evidence-tracing projection} is the map $\pi$ sending $G$ to the evidence view $\pi(G) = (V', E', \tau', \lambda', \alpha')$. All five clauses below are evaluated against the \emph{input} graph $G$; in particular clause~3 reads procedural edges of $G$ before clause~2 removes them, so \blue{clause~3 does not depend on the output of clause~2; clauses~4 and~5 are then computed from the edge sets that clauses~1 and~3 produce}.

\begin{enumerate}
\item \textbf{Edges retained verbatim.} $E_{\mathrm{keep}} = \{\, e \in E \;:\; \lambda(e) \in \mathcal{R}_{\mathrm{sem}} \cup \mathcal{R}_{\mathrm{der}} \cup \{\textsc{Support-candidate}\} \,\}$. The last disjunct is vacuous on an instrumented graph; together with clause~3's crossing requirement it is what makes $\pi$ idempotent.
\item \textbf{Edges discarded.} No edge with $\lambda(e) \in \mathcal{R}_{\mathrm{proc}}$ appears in $E'$.
\item \textbf{Contracted candidate paths.} Call a path of $G$ a \emph{content-flow path} when every one of its edges is a \textsc{Derive} edge traversed forwards, a \textsc{Generate} edge traversed forwards from an activity to the artifact it produced, or a \textsc{Use} edge traversed backwards from a consumed artifact to the activity that consumed it, \emph{and at least one of its edges is a \textsc{Use} or \textsc{Generate} edge}, so that the path passes through the execution layer. For $d \in \mathcal{E}$ and $c \in \mathcal{C}$, set $E_{\mathrm{cand}}$, a set of fresh edge identities disjoint from $E$, to contain a derived edge $\textsc{Support-candidate}(d, c)$ whenever $G$ contains a content-flow path from $d$ to $c$ \emph{and} no \blue{\textsc{Support} or \textsc{Invalidate}} edge already runs from $d$ to $c$ (a \textsc{Contradict} \blue{edge} between them does not block, since contradicting is not adjudicating support). A candidate exists to flag content flow that has not yet \blue{received a support judgment, so a pair that already carries one, whether inferred or declared, needs none; the edge's origin records which}. Such an edge records that $d$ \emph{could have} contributed content to $c$; it is an availability statement, and it is deliberately \emph{not} a \textsc{Support} edge. Requiring a content-flow path to cross the execution layer is the second half of what makes $\pi$ idempotent: a chain of \textsc{Derive} edges alone is already in the evidence view and must generate no candidate.
\item \textbf{Nodes retained.} $V' = \mathcal{E} \cap \big(\text{endpoints of } E_{\mathrm{keep}} \cup E_{\mathrm{cand}}\big)$, together with any execution unit that is the target of a retained \textsc{Support} or \textsc{Invalidate} edge, since justifying an action, and recording that its precondition no longer holds, are both evidence-tracing questions. No retained edge is therefore left dangling.
\item \textbf{Attributes retained.} Content, content hash, timestamps, trust labels, origin, assurance, and, on \textsc{Support-candidate} edges, a witness identifier are kept. Scoping the witness by \emph{label} rather than by which clause admitted the edge is what keeps $\pi$ idempotent on attributes as well as on structure: on a second application such an edge arrives through clause~1, not clause~3, and must not lose its witness. Execution-specific attributes are dropped: actor identity, latency, permission scopes, retry counts, and tool manifests. The witness identifier names one content-flow path of $G$ that licenses the edge; when several paths license the same edge the witness names one of them, so recovery of $G$ from $\pi(G)$ is partial by design. A contracted edge takes $\mathrm{origin} = \textsc{Derived}$, the fourth origin value of Section~\ref{app:relation-observability}, since it is computed from $G$ rather than logged, judged, or asserted.
\end{enumerate}

The components of the output are then $E' = E_{\mathrm{keep}} \cup E_{\mathrm{cand}}$, with \blue{$\tau'$ and $\lambda'$ the restrictions of $\tau$ and $\lambda$ to $V'$ and $E'$, and $\alpha'$ the restriction of $\alpha$ with the execution-specific attributes of clause~5 removed,} extended on $E_{\mathrm{cand}}$ by $\lambda'(e) = \textsc{Support-candidate}$ and the attributes just listed.

\subsection{Properties}

\begin{description}
\item[P1 (Information loss; $\pi$ is not injective).] Many distinct execution graphs have the same evidence view. Two runs that reach the same claims from the same evidence through different tool calls, different orderings, or different agents are indistinguishable under $\pi$. Consequently \textbf{evidence tracing cannot recover execution provenance}: the support view is strictly weaker than the process view, which is why a survey of one is incomplete without the other.
\item[P2 (Groundedness).] Every edge of $\pi(G)$ is licensed by $G$: each retained edge \emph{is} an edge of $G$, and each contracted edge is witnessed by at least one content-flow path of $G$. In particular $\pi$ asserts no \textsc{Support} edge that is not already asserted in $G$; the edges it adds carry the distinct label \textsc{Support-candidate} precisely so that availability is never read as support.
\item[P3 (Not all influence is evidence).] No edge labeled in $\mathcal{R}_{\mathrm{proc}}$ survives $\pi$. The consequential case is the two \emph{influence} relations, \textsc{Depend-on} and \textsc{Trigger}: this is the formal content of the distinction between \emph{supporting} a conclusion and merely \emph{influencing} it, since influence without semantic support is present in $G$ and absent from $\pi(G)$ by construction. \textsc{Use}, \textsc{Generate}, and \textsc{Update} are also discarded, but they are bookkeeping rather than influence claims, and \textsc{Use} and \textsc{Generate} survive indirectly as witnesses of contracted edges. \blue{The exclusion is deliberate}: \blue{the evidence view addresses evidential justification, while the process view records execution history and dependencies, including causal influence where established.} A system that reports only $\pi(G)$ cannot answer safety questions \blue{that require procedural or causal influence information}, and a system that reports only $\mathcal{R}_{\mathrm{proc}}$ cannot answer grounding questions.
\item[P4 (Idempotence).] $\pi(\pi(G)) = \pi(G)$. Clause~1 retains $\mathcal{R}_{\mathrm{sem}}$, $\mathcal{R}_{\mathrm{der}}$ and \textsc{Support-candidate} edges, which is every label present in an evidence view, so nothing is lost; clause~2 removes nothing, since no $\mathcal{R}_{\mathrm{proc}}$ edge survives the first application; and clause~3 adds nothing, since a content-flow path requires a \textsc{Use} or \textsc{Generate} edge and the evidence view contains none. This last point is why clause~3 requires the path to cross the execution layer: without that requirement a bare chain of \textsc{Derive} edges would generate a fresh candidate on every application and $\pi$ would not converge.
\end{description}

\noindent
We use \emph{projection} in this structural sense: $\pi$ is a total, idempotent, information-losing map from typed provenance graphs to typed evidence graphs that preserves the semantic sub-structure. We do not claim it is a projection in a categorical sense, and no result in this survey depends on stronger properties.

\subsection{Worked Example}

Table~\ref{tab:app-projection-example} applies the definition to the running example of Figure~\ref{fig:example-provenance-graph}. We list the twenty edges on which the projection acts. The generator of the recovery-phase artifact $a'$, and the memory item that the write $w$ produces, are elided; note that this is a simplification of the example rather than of the definition, since supplying $a'$ with a generator and an input would add further candidate edges into $a'$. Of those twenty the evidence view retains seven, discards thirteen, and adds four contracted candidates.

\begin{table}[!t]
\centering
\footnotesize
\setlength{\tabcolsep}{3.5pt}
\renewcommand{\arraystretch}{1.1}
\caption{The projection applied to the running example of Figure~\ref{fig:example-provenance-graph}. Nodes: $q$ user query, $r$ retrieval call, $p$ retrieved passage, $m$ memory item, $z_1$ claim-forming reasoning step, $c$ intermediate claim, $g$ generated tool parameter, $k$ tool call, $o$ tool output, $z_2$ answer-forming reasoning step, $a$ final answer, $n$ new external evidence, $w$ memory write, $a'$ revised answer.}
\label{tab:app-projection-example}
\begingroup
\rowcolors{2}{covRow}{white}
\begin{tabularx}{\textwidth}{>{\raggedright\arraybackslash}p{4.5cm}
                            >{\centering\arraybackslash}p{2.6cm}
                            >{\raggedright\arraybackslash}X}
\toprule
\textbf{Edge in $G$} & \textbf{In $\pi(G)$?} & \textbf{Why} \\
\midrule
\textsc{Trigger}$(q, r)$ & Dropped & Procedural: governs whether retrieval happened. \\
\textsc{Generate}$(r, p)$ & Dropped & Activity-to-artifact bookkeeping. It carries no edge into $\pi(G)$, but may appear inside the witness path of a contracted edge. \\
\textsc{Use}$(z_1, p)$, \textsc{Use}$(z_1, m)$ & Dropped & Availability, not support. \\
\textsc{Generate}$(z_1, c)$ & Dropped & Activity-to-artifact. \\
\textsc{Derive}$(p, c)$ & \textbf{Kept} & Content of the claim is a transformation of the passage. \\
\textsc{Support}$(p, c)$ & \textbf{Kept} & Passage content warrants the claim. \\
\textsc{Support}$(m, c)$ & \textbf{Kept} & Memory content warrants the claim; later invalidated, see below. \\
\textsc{Derive}$(c, g)$ & \textbf{Kept} & Tool argument derived from the claim's content. \\
\textsc{Trigger}$(c, k)$ & Dropped & Governs whether the tool was called. \\
\textsc{Use}$(k, g)$ & Dropped & The call consumed the parameter; recorded, not judged. \\
\textsc{Depend-on}$(k, g)$ & Dropped & Governs how the call was parameterized. This edge is essential for the tool-use safety question and invisible in the evidence view. \\
\textsc{Generate}$(k, o)$ & Dropped & Activity-to-artifact. \\
\textsc{Use}$(z_2, o)$, \textsc{Generate}$(z_2, a)$ & Dropped & Answer construction: procedural bookkeeping. \\
\textsc{Support}$(o, a)$ & \textbf{Kept} & Tool output warrants the final answer. \\
\textsc{Contradict}$(n, m)$ & \textbf{Kept} & New evidence conflicts with the memory item. \\
\textsc{Invalidate}$(n, c)$ & \textbf{Kept} & New evidence removes the claim's warrant. \\
\textsc{Trigger}$(n, w)$, \textsc{Update}$(a', a)$ & Dropped & Procedural recovery and state succession. \\
\midrule
\textsc{Support-candidate}$(p, a)$ & \textbf{Added} & Witness: $p \xrightarrow{\textsc{Derive}} c \xrightarrow{\textsc{Derive}} g \xleftarrow{\textsc{Use}} k \xrightarrow{\textsc{Generate}} o \xleftarrow{\textsc{Use}} z_2 \xrightarrow{\textsc{Generate}} a$. The passage is a possible content contributor to the answer; whether it \emph{supports} the answer is a separate judgment. \\
\textsc{Support-candidate}$(m, a)$ & \textbf{Added} & Same shape, witnessed through $m \rightarrow z_1 \rightarrow c \rightarrow \cdots \rightarrow a$. Read together with \textsc{Invalidate}$(n, c)$, this is the edge that tells an auditor a discredited memory item may have reached the answer. \\
\textsc{Support-candidate}$(c, a)$, \textsc{Support-candidate}$(g, a)$ & \textbf{Added} & The claim and the tool parameter derived from it are also content contributors to the answer. \\
\addlinespace[2pt]
\textsc{Support-candidate}$(p, c)$, \textsc{Support-candidate}$(m, c)$, \textsc{Support-candidate}$(o, a)$ & Not added & Blocked by clause~3: \textsc{Support}$(p,c)$, \textsc{Support}$(m,c)$ and \textsc{Support}$(o,a)$ already adjudicate these pairs. \\
\bottomrule
\end{tabularx}
\endgroup
\end{table}

The example makes the division of labour concrete. The evidence view alone shows an auditor that the final answer \blue{may have drawn on} a claim supported partly by a memory item that was later contradicted and whose warrant was invalidated, an attribution failure. It does not show that the call was triggered by that claim, or that its parameterization depended on the argument derived from it, which is the fact a safety analyst needs. Neither view subsumes the other, and this is exactly why the survey treats execution provenance as the full graph and evidence tracing as its projection.


\section{\red{Benchmark Catalog and Coverage Details}}
\label{app:benchmark-catalog}

This appendix derives every cell of the coverage heatmap from the benchmark catalog of Table~\ref{tab:benchmarks-provenance}, and states the annotation protocol under which that catalog was coded. The main text uses Figure~\ref{fig:benchmark-coverage-heatmap} to emphasize coverage gaps; Table~\ref{tab:app-heatmap-derivation} shows the computation itself, so that \blue{any rating can be reproduced and checked against the catalog}. \red{The per-benchmark entries are not repeated here: the catalog is Table~\ref{tab:benchmarks-provenance} in Section~\ref{sec:benchmarks}, and this appendix adds only what the main text does not carry: the derivation and the protocol.}

\subsection{Derivation of the Coverage Heatmap}
\label{app:heatmap-derivation}

Figure~\ref{fig:benchmark-coverage-heatmap} is derived from Table~\ref{tab:benchmarks-provenance} rather than assigned by judgment. For the five capabilities that appear as columns in the catalog (evidence labels, tool calls, memory, multi-agent structure, and safety attacks), a family's rating is the maximum over its member benchmarks under the mapping Yes\,$\rightarrow$\,\textbf{S}, Partial / Task-dependent / State-based\,$\rightarrow$\,\textbf{P}, No\,$\rightarrow$\,\textbf{L}.

The maximum is the right aggregator because each cell states an \emph{existential} claim (does \emph{any} benchmark in this family evaluate this capability?), and it has two consequences a reader should keep in view. First, the heatmap is an optimistic upper bound: a family rated \textbf{S} typically contains members rated \textbf{L} for the same capability. Second, families differ in size, from two members to nine, so a maximum is easier to attain in a large family; each cell of Table~\ref{tab:app-heatmap-derivation} therefore also reports how many members attain the rating. The \textbf{L} cells are \blue{unaffected by family size}, and they are the ones the survey's argument rests on: an \textbf{L} means \emph{no} member covers the capability, which cannot be an artifact of family size.

The catalog's remaining column, trace availability, has no heatmap counterpart. It is a precondition for evaluating any of the seven capabilities rather than a capability in its own right, so we report it per benchmark in Table~\ref{tab:benchmarks-provenance} and do not aggregate it.

Two capabilities have no catalog column and are coded directly from the released artifacts by a countable criterion. \emph{Provenance relations} is rated by how many of the nine relations of Section~\ref{app:relation-signatures} a benchmark releases as ground-truth labels: \textbf{S} for three or more, \textbf{P} for one or two, \textbf{L} for none. Citation benchmarks label \textsc{Support} and nothing else; taint benchmarks label source-to-sink influence, which is evidence for \textsc{Depend-on} and nothing else; trace-debugging benchmarks label the responsible step, which is evidence for \textsc{Trigger} and nothing else. Relations recoverable from a call log but never labeled (\textsc{Use} and \textsc{Generate}) do not count, because an unlabeled relation supplies no ground truth. \emph{Recovery} is rated \textbf{S} only when a benchmark scores repair outcomes; it is \textbf{P} when the benchmark scores something adjacent to repair, namely superseding stale memory, localizing the failure that a repair would target, or blocking an unsafe action before it executes, and \textbf{L} otherwise. Under these criteria no family reaches \textbf{S} in either column, which are therefore the sharpest gaps the heatmap records.

Table~\ref{tab:app-heatmap-derivation} gives the resulting derivation. The determining member is the benchmark that attains the family maximum; where several members tie we name them, up to four, and otherwise indicate the count.

\begin{table}[!t]
\centering
\scriptsize
\setlength{\tabcolsep}{3pt}
\renewcommand{\arraystretch}{1.1}
\caption{Cell-by-cell derivation of Figure~\ref{fig:benchmark-coverage-heatmap} from Table~\ref{tab:benchmarks-provenance}. \blue{For the five catalog columns,} each entry gives the rating, the number of family members attaining it out of the family size, and the determining member\blue{; the provenance-relations and recovery columns, coded from released relation labels and scored repairs (Section~\ref{app:heatmap-derivation}), give the rating and its basis}. Entries marked ``---'' have no member above \textbf{L}.}
\label{tab:app-heatmap-derivation}
\resizebox{\textwidth}{!}{%
\rowcolors{2}{covRow}{white}
\begin{tabular}{p{2.4cm} p{2.5cm} p{2.4cm} p{2.5cm} p{2.3cm} p{2.3cm} p{2.2cm} p{1.9cm}}
\toprule
\textbf{Family} & \textbf{Evidence labels} & \textbf{Tool calls} & \textbf{Memory} & \textbf{Multi-agent} & \textbf{Safety attacks} & \textbf{Prov. relations} & \textbf{Recovery} \\
\midrule

RAG \emph{(7)}
& \textbf{S} 4/7 (FActScore, ALCE, SourceCheckup, RAGTruth)
& \textbf{L} 0/7 (---)
& \textbf{L} 0/7 (---)
& \textbf{L} 0/7 (---)
& \textbf{L} 0/7 (---)
& \textbf{P} (\textsc{Support} only)
& \textbf{L} (---) \\

Tool-use \emph{(4)}
& \textbf{L} 0/4 (---)
& \textbf{S} 3/4 (WebArena, ToolBench, $\tau$-bench)
& \textbf{P} 2/4 (AgentBench task-dep.; $\tau$-bench state-based)
& \textbf{L} 0/4 (---)
& \textbf{P} 1/4 ($\tau$-bench policy checks)
& \textbf{L} (none labeled)
& \textbf{L} (---) \\

Memory \emph{(2)}
& \textbf{P} 2/2 (LoCoMo, MemoryArena)
& \textbf{P} 1/2 (MemoryArena task-dep.)
& \textbf{S} 2/2 (LoCoMo, MemoryArena)
& \textbf{L} 0/2 (---)
& \textbf{L} 0/2 (---)
& \textbf{P} (session-level \textsc{Support} references)
& \textbf{P} (superseding stale items; repair unscored) \\

Multi-agent \emph{(2)}
& \textbf{P} 1/2 (AgenTracer/Aegis)
& \textbf{P} 2/2 (MAST, AgenTracer task-dep.)
& \textbf{P} 2/2 (MAST, AgenTracer task-dep.)
& \textbf{S} 2/2 (MAST, AgenTracer/Aegis)
& \textbf{L} 0/2 (---)
& \textbf{P} (responsible step, evidencing \textsc{Trigger})
& \textbf{L} (---) \\

Trace-debugging \emph{(2)}
& \textbf{P} 1/2 (TRAIL)
& \textbf{P} 2/2 (TRAIL, MAST)
& \textbf{P} 2/2 (TRAIL, MAST)
& \textbf{S} 1/2 (MAST)
& \textbf{L} 0/2 (---)
& \textbf{P} (error location on trace steps)
& \textbf{P} (localization only; repair unscored) \\

Provenance / security \emph{(9)}
& \textbf{P} 3/9 (OpenAgentSafety, Agent-Sentry, NeuroTaint)
& \textbf{S} 9/9 (all members)
& \textbf{S} 1/9 (NeuroTaint/TaintBench)
& \textbf{P} 4/9 (OpenAgentSafety, MCP-SafetyBench, NeuroTaint, AgentSpec)
& \textbf{S} 9/9 (all members)
& \textbf{P} (source-to-sink influence, evidencing \textsc{Depend-on})
& \textbf{P} (blocking, not repair) \\

\bottomrule
\end{tabular}%
}
\end{table}

Three observations follow from the derivation that are not visible in the heatmap itself. First, \emph{five of the six families reach} \textbf{S} \emph{on exactly one capability} and are \textbf{P} or \textbf{L} everywhere else. For four of the five that capability is the one the family was built for; the exception is the trace-debugging family, whose only \textbf{S} is in the multi-agent column and is inherited from MAST, which belongs to both families. \blue{Second, the remaining family, provenance/security,} reaches \textbf{S} on three capabilities and is the only family with no \textbf{L} cell; but it is also the largest at nine members, two of its three \textbf{S} ratings are unanimous while the third rests on a single member, and it never reaches \textbf{S} on evidence labels or recovery. Third, and most consequentially, \emph{no family reaches} \textbf{S} \emph{on provenance relations or on recovery}. Every rating above \textbf{L} in those two columns is carried by a benchmark labeling exactly one relation (\textsc{Support} in citation benchmarks, \textsc{Depend-on} in taint benchmarks, \textsc{Trigger} in trace-debugging benchmarks), or by an artifact adjacent to repair rather than repair itself. No benchmark in the catalog releases the typed relation vocabulary as ground truth, and none scores recovery. This is the concrete form of the gap that Section~\ref{sec:open-problems} argues future benchmarks should close.

\subsection{\red{Annotation Protocol for the Coverage Ratings}}
\label{app:annotation-protocol}

\red{Three rating schemes appear in this survey: the \textbf{S}/\textbf{P}/\textbf{L} coverage ratings of Figure~\ref{fig:benchmark-coverage-heatmap}, the memory-capability symbols of Table~\ref{tab:memory-provenance}, and the benchmark-capability symbols of Table~\ref{tab:benchmarks-provenance}. They share one annotation protocol, stated here so that }\blue{any cell can be re-derived and checked against the artifact.} \red{The heatmap }\blue{is not coded directly}\red{: it is computed from the catalog by the rule of Section~\ref{app:heatmap-derivation}, so the protocol below governs only the per-artifact tables from which it is computed, and correcting a catalog cell propagates to the heatmap automatically.}

\paragraph{\red{Unit of coding and evidence admitted.}}
\red{The unit is a single (artifact, capability) pair: one benchmark or system, one column. Evidence is admitted in a fixed order of precedence, and the first level that decides the cell ends the search: (i) the released dataset, repository, or evaluation harness; (ii) the artifact's own paper, including its appendices; (iii) its documentation. The ordering is what makes the ratings checkable, and it has one }\blue{key consequence}\red{: \emph{a capability claimed in prose but with no released label or trace never yields the top rating}. We code what an artifact makes available to a third party, not what it reports about itself. Where an artifact was not publicly released at the time of coding, it is coded from its paper alone and can attain at most the middle rating\blue{; this is the one case in which the middle rating rests on a description rather than on a release}.}

\paragraph{\red{Decision procedure.}}
\red{Each cell is decided by the following four steps in order}\blue{, so that the same evidence always yields the same rating}\red{:}
\begin{enumerate}
\itemsep2pt
\item \red{\emph{Ground truth?} Does the artifact release per-instance labels for the capability and report a metric computed over them? If yes, the cell takes the top rating (\BenchYes{} / \MemYes{} / \textbf{S}) and coding stops.}
\item \red{\emph{Present but unlabeled?} Does the capability appear in the released traces or the system's behavior without ground-truth labels, or is it labeled for only a subset of instances, only at run level, or only for some memory types\blue{, or labeled for all instances with no metric computed over the labels}? If yes, the cell takes the middle rating (\BenchPartial{} / \MemPartial{} / \textbf{P}).}
\item \red{\emph{Structural qualifier?} \blue{Three} values are not degrees of coverage but statements about \emph{how} the capability is realized, and are applied before the negative rating: \BenchTask{} when coverage varies across the artifact's own task suite rather than across instances, and \BenchState{} when the capability exists as external environment state rather than as an agent-side artifact. \MemAttack{} is the corresponding value for a work that studies a capability adversarially rather than providing it.}
\item \red{\emph{Otherwise} the cell takes the negative rating (\BenchNo{} / \MemLimited{} or \MemNo{} / \textbf{L}).} \blue{\MemLimited{} marks a system that exposes the information from which the capability could be built but does not implement it.}
\end{enumerate}

\paragraph{\red{Where the middle and negative ratings divide.}}
\red{The step that decides most contested cells is the boundary between the middle rating (\BenchPartial{} / \MemPartial{} / \textbf{P}) and the negative one (\BenchNo{} / \MemLimited{} or \MemNo{} / \textbf{L}), }\blue{and it needs separate treatment because the two labels are easy to misread as degrees of the same thing.} \red{The middle rating is a claim that the capability is \emph{present in what the artifact releases} but is not ground truth: the traces contain it unlabeled, or it is labeled for only some instances, or realized in a form the column does not name. The negative rating is a claim that it is \emph{not there to be found}: for a benchmark, neither labels nor recoverable traces are released; for a system, the capability is not implemented. The operational test is therefore not ``how much coverage is there'' but ``could a third party extract this from the release at all''. One consequence is that \textbf{L} is the falsifiable direction: a single released file showing the capability overturns an \textbf{L}, whereas overturning a \textbf{P} requires showing that a metric is computed over labels, which is the step~1 test. This is why Section~\ref{app:heatmap-derivation} treats the \textbf{L} cells as the }\blue{signal unaffected by family size} \red{on which the survey's argument rests, and why \MemLimited{} in Table~\ref{tab:memory-provenance} is worded as ``exposes the information but does not implement it'' rather than as weak support.}

\paragraph{\red{Two worked cells.}}
\red{Applying the procedure to two cells that could plausibly have been coded either way shows }\blue{how the steps decide them}\red{.}

\red{\emph{ALCE, evidence labels} $\rightarrow$ \BenchYes{}. Step~1 asks for per-instance ground truth }\emph{\red{and}} \red{a metric over it. ALCE releases per-claim citation annotations and reports citation precision and recall computed from them~\citep{gao2023alce}, so both halves hold and coding stops at step~1. Had it released the annotations without scoring them, step~1 would fail and step~2 would give \BenchPartial{}, since the distinction is the metric, not the labels.}

\red{\emph{$\tau$-bench, memory} $\rightarrow$ \BenchState{}. The benchmark carries information across turns, so a coder might record \BenchYes{} or \BenchPartial{}. Step~1 fails: nothing about memory is labeled per instance. Step~2 is the tempting one, but what persists is the environment database the tools read and write, not an agent-side memory item~\citep{yao2025taubench}; the capability exists, yet not as the artifact the column is about. Step~3 therefore applies and the qualifier \BenchState{} records both facts at once: present, but realized as external state. It maps to }\textbf{\red{P}} \red{in the heatmap, so the family rating is unaffected; what the qualifier preserves is }\emph{\red{why}}\red{, which a bare }\textbf{\red{P}} \red{would discard.}

\paragraph{\red{Scope and known limits of the protocol.}}
\red{The protocol codes the \emph{presence} of a capability, never its quality: a benchmark with weak evidence labels and one with strong labels are both \BenchYes{}, and the survey makes no claim about which is better. Three limits follow. First, coding was performed by the authors jointly, with disagreements resolved by discussion against the released artifact; there was no independent second coder and we therefore report no inter-coder agreement, which is the same limitation recorded for the corpus in Section~\ref{subsec:methodology}. Second, the ratings are a snapshot: they reflect the artifact versions available at the time of writing, and a subsequent release can change a cell. Third, step~1 turns on whether a metric is \emph{computed over} the labels, which is the step at which we expect residual disagreement, and it is the step }\blue{to check first when a cell is in doubt.}\red{}

\blue{The protocol's main benefit is that every rating becomes a claim about a specific released artifact, which a single counterexample, such as a labeled file or a metric in a results table, can overturn.}


\section{Additional Details for Open Problems}
\label{app:open-problems-details}

This appendix expands the open problems discussed in Section~\ref{sec:open-problems}. \blue{It does not restate the main text}\red{: the main text makes the argument for each open problem and cites the work that motivates it, while this appendix carries in table form only the artifact the argument implies: a requirements checklist an implementer or benchmark designer can work from. Where a requirement here }\blue{corresponds to} \red{a sentence of Section~\ref{sec:open-problems}, the table row is the operational version of it, and a reader following the argument alone can skip this appendix without loss.} The main text presents the research agenda; this appendix provides more detailed design requirements, representative foundations, and evaluation dimensions.

\subsection{Trace Schema Requirements}
\label{app:trace-schema-requirements}

A unified LLM-agent provenance schema should combine operational execution records with semantic evidence relations. Table~\ref{tab:trace-schema-requirements} summarizes the main schema requirements.

\begin{table}[t]
\centering
\footnotesize
\setlength{\tabcolsep}{3.2pt}
\renewcommand{\arraystretch}{1.08}
\caption{Design requirements for unified LLM-agent provenance schemas.}
\label{tab:trace-schema-requirements}
\rowcolors{2}{covRow}{white}
\begin{tabularx}{\linewidth}{p{2.7cm} X X}
\toprule
\textbf{Requirement} & \textbf{What should be represented} & \textbf{Why it matters} \\
\midrule
Trace sources 
& User instructions, model outputs, retrieval calls, tool calls, memory operations, environment observations, inter-agent messages, and final responses. 
& Enables end-to-end reconstruction of agent execution. \\

Evidence units 
& User queries and instructions, documents, passages, observations, tool outputs, memory items, generated tool parameters, policies, tool manifests, inter-agent messages, intermediate claims, and final answers. 
& Supports claim-level verification and attribution. \\

Execution units 
& Reasoning steps, retrieval calls, tool invocations, memory reads and writes, environment actions, and delegation steps. 
& Supports debugging, safety enforcement, and audit. \\

Typed relations 
& \textsc{Support}, \textsc{Contradict}, \textsc{Depend-on}, \textsc{Derive}, \textsc{Trigger}, \textsc{Update}, \textsc{Invalidate}, \textsc{Use}, and \textsc{Generate}. 
& Turns raw logs into provenance graphs that expose influence and responsibility. \\

Trust metadata 
& Source trust, permission scope, confidentiality/integrity labels, timestamps, actor identity, and policy constraints. 
& Enables runtime enforcement, privacy-aware audit, and governance. \\

Replay and recovery fields 
& Checkpoints, rollback targets, invalidation events, repair steps, human approvals, and state-change records. 
& Supports intervention and recovery, not only post-hoc explanation. \\
\bottomrule
\end{tabularx}
\end{table}

Existing standards and frameworks provide partial foundations. W3C PROV-DM defines general provenance concepts such as entities, activities, agents, generation, usage, derivation, and invalidation~\citep{w3c2013provdm}. OpenTelemetry provides distributed tracing abstractions, while OpenLineage provides data workflow lineage~\citep{opentelemetry2026traces,openlineageSpec2026}. AgentOps and AgentTrace move closer to LLM-agent observability by recording structured agent artifacts and execution context~\citep{dong2024agentops,alsayyad2026agenttrace}. PROV-AGENT adapts provenance modeling to agentic workflows~\citep{souza2025provagent}. However, LLM-agent provenance requires additional semantic units and relations, especially generated claims, evidence support, contradiction, memory invalidation, tool-argument influence, and inter-agent responsibility.

\subsection{Detailed Agenda for Semantic and Claim-Level Provenance}
\label{app:semantic-provenance-agenda}

Claim-level provenance should distinguish at least five conditions that are often conflated in current systems: citation presence, source relevance, claim support, contradiction, and evidence omission. Only support and contradiction are taxonomy relations; citation presence and omission are the derived predicates of Appendix~\ref{app:derived-predicates}, and source relevance is weaker than any of them. ALCE, FActScore, RAGTruth, RAGChecker, SourceCheckup, and related attribution surveys provide foundations for this direction~\citep{gao2023alce,min2023factscore,niu2024ragtruth,ru2024ragchecker,wu2024sourcecheckup,schreieder2025attribution}. However, agentic settings introduce additional challenges because evidence may be transformed by reasoning, retrieval, tools, memory, reflection, or inter-agent communication before it influences the final answer or action.

Future work should therefore evaluate not only whether a final claim is supported, but also how the supporting information traveled through the execution. Important cases include: a claim derived from multiple retrieved passages; a tool output that contradicts a memory item; a memory item summarized from a previous conversation; a policy constraint that justifies or blocks an action; and an inter-agent message that propagates an unsupported assumption. Semantic taint and causal influence methods such as NeuroTaint suggest one path toward tracing transformed influence, but robust semantic provenance remains open~\citep{cai2026neurotaint}.

\subsection{Memory and Multi-Agent Provenance Checklist}
\label{app:memory-multiagent-checklist}

Table~\ref{tab:memory-multiagent-checklist} lists provenance fields that future memory and multi-agent systems should expose.

\begin{table}[t]
\centering
\footnotesize
\setlength{\tabcolsep}{3.2pt}
\renewcommand{\arraystretch}{1.08}
\caption{Checklist for memory and multi-agent provenance.}
\label{tab:memory-multiagent-checklist}
\rowcolors{2}{covRow}{white}
\begin{tabularx}{\linewidth}{p{2.8cm} X X}
\toprule
\textbf{Setting} & \textbf{Provenance fields} & \textbf{Open question} \\
\midrule
Memory write 
& Source, timestamp, writer, evidence basis, confidence, and policy context. 
& Was the memory created from trusted evidence or from unsupported/injected content? \\

Memory update 
& Previous version, update source, update reason, conflict status, and invalidation marker. 
& Does the update revise, contradict, or invalidate earlier memory? \\

Memory retrieval 
& Retrieval query, retrieved memory items, ranking score, context of reuse, and downstream consumer. 
& Why was this memory reused, and how did it affect the current execution? \\

Memory deletion or invalidation 
& Deletion trigger, invalidating evidence, affected downstream claims/actions, and recovery steps. 
& Can stale or poisoned memory be removed without breaking auditability? \\

Inter-agent communication 
& Sender, receiver, role, message content, cited evidence, delegated task, and verification status. 
& Which agent introduced, verified, reused, or amplified a claim? \\

Multi-agent failure 
& Responsible agent, responsible step, error mode, propagated message, and counterfactual trajectory. 
& Where did the failure originate, and how did it propagate across agents? \\
\bottomrule
\end{tabularx}
\end{table}

Memory systems should not treat long-term memory as a generic context store. Memory items should carry lineage, validity, trust, and downstream-use metadata, especially when they affect tool calls, personalization, or long-horizon decisions. Memory poisoning and memory-injection attacks highlight the need to trace how compromised memory influences later executions~\citep{chen2024agentpoison,dong2025memoryinjection}. Multi-agent systems require a parallel form of message-level provenance. In workflows built from delegation, critique, verification, and role specialization, provenance should identify not only which final answer was wrong, but which agent introduced the unsupported claim, which agent failed to verify it, and which message propagated it~\citep{wu2023autogen,li2023camel,cemri2025mast,zhang2025agentracer,kong2025aegis}.

\subsection{Runtime Safety and Recovery Requirements}
\label{app:runtime-safety-recovery}

Runtime provenance should support both enforcement and repair. Table~\ref{tab:runtime-recovery-requirements} summarizes the distinction.

\begin{table}[t]
\centering
\footnotesize
\setlength{\tabcolsep}{3.2pt}
\renewcommand{\arraystretch}{1.08}
\caption{Runtime safety and recovery requirements for provenance-aware agents.}
\label{tab:runtime-recovery-requirements}
\rowcolors{2}{covRow}{white}
\begin{tabularx}{\linewidth}{p{2.7cm} X X}
\toprule
\textbf{Capability} & \textbf{Provenance requirement} & \textbf{Representative foundations} \\
\midrule
\blue{Prompt-injection and tool-risk detection} 
& Identify whether untrusted external content influences privileged instructions, tool calls, or memory updates. 
& InjecAgent, AgentDojo, ToolEmu \citep{zhan2024injecagent,debenedetti2024agentdojo,ruan2024toolemu} \\

Information-flow control 
& Track confidentiality and integrity labels across inputs, tools, memory, and outputs. 
& CaMeL, Fides~\citep{debenedetti2025camel,costa2025fides} \\

Semantic taint tracking 
& Trace influence through summarization, paraphrasing, reasoning, and memory consolidation. 
& NeuroTaint~\citep{cai2026neurotaint} \\

Argument verification 
& Check whether sensitive tool arguments derive from trusted sources or explicit user intent. 
& Agent-Sentry~\citep{sequeira2026agentsentry} \\

Execution bounding 
& Enforce tool permissions, access boundaries, and policy constraints at runtime. 
& AgentSpec, AgentBound~\citep{wang2025agentspec,buhler2025agentbound} \\

Recovery 
& Invalidate stale memory, quarantine contaminated evidence, retry tool calls, request approval, roll back state changes, or replay from safe checkpoints. 
& Open problem building on provenance enforcement and trace debugging. \\
\bottomrule
\end{tabularx}
\end{table}

Most current systems focus on detection, blocking, or policy enforcement. A stronger provenance-aware runtime should support intervention. For example, if a tool argument is derived from untrusted web content, the system should not only block the action but also identify the contaminated source, remove the derived parameter, request confirmation, and prevent the same contamination from being written into memory. If a memory item is later contradicted by new evidence, the system should invalidate the memory and identify downstream claims or actions that depended on it. These recovery operations require provenance graphs that preserve dependencies across evidence, tools, memory, state changes, and actions.

\subsection{Benchmark and Governance Requirements}
\label{app:benchmark-governance}

Future benchmarks should evaluate provenance as a first-class object rather than as an implicit by-product of task success. Table~\ref{tab:benchmark-governance-requirements} summarizes the main benchmark and governance requirements.

\begin{table}[t]
\centering
\footnotesize
\setlength{\tabcolsep}{3.2pt}
\renewcommand{\arraystretch}{1.08}
\caption{Requirements for realistic provenance benchmarks and governance.}
\label{tab:benchmark-governance-requirements}
\rowcolors{2}{covRow}{white}
\begin{tabularx}{\linewidth}{p{3.0cm} X X}
\toprule
\textbf{Requirement} & \textbf{Current limitation} & \textbf{Future benchmark target} \\
\midrule
Full execution traces 
& Existing benchmarks often capture only answers, tool calls, attacks, memory retrieval, or failure labels. 
& Capture retrieval, tools, memory, environment states, inter-agent messages, claims, actions, and state changes. \\

Relation annotations 
& Provenance relations are rarely labeled directly. 
& Annotate the nine typed relations, and separately flag unsafe influence and recovery intervention, which are properties of an execution rather than relations. \\

Cross-component coverage 
& RAG, tool-use, memory, and multi-agent benchmarks usually evaluate isolated capabilities. 
& Evaluate how evidence, tools, memory, messages, and state changes jointly affect final outputs and actions. \\

Recovery evaluation 
& Benchmarks usually test success, failure, hallucination, or unsafe action. 
& Test whether agents can repair traces by invalidating memory, quarantining evidence, retrying tools, requesting approval, or rolling back unsafe actions. \\

Privacy-aware tracing 
& Traces may expose confidential documents, credentials, API arguments, memory items, and internal reasoning artifacts. 
& Require minimization, anonymization, access control, encryption, retention policies, and selective disclosure. \\

Audit and governance 
& Trace collection alone does not guarantee accountability. 
& Evaluate whether traces are inspectable, policy-compliant, comparable, and usable for audit. \\
\bottomrule
\end{tabularx}
\end{table}

Existing benchmark families provide partial foundations. RAG and attribution benchmarks support evidence grounding; tool-use and safety benchmarks expose tool calls, unsafe actions, and prompt-injection risks; memory benchmarks evaluate recall and personalization; multi-agent benchmarks evaluate coordination and failure modes; and trace-debugging benchmarks support failure localization~\citep{gao2023alce,ru2024ragchecker,wu2024sourcecheckup,liu2024agentbench,zhou2024webarena,ruan2024toolemu,debenedetti2024agentdojo,yao2025taubench,vijayvargiya2026openagentsafety,zong2025mcpsafetybench,maharana2024locomo,he2026memoryarena,deshpande2025trail,cemri2025mast,zhang2025agentracer,kong2025aegis}. The remaining gap is end-to-end evaluation of provenance chains that connect evidence and tool outputs to memory updates, intermediate claims, final answers, external state changes, and recovery actions.

Privacy and governance should be evaluated together with provenance quality. Provenance traces can contain sensitive user information, confidential documents, credentials, API arguments, internal reasoning artifacts, and long-term memory items. Observability systems such as AgentOps and AgentTrace provide foundations for trace collection~\citep{dong2024agentops,alsayyad2026agenttrace}, while Fides highlights confidentiality and integrity constraints~\citep{costa2025fides}. Future systems should additionally support trace minimization, access control, anonymization, retention policies, encryption, and selective disclosure so that provenance infrastructure does not itself become a privacy or compliance risk.

\end{document}